\documentclass[12pt]{article} 

\usepackage[latin1]{inputenc}
\usepackage{tabularx}
\usepackage[dvips]{graphicx}
\usepackage{amssymb}
\usepackage{amsmath}

\newcommand{\hepth}[1]{{\tt hep-th/#1}}

\newcommand{\wh}{\widehat}
\newcommand{\Ric}{\mathrm{R\!\!\ i\!\!\; c}}
\def\eps{\epsilon}
\def\p{\partial}

\newcommand{\be}{\begin{equation}}
\newcommand{\ee}{\end{equation}}
\newcommand{\bea}{\begin{eqnarray}}
\newcommand{\eea}{\end{eqnarray}}

\begin{document}

\begin{titlepage}

\begin{flushright}
UUITP-03/03\\
hep-th/0303202
\end{flushright}

\vspace{1cm}

\begin{center}
{\huge\bf Scalar Field Corrections\\[5mm]
to AdS$_4$ Gravity from \\[7mm]
Higher Spin Gauge Theory}
\end{center}
\vspace{5mm}

\begin{center}

{\large Fredric Kristiansson and  
Peter Rajan}

\vspace{5mm}

Institutionen f\"or Teoretisk Fysik, Box 803, SE-751 08 
Uppsala, Sweden

\vspace{5mm}

{\tt
fredric@teorfys.uu.se, peter.rajan@teorfys.uu.se}

\end{center}

\vspace{5mm}

\begin{center}
{\large \bf Abstract}
\end{center}
\noindent

We compute the complete contribution to the stress-energy
tensor in the minimal bosonic higher spin theory in $D=4$ that is
quadratic in the scalar field. We find arbitrarily high derivative
terms, and that the total sign of the stress-energy tensor depends on
the parity of the scalar field.

\vfill
\begin{flushleft}
March 2003
\end{flushleft}
\end{titlepage}

\newpage 

\section{Introduction}

By now there is plenty of evidence for the remarkable correspondence
between field theories in $d$ dimensions and string/M theory on
AdS$_{d+1}\times M$ spacetimes. In the conformal cases with maximal
number of supersymmetries the correspondence relates the low-energy
limits of two complementary descriptions of the sector of the theory
with $N$ units of D$3$-brane or M$2/5$-brane charge \cite{malda97}. To
be more precise, the correspondence relates the $1/N$ expansions of
the generating functionals on both sides. This is a remarkable
relation, in the sense that on the CFT side $N$ is, roughly, the
number of colours of the sigma-model living on the stack of branes,
while on the bulk side $1/N$ plays the role of Planck's constant. A
crucial property of the $1/N$ expansion of the CFT is that its
correlators factorise in the limit $N \rightarrow \infty$, such that
it makes sense to identify the connected part of the correlators with
the connected Feynman diagrams of the bulk theory with external
bulk-to-boundary propagators.

In fact, starting from any CFT in $d$ dimensions that factorises in
some limit and that has a well-defined generating functional (which is
a quite non-trivial condition), it should be possible to reconstruct
an effective action in $d+1$ dimensions which has an AdS vacuum and
which reproduces the correlators as described above. Moreover, within
this context there is a correspondence between the global, continuous
symmetries of the CFT and the local symmetries of the bulk theory. In
particular, this implies that the bulk theory necessarily contains
gravity. It is then rather gratifying from a string theorist's point
of view that AdS/CFT correspondence arises naturally within string/M
theory.

The most studied example relates $d=4$, $\mathcal{N}=4$ Yang-Mills
theory with SU$(N)$ gauge group to the Type IIB string theory on
AdS$_5\times S^5$ of radius $R$ with string coupling $g_s$, string
tension $T_s$ and $N$ units of five-form flux, which is an exact
solution to the string theory provided that $R^2T_s = \sqrt{4\pi g_s
N}$. The bulk string theory is notoriously difficult to quantise
starting from the worldsheet formulation. This correspondence has
therefore been tested primarily by comparing the weak coupling limit
of the worldsheet theory, i.e. the supergravity limit, to strong
coupling results in SYM obtained either by studying correlators
protected by some symmetries or by summing up the four-dimensional
perturbation series for weak 't Hooft coupling $\lambda=Ng_{\rm YM}^2
\ll 1$ and continuing the result to $\lambda\sim R^4T_s^2 \gg 1$.

However, as argued above, it should be possible to test the
correspondence directly order by order in the $1/N$ expansion on both
sides. In particular, it is interesting to consider the free limit
$\lambda\rightarrow 0$ of the SYM theory.  The generating functional
of composite SU$(N)$ invariant operators remains highly non-trivial in
this limit and has two remarkable properties: (i) it admits a
consistent truncation to the generating functional of bilinear
operators (which is important since there is no mass-gap); (ii) the
primary bilinear operators are superfields which contain conserved
currents of arbitrarily high spin. Hence, the free limit of SYM
corresponds to a limit of the Type IIB theory in which it develops
higher spin gauge symmetry and admits a consistent truncation to the
massless sector. These features should be sufficient to determine the
effective five-dimensional bulk action in the massless sector up to
some number of interaction ambiguities.  The first steps towards this
have been taken in \cite{sundb00,sezsun01,vasil01:1,witt60,mikh02},
where the precise higher spin algebra, the massless spectrum and
linearised field equations have been given, and in \cite{vasil01:2}
where certain cubic interactions have been constructed.

Massless higher spin theories are further developed in four
dimensions, where the full field equations are known
\cite{fradvas87,vasil96,sezsun_anal,engsezsun02}. The unbroken phase
of the higher spin gauge theory corresponds to the generating
functional of bilinear operators in free singleton field theory. By
various deformations which breaks the higher spin symmetry while
preserving some colour symmetry group one can flow to strongly coupled
interacting fixed points with some $1/N$ expansion. In the case of 32
supersymmetries in the bulk, by considering SU($N$) colour symmetry
and extending the bulk theory with massive fields and the generating
functional of the field theory with multi-linear single trace
operators the theory has been conjectured to flow to the IR fixed
point of the Yang-Mills theory dual to 11 dimensional
M-theory/supergravity in the bulk. This flow is analogous to the
$\lambda \rightarrow \infty$ in the Type IIB case.

Another interesting deformation, which preserves maximal O($N$) colour
symmetry, is generated by perturbing the free non-supersymmetric
scalar singleton theory with a double trace operator $fJ^2$, where
$J=\varphi^a\varphi^a$ is the scalar operator and $f$ is a parameter
of dimension energy, which results in a flow to the interacting fixed
point of the O($N$) model (in the limit of large $f$ in units of some
fixed length \cite{gukl02}). It was recently proposed in \cite{klpo02}
that this flow has a dual description in terms of the minimal bosonic
higher spin gauge theory in four dimensions (with parity even scalar
field), such that the two fixed points correspond to the $\Delta_\pm$
boundary conditions of the scalar field. This proposal has been
investigated further in \cite{gipoza02,pet03}.

The simplest way of testing the correspondence is to compare free
field theory correlation functions with the bulk amplitudes of the
higher spin gauge theory in the vacuum with $\Delta_-$ boundary
condition \cite{vasil01:1,sezsun98,sezsun02_hol}.  The closed form of
the full field equations involves many auxiliary fields. The physical
field equations can be obtained by eliminating the auxiliary fields
order by order in a curvature expansion scheme. No action reproducing
any of these forms of the field equations is known, however. In this
paper we take the first steps towards finding the cubic action for a
subset of the physical fields.

As we will see in the introductory section 2, the minimal bosonic
higher spin theory in AdS$_4$ is based on an algebra which is a
minimal extension of the AdS$_4$ algebra, here called $hs(4)$. A
unitary irreducible representation of $hs(4)$ can be constructed from
the symmetric product of two spin 0 singletons, and consists of
massless fields of spins $s=0,2,4,\ldots$. The scalar, which we denote
by $\phi$, is even under parity and its ground state has AdS energy
$E_0=1$. This model can be obtained by truncating various
supersymmetric theories \cite{sezsun01,engsezsun02}.

In this paper we expand the minimal bosonic higher spin gauge theory
by treating the scalar field $\phi$ and the higher spin fields as well
as all curvatures as weak fields, and calculate contributions to the
Einstein equation from the scalar $\phi$.  In the leading order $s=2$
field equation coincides with the Einstein equation with a
cosmological constant.  The first non-trivial corrections from the
scalar field are contributions to the stress-energy tensor according
to
\begin{eqnarray}
\Ric_{\mu\nu} - \frac12 R g_{\mu\nu} + \Lambda g_{\mu\nu} & = &
\mbox{Re}\{b_1^2\} \bigg[\sum_k \frac{2^k}{(k!)^2}
\bigg( \xi(k) \, g_{\mu\nu} \nabla_{\rho\{k+1\}}  \, \phi \  
\nabla^{\rho\{k+1\}} \, \phi + \nonumber \\
& & \qquad  
+ \ \eta(k) \nabla_{\rho\{k\}\mu}  \, \phi \  
\nabla^{\rho\{k\}}{}_\nu \, \phi  + \nonumber \\
& & \qquad 
+ \ \zeta(k) \nabla_{\rho\{k\}\mu\nu}  \, \phi \  
\nabla^{\rho\{k\}} \, \phi \bigg)
\ - \ \frac49 \,g_{\mu\nu} \,\phi\,\phi \bigg], \label{eq:smakprov}
\end{eqnarray}
where 
\begin{equation}
\nabla_{\mu\{k\}}\equiv \nabla_{(\mu_1} \ldots \nabla_{\mu_k)} -
\mbox{traces,} 
\qquad \nabla_{\mu\{k\}\nu}\equiv 
\nabla_{(\mu_1} \ldots \nabla_{\mu_k}\nabla_{\nu)}  -
\mbox{traces,} 
\label{eq:traceless}
\end{equation}
and $b_1$ is a complex constant which enters the closed form of the
field equations. The functions $\xi$, $\eta$ and $\zeta$ are shown
explicitly in eqs. (\ref{eq:xi})--(\ref{eq:zeta}). There are two
qualitatively interesting properties, namely the higher derivative
nature of the stress energy tensor, and the potential sign ambiguity
in $\mbox{Re}\{b_1^2\}$, which we shall discuss further in the last
section. We shall also discuss what further computations needs to be
done in order to find the cubic action in the spin $s=0,2$ sector.

The papers is organised as follows. In section 2 we briefly review the
higher spin theory in AdS$_4$.  In section 3 we use the weak field
expansion scheme to work out the second order scalar corrections to
the spin $s=2$ field equation. Section 4 concludes with a summary and
a discussion about the relevance of the result. Details on the actual
calculation can be found in the Appendices. We have made an effort to
present the calculation in such a way that the interested reader may
follow it step by step.

%=============================================================

\section{Higher spin formalism in $D=4$} \label{prelsec}

Below we review the framework of higher spin theory in $D=4$. It is
instructive to first examine pure gravity, then extending to the field
content of the minimal bosonic higher spin gauge theory, which in
particular contains the scalar field whose stress energy tensor will
be examined in the next section.

%----------------------------------------------------------

\subsection{Pure AdS$_4$ Gravity}

We start by formulating gravity with a negative cosmological constant
as a constrained system of 0-forms and 1-forms based on the AdS
algebra SO(3,2),
\begin{eqnarray}
\left[M_{ab},M_{cd}\right] & = & i\eta_{bc} M_{ad} - i\eta_{bd} M_{ac}
+ i\eta_{ad} M_{bc}  - i\eta_{ac} M_{bd}, \\ 
\left[M_{ab},P_c\right] & = & i\eta_{bc} P_a - i\eta_{ac} P_b, \\ 
\left[P_a,P_b\right] & = & i M_{ab}.
\end{eqnarray}
Out of the generators above we can
define an SO(3,2)-valued connection 1-form $E$ as
\begin{equation}
E \equiv  -i \bigg(e^a P_a \, + \, \frac12 \, \omega^{ab}
M_{ab}\bigg) \equiv e + \omega = -E^\dagger.
\label{eq:E}
\end{equation}
where $e^a$ is the vierbein and $\omega^{ab}$ is the Lorentz
connection. The factor of $-i$ is inserted for later convenience. The
field strength of $E$ becomes
\begin{equation}
\mathcal{R} \equiv dE + E \wedge E = \mathcal{R}^a P_a 
+ \frac12 \mathcal{R}^{ab} M_{ab},
\end{equation}
where 
\begin{eqnarray}
\mathcal{R}^a & = & -i ( de^a + \omega^a{}_c  \wedge e^c) \ = \ 
-i T^a, \\
\mathcal{R}^{ab} & = & -i ( d\omega^{ab} +
\omega^{ac} \wedge \omega_c{}^b + 
e^a \wedge e^b ) \ = \ 
-i ( R^{ab} + e^a \wedge e^b ). \label{eq:krullR}
\end{eqnarray}
where $T^a$ is the torsion and $R^{ab}$ the Riemann tensor. The
equations for gravity follow from the following curvature constraints:
\begin{eqnarray}
\mathcal{R}^a & = & 0, \label{eq:torsion_zero} \\
\mathcal{R}^{ab} & = & -i \, e_c \wedge e_d \phi^{abcd}.  
\label{eq:riemann=weyl}
\end{eqnarray}
where $\phi^{abcd}$ is the Weyl tensor, which belongs to the 
\begin{picture}(4,3)
\put(1,0){\line(1,0){2}}
\put(1,1){\line(1,0){2}}
\put(1,2){\line(1,0){2}}
\put(1,0){\line(0,1){2}}
\put(2,0){\line(0,1){2}}
\put(3,0){\line(0,1){2}}
\end{picture} representation of SO(3,1). The Weyl tensor describes 
the traceless part of $R_{\mu\nu}{}^{ab}$. Eq.~(\ref{eq:torsion_zero})
fixes the Lorentz connection $\omega^{ab}$ in terms of the vierbein
$e^a$ and the trace of (\ref{eq:riemann=weyl}) gives the Einstein
equation,
\begin{equation}
\Ric_\mu{}^a - \frac12 R e_\mu{}^a + \Lambda e_\mu{}^a = 0,
\end{equation}
with a negative cosmological constant $\Lambda=-3$.

From the curvature identities we
conclude that the Weyl tensor must obey
\begin{eqnarray}
\nabla \phi^{abcd} & = & e_f \phi^{abcdf}, 
\label{eq:Weyl1} \\
\nabla \phi^{abcdf} & = & e_g \phi^{abcdfg} + 
4 e^{f}
\phi^{abcd} 
+ 5 e_{g}\phi^{abch}\phi_{h}{}^{dfg} \label{eq:Weyl2} \\
& \vdots & \nonumber
\end{eqnarray}
where $\phi^{abcdf}$ is 
\begin{picture}(5,3)
\put(1,0){\line(1,0){2}}
\put(1,1){\line(1,0){3}}
\put(1,2){\line(1,0){3}}
\put(1,0){\line(0,1){2}}
\put(2,0){\line(0,1){2}}
\put(3,0){\line(0,1){2}}
\put(4,1){\line(0,1){1}}
\end{picture},
$\phi^{abcdfg}$ is 
\begin{picture}(6,3)
\put(1,0){\line(1,0){2}}
\put(1,1){\line(1,0){4}}
\put(1,2){\line(1,0){4}}
\put(1,0){\line(0,1){2}}
\put(2,0){\line(0,1){2}}
\put(3,0){\line(0,1){2}}
\put(4,1){\line(0,1){1}}
\put(5,1){\line(0,1){1}}
\end{picture}, and so on.
The Bianchi identity $\nabla R_{ab} = 0$ holds
since (\ref{eq:Weyl1}) implies that $\nabla^f \phi^{abcd}$ is 
\begin{picture}(5,3)
\put(1,0){\line(1,0){2}}
\put(1,1){\line(1,0){3}}
\put(1,2){\line(1,0){3}}
\put(1,0){\line(0,1){2}}
\put(2,0){\line(0,1){2}}
\put(3,0){\line(0,1){2}}
\put(4,1){\line(0,1){1}}
\end{picture}. The integrability of (\ref{eq:Weyl1}) implies
(\ref{eq:Weyl2}) and so forth.  Eqs.~(\ref{eq:torsion_zero}) and
(\ref{eq:riemann=weyl}) together with (\ref{eq:Weyl1}) and
(\ref{eq:Weyl2}) are invariant under SO(3,2) gauge transformations,
under which $e^a$ and $\omega^{ab}$ transforms as gauge fields in the
adjoint representation. The Weyl tensor $\phi^{abcd}$ and all its
derivatives form another, infinite dimensional representation, which
generalises from spin two to arbitrary spin, as we shall discuss
below.

%----------------------------------------------------------

\subsection{Generalisation to higher spin gauge theory}

The framework outlined in the previous paragraph is suitable for
formulating gauge theories based on higher spin extensions of SO(3,2)
\cite{vasil96}. To facilitate these constructions one introduces
Grassmann even oscillators $y_\alpha$ and $\bar
y_{\dot\alpha}=(y_\alpha)^\dagger$, which are Weyl spinors and obey
the following algebra,
\begin{eqnarray}
y_\alpha \star y_\beta  =  y_\alpha y_\beta + i\epsilon_{\alpha\beta},
\quad & & \bar y_{\dot\alpha} \star \bar y_{\dot\beta}  =   
\bar y_{\dot\alpha} \bar y_{\dot\beta} +
i\epsilon_{\dot\alpha\dot\beta} \nonumber \\
y_\alpha \star \bar y_{\dot\beta} =  y_\alpha \bar y_{\dot\beta} \quad & & 
\bar y_{\dot\alpha} \star y_\beta = \bar y_{\dot\alpha} y_\beta,
\label{eq:y_comm}
\end{eqnarray}
where $\star$ denotes the associative product of oscillators and the
products on the right hand sides, written without stars, are Weyl
ordered (our spinor conventions are given in Appendix A). This extends
to general Weyl ordered functions as
\begin{equation}
f(y,\bar y) \star g(y,\bar y) = 
f(y,\bar y) e^{-i(\stackrel{\, _\leftarrow}{\partial}{\!}^\alpha   
\!\!\! \stackrel{\, _\rightarrow}{\partial}{\!}_\alpha 
+ \stackrel{\, _\leftarrow}{\partial}{\!}^{\dot\alpha}  
\!\!\! \stackrel{\, _\rightarrow}{\partial}{\!}_{\dot\alpha})} 
g(y,\bar y), \label{eq:fstarg}
\end{equation}
where $\partial_\alpha \equiv \partial/\partial y^\alpha$ and 
$\partial^\alpha \equiv \epsilon^{\alpha\beta} \partial_\beta = 
-\partial/\partial y_\alpha$. 
We can now realise the SO(3,2) algebra by writing
\begin{eqnarray}
M_{ab} & = & -\frac18 (\sigma_{ab})^{\alpha\beta} y_\alpha y_\beta 
         -\frac18 (\sigma_{ab})^{\dot\alpha\dot\beta} 
            \bar y_{\dot\alpha} \bar y_{\dot\beta} \\
P_{a} & = & \frac 14 (\sigma_a)^{\alpha\dot\alpha} 
            y_\alpha \bar y_{\dot\alpha},
\end{eqnarray}
and using the $\star$ product in the commutators. As a result the
SO(3,2) connection $E$ defined in (\ref{eq:E}) reads
\begin{equation}
E = \frac{i}2 \bigg[ e^{\alpha\dot\alpha} y_\alpha \bar y_{\dot\alpha} +
   \frac12 \Big( \omega^{\alpha\beta} y_\alpha y_\beta +
 \omega^{\dot\alpha\dot\beta} 
        \bar y_{\dot\alpha}\bar y_{\dot\beta} \Big) \bigg],
\label{eq:usefull2}
\end{equation}
where $e^{\alpha\dot\alpha}$ and $\omega^{\alpha\beta}$ are related to
$e^a$ and $\omega^{ab}$ via the relations (\ref{eq:vect_spin1}) and
(\ref{eq:vect_spin2}). Useful relations are
\begin{equation}
e_\mu=-\frac{i}4
(\sigma_\mu)^{\alpha\dot\alpha} y_\alpha \bar y_{\dot\alpha}, \qquad
e_\mu{}^{\alpha\dot\alpha} = -\frac12 (\sigma_\mu)^{\alpha\dot\alpha}.
 \label{eq:usefull}
\end{equation}
The set of arbitrary polynomials $U(y,\bar y)$ which obey projection
and reality conditions according to
\begin{equation}
\tau\{U(y,\bar y)\} \equiv U(iy,i\bar y) = -U(y,\bar y), \qquad 
U(y,\bar y)^\dagger = -U(y,\bar y), \label{eq:Aproj}
\end{equation}
form a Lie algebra with respect to the commutator $[U,V]_\star$,
denoted by $hs(4)$ in \cite{sezsun_anal}. The algebra closes since
$\tau(U \star V) = \tau(V) \star \tau(U)$, which follows from
(\ref{eq:fstarg}), and $(U \star V)^\dagger = V^\dagger \star
U^\dagger$. The $\tau$-projection restricts the polynomials to sums of
monomials of degree $2+4\ell$, $\ell=0,1,2\ldots$, containing
generators with spin $s=1+2\ell$. In order to gauge $hs(4)$, one
introduces an $hs(4)$-valued connection $A=A_\mu(x;y,\bar y) dx^\mu$
defined by
\begin{eqnarray}
A_\mu(x;y,\bar y) & = & \frac{i}2 \sum_{
\begin{array}{c} \\[-7mm] \scriptstyle m, n, \ell \,\ge \, 0 \\[-2mm]
                 \scriptstyle m+n=2+4\ell \end{array}} 
  \frac1{m!n!} 
  A_{\mu\,\alpha_1 \ldots \alpha_m \dot\alpha_1 \ldots \dot\alpha_n}(x)
  y^{\alpha_1} \ldots y^{\alpha_m} 
  \bar y^{\alpha_1} \ldots \bar y^{\dot\alpha_n}.
  \label{eq:A_expansion} \\
 & = & e + \omega + W,
\end{eqnarray} 
where $W$ contains the higher spin gauge fields. It is convenient to
define the SO(3,2) covariant field strength
\begin{equation}
\mathcal{F} \equiv dW + \{E,W\}_\star = \nabla W + \{e,W\}_\star.
\end{equation}
The higher spin algebra $hs(4)$ has a unitary irreducible
representation containing massless fields in AdS$_4$ with spins
$s=0,2,4, \ldots$, of which the spin $s\geq 2$ sector is realised as
the physical degrees of freedom in $A$.  In order to accommodate the
physical scalar field one introduces a 0-form $\Phi(x;y,\bar y)$ in
the quasi-adjoint representation, defined by
\begin{equation}
\tau\{\Phi(y,\bar y\} = \bar \pi\{\Phi(y,\bar y)\} \equiv 
\Phi(y,-\bar y), \qquad 
\Phi(y,\bar y)^\dagger = \pi \{\Phi(y,\bar y)\} \equiv \Phi(-y,\bar y),
\label{eq:Phiproj}
\end{equation}
and so has the structure
\begin{equation}
\Phi(x;y,\bar y) = \sum_{
\begin{array}{c} \\[-7mm] \scriptstyle \ell+1, m, n  \, \ge \, 0 \\[-2mm]
                 \scriptstyle |m-n|=4(\ell+1) \end{array}} 
  \frac1{m!n!} 
  \Phi_{\alpha_1 \ldots \alpha_m \dot\alpha_1 \ldots \dot\alpha_n}(x)
  y^{\alpha_1} \ldots y^{\alpha_m} 
  \bar y^{\alpha_1} \ldots \bar y^{\dot\alpha_n}.
\end{equation}
The SO(3,2) covariant derivative of $\Phi$ is defined as
\begin{equation}
\mathcal{D}\Phi \equiv d\Phi + E\star\Phi - \Phi\star\pi\{E\} = \nabla
\Phi + \{e,\Phi\}_\star.
\end{equation}
As we will see below, the level $\ell=0$ components are the Weyl
tensor and its derivatives, which we introduced in the previous
section, and at level $\ell>0$ reside higher spin generalisations
thereof. The physical scalar and its derivatives are contained at
level $\ell=-1$.

Assuming that $\Phi$ and $W$ are weak fields, the constraints on $A$
and $\Phi$ leading to spacetime dynamics have the following expansion
\cite{vasil96,sezsun_anal,vasil99}
\begin{eqnarray}
\mathcal{R}_{\mu\nu}  +  \mathcal{F}_{\mu\nu}  & = &
-2 W_{[\mu} \star W_{\nu]} - i\big[R_{\mu\nu}{}^{\alpha\beta} \wh A_\alpha \star \wh A_\beta
+ \mbox{h.c.} \big]_{z=0} \nonumber \\
& & \qquad - \ 2 \sum_{n=1}^{\infty}\sum_{j=0}^{n} \Big(
(\wh e+\wh W)_{[\mu}^{(j)} \star (\wh e+\wh W)_{\nu]}^{(n-j)} 
\Big)_{z=0}\  , 
\label{eq:F} \\
\nabla_\mu \Phi + \{e_\mu,\Phi\}_\star & = & 
\Phi\star\bar\pi\{W_\mu\} - W_\mu \star \Phi \nonumber \\
& + & \sum_{n=2}^{\infty}\sum_{j=1}^{n} \Big( 
\wh \Phi^{(j)}\star\bar\pi\{ (\wh e+\wh W)_\mu^{(n-j)}\} -
(\wh e+\wh W)_\mu^{(n-j)} \star 
\wh \Phi^{(j)}\Big)_{z=0} \label{eq:DPhi}
\end{eqnarray}
Here hatted quantities depend on $y,\bar y$ as well as an auxiliary
set of oscillators $z_\alpha, \bar z_{\dot\alpha}$ obeying the
following algebra
\begin{equation}
\wh f \star \wh g = 
\wh f e^{i(\stackrel{\, _\leftarrow}{\partial}{\!}_+^\alpha  
\!\!\! \stackrel{\, _\rightarrow}{\partial}{\!}^-_\alpha
+ \stackrel{\, _\leftarrow}{\partial}{\!}_-^{\dot\alpha}  
\!\!\! \stackrel{\, _\rightarrow}{\partial}{\!}^+_{\dot\alpha})} 
\wh g, \label{eq:hfstarhg}
\end{equation}
where $\partial_\pm = \partial^\pm = \partial_z \pm \partial_y$ and
$\wh f = \wh f(y,\bar y,z,\bar z)$ and $\wh g = \wh g(y,\bar y,z,\bar
z)$ are Weyl ordered functions (see Appendix \ref{sect:star_apdx}),
and the expansions $\wh e_\mu=\sum_{j=0}^{\infty} \wh e_\mu^{(j)}$ and
$\wh W_\mu=\sum_{j=0}^{\infty} \wh W_\mu^{(j)}$ are given by
\begin{equation}
\wh e_\mu = \frac1{1+\wh L^{(1)}+\wh L^{(2)}+\ldots} 
e_\mu, \qquad \mbox{and} \qquad 
\wh W_\mu = \frac1{1+\wh L^{(1)}+\wh L^{(2)}+\ldots} 
W_\mu,
\end{equation}
where
\begin{equation}
\wh L^{(n)}(\wh f) = i \int_0^1 \frac{dt}{t} \bigg( 
\wh A_\alpha^{(n)} \star \partial^\alpha_- \wh f + 
\partial^\alpha_+ \wh f \star \wh A_\alpha^{(n)}
+ \wh A_{\dot\alpha}^{(n)} \star \partial^{\dot\alpha}_+ \wh f +
\partial^{\dot\alpha}_- 
\wh f \star \wh A_{\dot\alpha}^{(n)} \bigg)_{z\rightarrow tz}.
\end{equation}
The quantities $\wh A_\alpha^{(n)}$ and $\wh \Phi^{(n)}$ are given by
\begin{eqnarray}
\wh A^{(0)}_\alpha & = & 0 \\
\wh \Phi^{(1)} & = & \Phi(y,\bar y) \\
\wh A^{(1)}_\alpha & = & - \frac{ib_1}2 z_\alpha \int_0^1 tdt
\Phi(-tz,\bar y) \kappa(tz,y) \\
\wh \Phi^{(n)} & = & z^\alpha \sum_{j=1}^{n-1} \int_0^1 dt
\Big(\wh \Phi^{(j)} \star \bar\pi\{ \wh A_\alpha^{(n-j)}\} -
\wh A_\alpha^{(n-j)} \star \wh \Phi^{(j)}\Big)_{z\rightarrow tz} + 
\nonumber \\
& & \qquad  + \ \bar z^{\dot\alpha} \sum_{j=1}^{n-1} \int_0^1 dt
\Big(\wh \Phi^{(j)} \star \pi\{ \wh A_{\dot\alpha}^{(n-j)}\} -
\wh A_{\dot\alpha}^{(n-j)} \star \wh \Phi^{(j)}\Big)_{z\rightarrow tz}
\\
\wh A^{(n)}_\alpha & = & z_\alpha \int_0^1 tdt \bigg( 
-\frac{i}2 \mathcal{V}^{(n)}(\wh \Phi \star \kappa ) +
\sum_{j=1}^{n-1} \wh A^{(j)\beta} \star \wh A^{(n-j)}_\beta 
\bigg)_{z\rightarrow tz} + \nonumber \\
& & \qquad + \ \bar z^{\dot\beta} \sum_{j=1}^{n-1} \int_0^1 tdt
\Big[\wh A^{(j)}_\alpha, \wh A^{(n-j)}_{\dot\beta} 
\Big]_{\star, z\rightarrow tz} \label{eq:A_alpha^(n)}
\end{eqnarray} 
and $\wh A_{\dot\alpha}^{(n)}=-(\wh A_\alpha^{(n)})^\dagger$.  The
function $\mathcal{V}(\wh \Phi \star \kappa)$, where
$\kappa(y,z)=\exp(iy^\alpha z_\alpha)$, has to be odd. Already the
simplest choice of $\mathcal{V}$, namely a linear function, leads to
highly nontrivial interactions in the right hand sides of (\ref{eq:F})
and (\ref{eq:DPhi}). Adding a $(2n+1)^{\mathrm{th}}$ order term to
$\mathcal{V}$ leads to modifications of the interactions starting at
the $(2n+1)^{\mathrm{th}}$ order. Whether these are genuine
interaction ambiguities, or can be removed by field redefinitions is
not known. In the former case we expect the ambiguity to be determined
once the theory is compared with its holographic dual, or some more
fundamental formulation of the theory in the bulk. Since we will focus
on quadratic contributions in the scalar field we can assume that
$\mathcal{V}(\wh \Phi \star \kappa)= b_1 \wh \Phi \star \kappa$, where
$b_1$ is the first order expansion coefficient.

The constraints (\ref{eq:F}) and (\ref{eq:DPhi}) follow from solving
an extended set of constraints on $\wh \Phi$ and
\begin{equation}
\wh A = \big(\wh A_\mu + \big[
i\omega_\mu^{\alpha\beta} \wh A_\alpha \star \wh A_\beta
- \mbox{h.c.} \big] \big)dx^\mu + \wh A_\alpha dz^\alpha + 
\wh A_{\dot\alpha} dz^{\dot\alpha}, \label{eq:A}
\end{equation}
which are forms living on spacetime times an internal manifold for
which $z,\bar z$ are coordinates. Besides being consistent with the
$\tau$ and reality conditions, the basic property of (\ref{eq:F}) and
(\ref{eq:DPhi}) is that they are integrable order by order in the weak
field expansion. Note that this ensures invariance under higher spin
gauge transformations and diffeomorphisms (which are incorporated into
the gauge group as field dependent gauge transformations).  The
rationale behind the expansion of the $\mu$ component in (\ref{eq:A})
is that it implies that the constraints are manifestly invariant under
local Lorentz transformations under which the component fields in $e$
and  $W$ transform as Lorentz tensors and $\omega$ as the Lorentz
connection \cite{vasil99,sezsun_anal}.

The linearised form of (\ref{eq:F}) contains the physical field
equations for spin $s=2,4,\ldots$ and algebraic equations for the
auxiliary gauge fields \cite{vasil96} and
$\Phi_{\alpha_1\ldots\alpha_{2s}}$, which are the Weyl tensor and its
higher spin generalisations. The linearised form of (\ref{eq:DPhi})
reads $\nabla_\mu\Phi + \{e_\mu,\Phi\}=0$ where $e_\mu$ is given by
(\ref{eq:usefull}), which can be written in components as
\begin{equation}
\nabla_\mu \Phi_{\alpha_1\ldots \alpha_m\dot \alpha_1\ldots
\dot\alpha_n} = \frac{i}2 mn (\sigma_\mu)_{\alpha_1\dot\alpha_1} 
\Phi_{\alpha_2\ldots \alpha_m\dot \alpha_2\ldots
\dot\alpha_n} - 
\frac{i}2 (\sigma_\mu)^{\beta\dot\beta} 
\Phi_{\beta\alpha_1\ldots \alpha_m\dot\beta \dot \alpha_1\ldots
\dot\alpha_n}, \label{eq:nablamuphi}
\end{equation}
where separate symmetrisation of the dotted and undotted indices on
the right hand side is assumed.  From (\ref{eq:nablamuphi}) it follows
that
$\Phi_{\alpha_1\ldots\alpha_{2s+k}\dot\alpha_1\ldots\dot\alpha_k}$,
$s=0,2,4,\ldots$, can be expressed in terms of $k$ derivatives of
$\Phi_{\alpha_1\ldots\alpha_{2s}}$. As an example, we see that $
\nabla \Phi_{\alpha_1\ldots\alpha_4} = 2ie^{\beta\dot\beta}
\Phi_{\alpha_1\ldots\alpha_4\beta\dot\beta}$, which is nothing but
equation (\ref{eq:Weyl1}).  For $s=0$ one finds
\begin{equation}
\Phi_{\alpha_1\ldots\alpha_k\dot\alpha_1\ldots\dot\alpha_k} =
(-i)^k (\sigma^{\mu_1})_{\alpha_1\dot\alpha_1}\ldots 
       (\sigma^{\mu_k})_{\alpha_k\dot\alpha_k}
       \nabla_{\mu\{k\}} \phi
\label{eq:deriv_scalar}
\end{equation}
where we use the notation defined in (\ref{eq:traceless}).  Note that
the dotted and undotted indices on the right hand side are
automatically symmetrised since the Lorentz vector indices are
traceless and symmetric.  From the linearised form of (\ref{eq:DPhi})
it also follows that $\phi$ is the physical scalar with the linearised
field equation
\begin{equation}
\nabla^\mu \nabla_\mu \phi = -2 \phi.
\end{equation}
The mass $m^2=-2$ corresponds to a scalar lowest weight state with AdS
energy $E=1$.

By working out the $\star$ products in the above relations, and
solving for the auxiliary fields order by order in the weak field
expansion, it is possible to extract the field equations describing
the full interacting massless higher spin theory to any desired order
by means of straightforward albeit increasingly tedious calculations.

%----------------------------------------------------------

\section{Scalar field terms in the Einstein equation}

We will now consider the expansion scheme to second order and
calculate all quadratic contributions to the Einstein equation from
the scalar field $\phi$ and its derivatives.

One can show that upon linearising the right hand side of (\ref{eq:F})
in the weak fields one obtains (\ref{eq:torsion_zero}) and
(\ref{eq:riemann=weyl}) provided that $b_1=1$. For a general complex
$b_1$ the right hand side of the $\mathcal{R}^{ab}$ constraint is
modified though its structure remains the same.  To the second order,
(\ref{eq:F}) implies that
\begin{eqnarray}
\mathcal{R}^{ab}(\omega,e)  & = & \frac12 (\sigma^{ab})^{\alpha\beta} J_{\alpha\beta} -
\mbox{h.c.} \label{eq:R=J1} \\
\mathcal{R}^{a}(\omega,e)   & = & 
(\sigma^a)^{\alpha\dot\alpha}
J_{\alpha\dot\alpha}, \label{eq:R=J2} 
\end{eqnarray}
where
\begin{equation}
J_{\alpha\beta} = \frac{\partial^2}{\partial y^\alpha \partial
y^\beta}\ J \big|_{y=0} \qquad \mbox{and} \qquad 
J_{\alpha\dot\alpha} = \frac{\partial^2}{\partial y^\alpha \partial
\bar y^{\dot\alpha}}\ J \big|_{y=0},
\end{equation}
and the 2-form $J=J_{\mu\nu}dx^\mu dx^\nu$ is given by
\begin{equation}
J_{\mu\nu} =  2i\big\{
e_{[\mu}{}^{\alpha\dot\alpha} e_{\nu]}{}^\beta{}_{\dot\alpha} 
\widehat A^{(1)}_\alpha
\star \widehat A^{(1)}_\beta +  \mbox{h.c.} \big\}_{z=0}
+ L_{\mu\nu}  +\mbox{$W$-terms}, \label{eq:Jdef}
\end{equation}
where
\begin{eqnarray}
L_{\mu\nu} & = & -\bigg(\Big[\wh L^{(1)}(e_\mu),\wh L^{(1)}(e_\nu)\Big]_\star -
2\Big[e_{[\mu},\wh L^{(2)}(e_{\nu]})\Big]_\star   \nonumber \\
 & &  \qquad \qquad \qquad \qquad \qquad \qquad \qquad 
+ \ 2\Big[e_{[\mu},\wh L^{(1)} \circ \wh L^{(1)}(e_{\nu]})\Big]_\star
\bigg)_{z=0}.  \label{eq:Ldef}
\end{eqnarray}
In (\ref{eq:Jdef}) we have used the background value of $R_{\mu\nu}$
which is given by $R_{\mu\nu}{}^{\alpha\beta} =
-2e_{[\mu}{}^{\alpha\dot\alpha} e_{\nu]}{}^\beta{}_{\dot\alpha}$, as
follows from (\ref{eq:krullR}).  Writing
$\omega^{ab}=\omega^{ab}(e)+\kappa^{ab}$, where $\omega^{ab}(e)$ is
the Levi-Civita connection obeying $\mathcal{R}^{a}(\omega(e),e)=0$,
and $\kappa^{ab}$ is contorsion, we have
\begin{eqnarray}
\mathcal{R}^{ab}(\omega,e) & = & \mathcal{R}^{ab}(\omega(e),e) 
- i \nabla \kappa^{ab} \\
\mathcal{R}^{a}(\omega,e) & = &  
-i e^b \wedge \kappa_b{}^a. \label{eq:C}
\end{eqnarray}
By taking the trace of (\ref{eq:R=J1}) one obtains a field
equation containing the Ricci tensor, 
\begin{eqnarray}
\Ric_{\mu\nu} + 3g_{\mu\nu} & = & 2\nabla_{[\rho} \kappa_{\mu]\nu}{}^{\rho} -
\frac12 \Big\{ (\sigma_\mu)^{\gamma\dot\alpha}(\sigma_\nu)^{\delta\dot\beta}
\Big(\widehat A^{(1)}_{\dot\alpha} \star \widehat A^{(1)}_{\dot\beta} 
\Big)_{\gamma\delta} + \mbox{h.c.} \Big\}_{z=0} + \nonumber \\
& & \qquad + \ \frac{i}2 \Big\{ (\sigma_{\mu}{}^{\rho})_{\alpha\beta}
L_{\nu\rho}{}^{\alpha\beta} - \mbox{h.c.} \Big\}, 
\label{eq:Ricci_contr}
\end{eqnarray}
where symmetrisation of $\mu$ and $\nu$ is understood in each term,
and we have used (\ref{eq:usefull}).  To obtain (\ref{eq:Ricci_contr})
we picked the symmetric part of the trace of eq. (\ref{eq:R=J1}). The
antisymmetric part should be identically satisfied, for the system not
to be overdetermined. We have not attempted to prove the identity.
Upon substituting (\ref{eq:C}) in (\ref{eq:R=J2}) and solving for the
contorsion, we find
\begin{equation}
\kappa_\nu{}^{ab} = \frac{i}2 \Big( 
(\sigma_\nu)^{\alpha\dot\beta} J^{ab}{}_{\alpha\dot\beta}
+(\sigma^a)^{\alpha\dot\beta} J_\nu{}^b{}_{\alpha\dot\beta}
-(\sigma^b)^{\alpha\dot\beta} J_\nu{}^a{}_{\alpha\dot\beta} \Big).
\label{eq:kappa_def}
\end{equation}

In order to obtain the contributions to the stress-energy tensor that
are quadratic in the scalar we henceforth drop the contributions to
$J$ from $W$ and all components in $\Phi$ which have different number
of dotted and undotted indices. The remaining scalar contributions to
the stress-energy tensor is a sum of terms containing the structures
$\Phi_{\alpha_1\ldots \alpha_k \dot\alpha_1 \ldots\dot\alpha_k}
\Phi_{\alpha_1\ldots \alpha_\ell \dot\alpha_1 \ldots\dot\alpha_\ell}$,
where $\Phi_{\alpha_1\ldots \alpha_k \dot\alpha_1 \ldots\dot\alpha_k}$
can be substituted using (\ref{eq:deriv_scalar}).  To compute the
various terms in the quantity $L_{\mu\nu}$ given in (\ref{eq:Ldef}) we
make use of (\ref{eq:hfstarhg})-(\ref{eq:A_alpha^(n)}). We first
obtain
\begin{equation}
\wh L^{(1)}(e_\mu) = -\frac{b_1}{2}z^\alpha 
e_{\mu,\alpha\dot\alpha}
\bar{\p}_y^{\dot{\alpha}}\int_0^1 \!\! \int_0^1 dt'tdt\Phi(-tt'z,\bar{y})
\kappa(tt'z,y) \ - \ \mbox{h.c.} \label{eq:e1}
\end{equation}
At this stage it is convenient to introduce
\begin{equation}
\wh{\mathcal{L}}^{(n)}(\wh f) \equiv -i \int_0^1 \frac{dt}{t} \bigg( 
\wh A_\alpha^{(n)} \star \partial^\alpha_- \wh f - 
\partial^\alpha_+ \wh f \star \wh A_\alpha^{(n)} \bigg)_{z\rightarrow tz}
\end{equation}
such that we can write,
\begin{equation}
\wh L^{(n)}=\wh{\mathcal{L}}^{(n)} +\wh{\mathcal{L}}^{\dagger(n)}.  
\end{equation}
We then find 
\begin{equation}
\Big[\wh L^{(1)}(e_\mu),\wh L^{(1)}(e_\nu)\Big]_{\star,z=0} =
\Big[\wh{\mathcal{L}}^{(1)}(e_\mu),
\wh{\mathcal{L}}^{(1)}(e_\nu)\Big]_{\star,z=0} + 
\Big[\wh{\mathcal{L}}^{\dagger(1)}(e_\mu),
\wh{\mathcal{L}}^{\dagger(1)}(e_\nu)\Big]_{\star,z=0},
\label{eq:L1L1}
\end{equation}
modulo the omitted terms, as explained above.  Furthermore, we have
\begin{equation}
\Big[e_\mu,\wh L^{(2)}(e_\nu) \Big]_{\star,z=0}  = 
\Big[e_\mu,\wh{\mathcal{L}}^{(2)}(e_\nu) \Big]_{\star,z=0} + 
\Big[e_\mu,\wh{\mathcal{L}}^{\dagger(2)}(e_\nu) \Big]_{\star,z=0}
\end{equation}
where
\begin{equation}
\Big[e_\mu,\wh{\mathcal{L}}^{(2)}(e_\nu) \Big]_{\star,z=0}
  =  
-\frac12 e_\mu{}^{\beta\dot{\beta}}
e_\nu{}^{\alpha\dot{\alpha}}\bigg[y_\beta \bar{y}_{\dot{\beta}},
\int_0^1\frac{dt}{t}\bar{\p}^{y}_{\dot{\alpha}}
\wh A_\alpha^{(2)}(z\to tz)\bigg]_{\star,z=0}
\label{eq:eL2}
\end{equation}
and
\begin{equation}
\wh A_\alpha^{(2)}  =  \frac{z_\alpha}2
\int_0^1 tdt \Big( \left[\wh A^{(1)\delta},\wh
A^{(1)}_\delta\right]_\star  -  
ib_1 \wh \Phi^{(2)}\star \kappa \Big) (z\to tz) \ + \ \wh B_\alpha, 
\end{equation}
denoting by $\wh B_\alpha$ terms in $\wh A_\alpha^{(2)}$ that do not
contribute once $z$ is set to zero. Finally, considering
\begin{equation}
\wh L^{(1)} \circ \wh L^{(1)}(e_\mu) =
- \int_0^1\frac{dt}{t}
\left(\left\{\wh A^{(1)\delta} ,
\p_\delta^{(z)}\wh L^{(1)}(e_\mu) \right\}_\star-\left[\wh A^{(1)\delta} ,
\p_\delta^{(y)}\wh L^{(1)}(e_\mu) \right]_\star\right)(z\to tz) \ + 
\ \mbox{h.c.},
\end{equation} 
we find that
\begin{equation}
\Big[e_\mu,\wh L^{(1)}\circ\wh L^{(1)}(e_\nu)\Big]_{\star,z=0} =
\Big[e_\mu,\wh{\mathcal{L}}^{(1)}\circ\wh{\mathcal{L}}^{(1)}(e_\nu)\Big]_{\star,z=0} + 
\Big[e_\mu,\wh{\mathcal{L}}^{\dagger(1)}\circ
 \wh{\mathcal{L}}^{\dagger(1)}(e_\nu)\Big]_{\star,z=0}.
\label{eq:eL1L1}
\end{equation}
From the above analysis we conclude that
\begin{equation}
J_{\mu\nu} = \mathcal{J}_{\mu\nu} - \mathcal{J}^\dagger_{\mu\nu},
\label{eq:Jnew}
\end{equation}
where 
\begin{equation}
\mathcal{J}_{\mu\nu}  =  2i\big\{
e_{[\mu}{}^{\alpha\dot\alpha} e_{\nu]}{}^\beta{}_{\dot\alpha} 
\widehat A^{(1)}_\alpha
\star \widehat A^{(1)}_\beta \big\}_{z=0} + \mathcal{L}_{\mu\nu}
\label{eq:curlJ}
\end{equation}
and
\begin{eqnarray}
\mathcal{L}_{\mu\nu} & = & -\bigg(\Big[\wh{\mathcal{L}}^{(1)}(e_\mu),
\wh{\mathcal{L}}^{(1)}(e_\nu)\Big]_\star -
2\Big[e_{[\mu},\wh{\mathcal{L}}^{(2)}(e_{\nu]})\Big]_\star   \nonumber \\
 & &  \qquad \qquad \qquad \qquad \qquad \qquad \qquad 
+ \ 2\Big[e_{[\mu},\wh{\mathcal{L}}^{(1)} \circ \wh{\mathcal{L}}^{(1)}(e_{\nu]})\Big]_\star
\bigg)_{z=0}.
\label{eq:curlL}
\end{eqnarray}
The explicit calculation of (\ref{eq:L1L1}), (\ref{eq:eL2}) and
(\ref{eq:eL1L1}) is straightforward but lengthy. The details are given
in Appendix \ref{sect:result_apdx}. The result is a sum of various
contractions of $\Phi_{\alpha_1\ldots \alpha_k \dot\alpha_1
\ldots\dot\alpha_k} \Phi_{\alpha_1\ldots \alpha_\ell \dot\alpha_1
\ldots\dot\alpha_\ell}$ for $|k-\ell|=0$ or $2$.  Finally, after
converting spinor indices to vector indices, as described in Appendix
\ref{sect:derivs_apdx}, we arrive at
\begin{eqnarray}
\Ric_{\mu\nu} -\frac12 R g_{\mu\nu} - 3 g_{\mu\nu} & = &
\mbox{Re}\{b_1^2\} \bigg[ \sum_k \frac{2^k}{(k!)^2}
\bigg( \xi(k) \, g_{\mu\nu} \nabla_{\rho\{k+1\}}  \, \phi \  
\nabla^{\rho\{k+1\}} \, \phi + \nonumber \\
& & \qquad   
+ \ \eta(k) \nabla_{\rho\{k\}\mu}  \, \phi \  
\nabla^{\rho\{k\}}{}_\nu \, \phi  + \nonumber \\
& & \qquad 
+ \ \zeta(k) \nabla_{\rho\{k\}\mu\nu}  \, \phi \  
\nabla^{\rho\{k\}} \, \phi \bigg)
\ - \ \frac49 \ g_{\mu\nu} \ \phi\,\phi \bigg], \label{eq:result}
\end{eqnarray}
where 
\begin{eqnarray}
\xi(k) & = & - \frac1{24} \{132 k^{10} +4169 k^9 
+57902 k^8 +464477 k^7 +2378336 k^6 +8109935 k^5  
\nonumber \\ & & \qquad \qquad 
+18627566 k^4 +28429503 k^3 +27570000 k^2 +15326604 k 
\nonumber \\ & & \qquad \qquad 
+3703824 \}\Big/\{(k+5)^2(k+4)^2(k+3)^2(k+2)^2(k+1)^2\}, 
\label{eq:xi}\\
\eta(k) & = & \frac13 \{42 k^9 +1234 k^8
 +15738 k^7 +114011 k^6 +515273 k^5 +1500759 k^4 
\nonumber \\ & & \qquad \qquad
+2804017 k^3 +3224520 k^2 +2060706 k +554772 \}\Big/
\nonumber \\ & & \qquad \qquad \qquad \qquad \qquad
\{(k+5)^2 (k+4)^2 (k+3)^2 (k+2)^2
(k+1) \}, \\
\zeta(k) & = & \frac1{6} \{ 12 k^8 +379 k^7 +5047 k^6 +36860 k^5
+161255 k^4 
+433379 k^3 +701764 k^2
\nonumber \\ & & \qquad \qquad
+629748 k +240912 \}\Big/\{(k+5)^2 (k+4)^2 (k+3)^2 (k+2)^2 \}.
\label{eq:zeta}
\end{eqnarray}
We note that for $k \gg 1$
\begin{eqnarray}
\xi(k) & \sim & - 11/2  \nonumber \\
\eta(k) & \sim &  14 \nonumber \\
\zeta(k) & \sim  & 2.
\end{eqnarray}

\section{Summary and Discussion}

We have calculated the scalar field content of the Einstein equation
in AdS$_4$ higher spin gauge theory. This constitutes a first step
towards finding the graviton-$\phi^2$ terms in the action. The details
of the calculation are given in the appendices, since we believe they
can be useful in making further calculations, perhaps implemented on a
computer.

In order to compute the cubic action in the spin $s=0,2$ sector we
would also have to calculate graviton-scalar terms in the scalar field
equation. Then these two contributions in the field equations should
have identical coefficients since they originate from the same
graviton-$\phi^2$ terms in the action. It is also possible that some
recombination of the spin $s=0,2$ field equations is required in order
to satisfy the above integrability conditions for the action. The
existence of an action is not necessarily contradicted by the higher
derivative terms in (\ref{eq:result}), since these terms may arise
from a term of the form $G^{\mu\nu}(g) \partial_\mu \phi\partial_\nu
\phi$, where $G_{\mu\nu}(g)$ depends on higher derivatives of the
metric while $G_{\mu\nu}(g)$ and $\nabla^\mu G_{\mu\nu}(g)$ vanishes
in the AdS background.

Starting from a general $\mathcal{V}(X)$ we have found that the
contribution to the stress-energy tensor which is quadratic in the
scalar field is proportional to $\mathrm{Re}\{b_1^2\}$. This raises
the issue of the positivity of the Killing energy associated with the
stress-energy tensor, since it is in fact possible for
$\mathrm{Re}\{b_1^2\}$ to be either positive or negative depending on
the parity of the scalar field \cite{sezsun_new}. Moreover, the
analysis of the Killing energy functional might depend on the choice
of the boundary condition for the scalar field.  We defer these issues
to a future publication.

It would be interesting to consider the influence of this scalar field
in a cosmological context. Close to the Big Bang more symmetries were
realised, and it might be possible that a higher spin theory is needed
to understand the dynamics. In this point of view the scalar field
$\phi$ investigated here may have connections to the inflation. For
this reason and also for understanding better the bulk description of
the O($N$) model RG-flows, it would be interesting to find domain wall
solutions to the full field equations in which all fields are
depending on one space or time coordinate.

\section*{Acknowledgements}

We are very grateful to Per Sundell for invaluable discussions and
comments. We also thank Ulf Danielsson, Ergin Sezgin and Johan
Engquist for useful conversations.

%XXXXXXXXXXXXXXXXXXXXXXXXXXXXXXXXXXXXXXXXXXXXXXXXXXXXXXXXXXX

\appendix

%==========================================================

\section{Conventions and Useful Relations}

%----------------------------------------------------------

\subsection{Spinor conventions}

We always use symmetrisations and anti-symmetrisations with unit strength. 
We define the $SL(2,C)$ invariant
$\epsilon_{\alpha\beta}$ by
\begin{equation}
\epsilon_{\alpha\beta} = -\epsilon_{\beta\alpha} = 
(\epsilon_{\dot\alpha \dot\beta})^\dagger, \qquad 
\epsilon_{\alpha\beta}\epsilon^{\delta\gamma} = 
2 \delta^{\delta\gamma}_{\alpha\beta}.
\end{equation}
Spinor index contraction is according to the north-west south-east
rule. In particular,
\begin{equation}
\psi^\alpha=\epsilon^{\alpha\beta} \psi_\beta, \qquad \mbox{and}
\qquad 
\psi_\alpha= \psi^\beta \epsilon_{\beta\alpha}.
\end{equation}
The van der Waerden symbols
$(\sigma^\mu)_{\alpha\dot{\beta}}$ are defined as
\begin{equation} \label{eq:sigma_def}
{(\sigma^{(\mu})_\alpha}^{\dot\alpha}
(\sigma^{\nu)})_{\beta \dot\alpha} = \eta^{\mu\nu}\epsilon_{\alpha\beta},
\qquad 
\Big((\sigma^\mu)_{\alpha \dot\beta}\Big)^\dagger = 
     (\sigma^\mu)_{\beta \dot\alpha}.
\end{equation}
We also define the following matrices
\begin{eqnarray}
(\sigma^{\mu\nu})_{\alpha\beta} & = &
{(\sigma^{[\mu})_\alpha}^{\dot\alpha}
(\sigma^{\nu]})_{\beta \dot\alpha}, \label{eq:sigma2} \\
(\sigma^{\mu\nu\rho})_{\alpha \dot\beta} & = &
{(\sigma^{[\mu})_\alpha}^{\dot\alpha}
{(\sigma^{\nu})^\beta}_{\dot\alpha} 
(\sigma^{\rho]})_{\beta \dot\beta}, \label{eq:sigma3} \\
(\sigma^{\mu\nu\rho\tau})_{\alpha\beta} & = &
{(\sigma^{[\mu})_\alpha}^{\dot\alpha}
{(\sigma^{\nu})^\gamma}_{\dot\alpha} 
{(\sigma^{\rho})_\gamma}^{\dot\gamma}
(\sigma^{\tau]})_{\beta \dot\gamma} \label{eq:sigma4} \\
(\sigma^{\mu\nu})_{\dot\alpha \dot\beta} & = &
    \Big( (\sigma^{\mu\nu})_{\alpha\beta} \Big)^\dagger, \\
(\sigma^{\mu\nu\rho\tau})_{\dot\alpha \dot\beta} & = &
    \Big( (\sigma^{\mu\nu\rho\tau})_{\alpha\beta} \Big)^\dagger.
\end{eqnarray}
One can show that
\begin{eqnarray}
\Big( (\sigma^{\mu\nu\rho})_{\alpha \dot\beta} \Big)^\dagger & = &
    - (\sigma^{\mu\nu\rho})_{\beta \dot\alpha}, \\
(\sigma^{\mu\nu\rho\tau})_{\alpha\beta} & = &
i\epsilon^{\mu\nu\rho\tau}\epsilon_{\alpha\beta}.
\end{eqnarray}
From the defining relation (\ref{eq:sigma_def}) it follows that
\begin{eqnarray}
\sigma^{\mu_1\ldots\mu_m}\sigma_{\nu_1\ldots\nu_n} & = & 
{\sigma^{\mu_1\ldots\mu_m}}_{\nu_1\ldots\nu_n} \ +  \ 
mn \, \delta^{[\mu_m}_{[\nu_1} {
\sigma^{\mu_1\ldots\mu_{m-1}]}}_{\nu_2\ldots\nu_n]} \ + \ \ldots \ +
\nonumber \\ & &   + \ 
\Big( \!\! \begin{array}{c} m \\ k \end{array} \!\! \Big) 
\Big( \!\! \begin{array}{c} n \\ k \end{array} \!\! \Big) 
k! \ \delta^{[\mu_m\ldots\mu_{m-k+1}}_{[\nu_1\ldots\nu_k} 
{\sigma^{\mu_1\ldots\mu_{m-k}]}}_{\nu_{k+1}\ldots\nu_n]} 
\ + \ \nonumber \\ & &  + \ \ldots \label{eq:gammautv}
\end{eqnarray}
Vectors and antisymmetric tensors are expressed in spinor indices
according to
\begin{eqnarray}
 & & V^\mu = (\sigma^\mu)_{\alpha\dot\alpha} V^{\alpha\dot\alpha}, \qquad
\qquad \qquad \qquad \qquad \ V^{\alpha\dot\alpha} = -\frac12
(\sigma_\mu)^{\alpha\dot\alpha}V^\mu, \label{eq:vect_spin1} \\
 & & A^{\mu\nu} = \frac12 \, \Big(
(\sigma^{\mu\nu})_{\alpha\beta} A^{\alpha\beta} +
(\sigma^{\mu\nu})_{\dot\alpha\dot\beta} A^{\dot\alpha\dot\beta} \Big),
\qquad
A^{\alpha\beta} = \frac14 (\sigma_{\mu\nu})^{\alpha\beta} A^{\mu\nu}.
\label{eq:vect_spin2}
\end{eqnarray}

%----------------------------------------------------------

\subsection{Notation for symmetrised spinor indices}

In the following we shall use a condensed notation for symmetrisation
of spinor indices, defined by
\begin{equation}
f_{\alpha(m)}=f_{\alpha_1\ldots\alpha_m}
=\frac{1}{m!}\sum_Pf_{\alpha_{P(1)}\ldots\alpha_{P(m)}},
\end{equation}
and
\begin{equation}
f_{\alpha(m_1)}g_{\alpha(m_2)}=\frac{1}{(m_1+m_2)!}\sum_P 
f_{\alpha_{P(1)}\ldots
\alpha_{P(m_1)}}g_{\alpha_{P(m_1+1)}\ldots\alpha_{P(m_1+m_2)}},
\end{equation}
where the right hand sides are summed over all permutations $P$ of
indices. Note that within this notation there is no symmetrisation in
$f_{\alpha_1(m_1)}g_{\alpha_2(m_2)}$ between the indices of type
$\alpha_1$ and those of type $\alpha_2$.

%----------------------------------------------------------

\subsection{Useful relations}

From (\ref{eq:gammautv}) we can derive the following relations.
\begin{eqnarray}
(\sigma^\mu)_{\alpha\dot\alpha}(\sigma^\nu)_{\beta\dot\beta} & = &
-\frac12 \Big( \eta^{\mu\nu} \epsilon_{\alpha\beta}
\epsilon_{\dot\alpha\dot\beta} + 
(\sigma^{\mu\nu})_{\alpha\beta}\epsilon_{\dot\alpha\dot\beta} +
\nonumber \\ & & 
\qquad + \ (\sigma^{\mu\nu})_{\dot\alpha\dot\beta}\epsilon_{\alpha\beta} +
(\sigma^{\mu\rho})_{\alpha\beta}(\sigma_\rho{}^\nu)_{\dot\alpha\dot\beta} 
\Big)
\\
(\sigma^\mu)_{\alpha\dot\alpha}(\sigma_\mu)_{\beta\dot\beta} & = &
-2 \epsilon_{\alpha\beta}\epsilon_{\dot\alpha\dot\beta} \\
(\sigma^{\mu\nu})_{\alpha\beta}(\sigma_{\mu\nu})^{\delta\gamma} & = &
 8 \epsilon_\alpha{}^{(\delta} \epsilon_\beta{}^{\gamma)} \\
(\sigma^{\mu\nu})_{\alpha\beta}(\sigma_{\mu\nu})^{\dot\delta\dot\gamma} 
& = & 0 \\
(\sigma^{\mu\nu})_{\alpha\beta}(\sigma_\mu)^{\delta\dot\delta}
(\sigma_\nu)^{\gamma\dot\gamma} & = & 
-4 \epsilon^{\dot\delta\dot\gamma}
\epsilon_\alpha{}^{(\delta} \epsilon_\beta{}^{\gamma)} \\
(\sigma^{\mu\nu})_{\alpha\beta}(\sigma_\mu{}^\rho)^{\gamma\delta}
 & = & 2 \epsilon_{(\alpha}{}^{(\gamma}
(\sigma^{\nu\rho})_{\beta)}^{\delta)}.
\end{eqnarray}

%============================================================

%============================================================

\section{Evaluations of $\star\;$--products }
\label{sect:star_apdx}

We begin by observing the following useful formulae
\begin{eqnarray}
\partial_\alpha y_\beta = \epsilon_{\alpha\beta} 
&\qquad & 
\partial^\alpha y^\beta = \epsilon^{\alpha\beta}  \\
\partial_\alpha y^\beta = \delta_\alpha^\beta 
&\qquad & 
\partial^\alpha y_\beta = -\delta^\alpha_\beta.
\end{eqnarray}
The star product (\ref{eq:hfstarhg}) is equivalent to the following
contraction rules between the $y$ and $z$ oscillators
\begin{eqnarray}
y_{\alpha} \star y_{\beta} = y_{\alpha}y_{\beta} +i\eps_{\alpha\beta} 
& \qquad &
y_{\alpha} \star z_{\beta} = y_{\alpha}z_{\beta} -i\eps_{\alpha\beta} 
\nonumber \\
z_{\alpha} \star y_{\beta} = z_{\alpha}y_{\beta} +i\eps_{\alpha\beta} 
& \qquad &
z_{\alpha} \star z_{\beta} = z_{\alpha}z_{\beta} -i\eps_{\alpha\beta}
\nonumber\\[4mm]
\bar y_{\dot{\alpha}} \star \bar y_{\dot{\beta}} = 
\bar y_{\dot{\alpha}} \bar y_{\dot{\beta}} + i\eps_{\dot{\alpha}\dot{\beta}} 
& \qquad &
\bar z_{\dot{\alpha}} \star \bar y_{\dot{\beta}} =
\bar z_{\dot{\alpha}}\bar y_{\dot{\beta}}
- i\eps_{\dot{\alpha}\dot{\beta}} \nonumber \\
\bar y_{\dot{\alpha}} \star \bar z_{\dot{\beta}} =
\bar y_{\dot{\alpha}} \bar z_{\dot{\beta}} + i\eps_{\dot{\alpha}\dot{\beta}} 
& \qquad &
\bar z_{\dot{\alpha}} \star \bar z_{\dot{\beta}} =
\bar z_{\dot{\alpha}} \bar z_{\dot{\beta}} 
-i\eps_{\dot{\alpha}\dot{\beta}}.
\label{eq:yzcontr}
\end{eqnarray}
Note that $[z_\alpha,y_\beta]_\star=0$.  The Weyl ordered product is
denoted by
\begin{equation}
y_{\alpha(m)}=\frac{1}{m!}\sum_{P} y_{\alpha_{P(1)}}\star \cdots \star
y_{\alpha_{P(m)}} = \underbrace{y_{\alpha(1)} \star \cdots
\star y_{\alpha(1)}}_{\mbox{$m$ factors}}.
\end{equation}
and equivalently for $z$.  From the contraction rules above it follows
that
\begin{equation}
y_{\alpha_1(m_1)}\star y_{\alpha_2(m_2)}=
\sum_k i^k k!\left(\begin{array}{c}m_1\\k\end{array}\right)
\left(\begin{array}{c}m_2\\k\end{array}\right)
y_{\alpha_1(m_1-k)} y_{\alpha_2(m_2-k)}\eps_{\alpha_1(k)\alpha_2(k)}
\end{equation}
where
\begin{equation}
\eps_{\alpha(k)\beta(k)}=\frac{1}{k!}\sum_P
\eps_{\alpha_1\beta_{P(1)}}...
\eps_{\alpha_k\beta_{P(k)}}.
\end{equation}
Indeed, from this we can verify that the $\star$ product is equivalent
to the differential operator given in (\ref{eq:fstarg}),
\begin{equation}
\star=1+i\eps_{\alpha\beta}
\overleftarrow{\p}^\alpha\overrightarrow{\p}^\beta+
\frac{i^2}{2!}\eps_{\alpha(2)\beta(2)}
\overleftarrow{\p}^{\alpha(2)}\overrightarrow{\p}^{\beta(2)}+...=
\exp\{-i\overleftarrow{\p}^\alpha\overrightarrow{\p}_\alpha\}.
\end{equation}
As an example, consider the product 
\begin{eqnarray}
y_{\alpha_1(2)}\star y_{\alpha_2(2)}&=&
\sum_k i^k k!\left(\begin{array}{c}2\\k\end{array}\right)
\left(\begin{array}{c}2\\k\end{array}\right)
y_{\alpha_1(2-k)} y_{\alpha_2(2-k)}\eps_{\alpha_1(k)\alpha_2(k)}
\nonumber \\
&=&-2\eps_{\alpha_1\alpha_2(2)}+4iy_{\alpha_1(1)}y_{\alpha_2(1)}
\eps_{\alpha_1(1)\alpha_2(1)}+y_{\alpha_1(2)}y_{\alpha_2(2)}
\nonumber \\&=&-
(\eps_{\alpha_{11}\alpha_{21}}\eps_{\alpha_{12}\alpha_{22}}+
\eps_{\alpha_{11}\alpha_{22}}\eps_{\alpha_{12}\alpha_{21}})
\nonumber \\
&+&i(y_{\alpha_{11}}y_{\alpha_{22}}\eps_{\alpha_{12}\alpha_{21}}
+y_{\alpha_{11}}y_{\alpha_{21}}\eps_{\alpha_{12}\alpha_{22}}
+y_{\alpha_{12}}y_{\alpha_{22}}\eps_{\alpha_{11}\alpha_{21}}
+y_{\alpha_{12}}y_{\alpha_{21}}\eps_{\alpha_{11}\alpha_{22}})
\nonumber \\
&+&y_{\alpha_{11}}y_{\alpha_{12}}y_{\alpha_{21}}y_{\alpha_{22}}.
\end{eqnarray}
Expanding a  Weyl ordered polynomial in $y$ as
\be
F(y)=\sum_m \frac{1}{m!}F_{\alpha(m)}y^{\alpha(m)}.
\ee
and using
\be
\frac{\p^{\alpha(m)}}{m!}y_{\beta(m_1)}y_{\gamma(m_2)}|_{y=0} = 
(-1)^m \delta^{\alpha(m)}_{\beta(m_1)\gamma(m_2)}
\qquad \mbox{and} \qquad
\p^{\alpha(m)}F(y)|_{y=0}=F^{\alpha(m)},
\ee
where $m_1+m_2=m$,
we compute
\begin{eqnarray}
F\star G &=& \sum_{m_1,m_2}(-1)^{m_1+m_2}\frac{F^{\alpha_1(m_1)}}{m_1!}y_{\alpha_1(m_1)}\star
\frac{G^{\alpha_2(m_2)}}{m_2!}y_{\alpha_2(m_2)} \nonumber \\
&=&\sum_{m_1,m_2}(-1)^{m_1+m_2}
\frac{F^{\alpha_1(m_1)}G^{\alpha_2(m_2)}}{m_1!m_2!}
\nonumber\\&\times&\sum_k i^k k!\left(\begin{array}{c}m_1\\k\end{array}\right)
\left(\begin{array}{c}m_2\\k\end{array}\right)
y_{\alpha_1(m_1-k)} y_{\alpha_2(m_2-k)}
\eps_{\alpha_1(k)\alpha_2(k)}, \nonumber \\[4mm]
\left(F\star G\right)^{\alpha(m)}
&=& \sum_{m_1,m_2}\frac{F^{\alpha_1(m_1)} G^{\alpha_2(m_2)}}
{m_1!m_2!}\sum_k i^k
k!\left(\begin{array}{c}m_1\\
k\end{array}\right)\left(\begin{array}{c}m_2\\k\end{array}\right)
m!\delta^{\alpha(m)}_{\alpha_1(m_1-k)\alpha_2(m_2-k)}
\eps_{\alpha_1(k)\alpha_2(k)} \nonumber \\
&=&
\!\!\sum_{m_1,m_2,k}
\frac{i^k m!k!}{(m_1+k)!(m_2+k)!}\left(\begin{array}{c}m_1+k\\
k\end{array}\right)\left(\begin{array}{c}m_2+k\\k\end{array}\right)
\nonumber \\[-2mm] 
& & \qquad \qquad \qquad \qquad \qquad 
\times \ F^{\alpha_1(m_1+k)}G^{\alpha_2(m_2+k)}
\delta^{\alpha(m)}_{\alpha_1(m_1)\alpha_2(m_2)}
\eps_{\alpha_1(k)\alpha_2(k)} \nonumber \\&=&
\sum_{{\begin{array}{c}\\[-7mm] \scriptstyle k\\[-2mm]
                 \scriptstyle m_1+m_2=m\end{array}}}
\frac{i^k m!}{k!m_1!m_2!}
F^{\alpha(m_1)}_{\gamma(k)}G^{\alpha(m_2)\gamma(k)}
\end{eqnarray}
This is easily extended to general polynomials in
$y,\bar{y},z,\bar{z}$, which we expand as
\begin{equation}\label{expansion}
F(Y,Z)=\sum_{m,\tilde{m},n,\tilde{n}}
\frac{F_{\alpha(m)\dot{\alpha}(\tilde{m}),
\beta(n)\dot{\beta}(\tilde{n})}}
{m!\tilde{m}!n!\tilde{n}!}
y^{\alpha(m)}\bar{y}^{\dot{\alpha}(\tilde{m})}z^{\beta(n)}
\bar{z}^{\dot{\beta}(\tilde{n})}
\end{equation}
where commas are used to separate the indices belonging to $y$'s and
$z$'s.  Using (\ref{eq:yzcontr}) we compute
\begin{eqnarray}\label{eq:gen_contr}
&&(F\star G)^{\alpha(m)\dot{\alpha}(\tilde{m}),\beta(n)\dot{\beta}(\tilde{n})}=
\sum_{K(m,\tilde{m},n,\tilde{n})}C(K(m,\tilde{m},n,\tilde{n}))
\nonumber \\
&& \qquad \times
F^{\alpha_1(m_1+k_{\alpha\alpha}+k_{\alpha\beta})
\dot{\alpha}_1(\tilde{m}_1+\tilde{k}_{\alpha\alpha}+\tilde{k}_{\alpha\beta}),
\beta_1(n_1+k_{\beta\alpha}+k_{\beta\beta})
\dot{\beta}_1(\tilde{n}_1+\tilde{k}_{\beta\alpha}+\tilde{k}_{\beta\beta})}
\nonumber \\
&&\qquad \times G^{\alpha_2(m_2+k_{\alpha\alpha}+k_{\beta\alpha})
\dot{\alpha}_2(\tilde{m}_2+\tilde{k}_{\alpha\alpha}+\tilde{k}_{\beta\alpha}),
\beta_2(n_2+k_{\alpha\beta}+k_{\beta\beta})
\dot{\beta}_2(\tilde{n}_2+\tilde{k}_{\alpha\beta}+\tilde{k}_{\beta\beta})}
\nonumber \\
&&\qquad \times\delta^{\alpha(m)}_{\alpha_1(m_1)\alpha_2(m_2)}
\delta^{\dot{\alpha}(\tilde{m})}_{\dot{\alpha}_1(\tilde{m}_1)
\dot{\alpha}_2(\tilde{m}_2)}
\delta^{\beta(n)}_{\beta_1(n_1)\beta_2(n_2)}
\delta^{\dot{\beta}(\tilde{n})}_
{\dot{\beta}_1(\tilde{n}_1)\dot{\beta}_2(\tilde{n}_2)} 
\nonumber \\
&&\qquad \times\eps_{\alpha_1(k_{\alpha\alpha})\alpha_2(k_{\alpha\alpha})}
\eps_{\alpha_1(k_{\alpha\beta})\beta_2(k_{\alpha\beta})}
\eps_{\beta_1(k_{\beta\alpha})\alpha_2(k_{\beta\alpha})}
\eps_{\beta_1(k_{\beta\beta})\beta_2(k_{\beta\beta})}
\nonumber \\ 
&&\qquad \times\eps_{\dot{\alpha}_1(\tilde{k}_{\alpha\alpha})
\dot{\alpha}_2(\tilde{k}_{\alpha\alpha})}
\eps_{\dot{\alpha}_1(\tilde{k}_{\alpha\beta})
\dot{\beta}_2(\tilde{k}_{\alpha\beta})}
\eps_{\dot{\beta}_1(\tilde{k}_{\beta\alpha})
\dot{\alpha}_2(\tilde{k}_{\beta\alpha})}
\eps_{\dot{\beta}_1(\tilde{k}_{\beta\beta})
\dot{\beta}_2(\tilde{k}_{\beta\beta})} \nonumber \\[4mm]
&& =
\sum_{K(m,\tilde{m},n,\tilde{n})}C(K(m,\tilde{m},n,\tilde{n}))
\nonumber \\
&& \qquad \times
F^{\alpha(m_1)\dot{\alpha}(\tilde{m}_1),\beta(n_1)\dot{\beta}(\tilde{n}_1)}_
{\gamma_1(k_{\alpha\alpha})\gamma_2(k_{\alpha\beta})\dot{\gamma}_1(\tilde{k}_{\alpha\alpha})
\dot{\gamma}_2(\tilde{k}_{\alpha\beta}),
\gamma_1(k_{\beta\alpha})\gamma_2(k_{\beta\beta})\dot{\gamma}_1(\tilde{k}_{\beta\alpha})
\dot{\gamma}_2(\tilde{k}_{\beta\beta})}
\nonumber \\&& \qquad 
\times G^{\alpha(m_2)\dot{\alpha}(\tilde{m}_2)
\gamma_1(k_{\alpha\alpha}+k_{\beta\alpha})
\dot{\gamma}_1(\tilde{k}_{\alpha\alpha}+\tilde{k}_{\beta\alpha}),
\beta(n_2)\dot{\beta}(\tilde{n}_2)\gamma_2(k_{\alpha\beta}+k_{\beta\beta})
\dot{\gamma}_2(\tilde{k}_{\alpha\beta}+\tilde{k}_{\beta\beta})}
\end{eqnarray}
where 
\begin{eqnarray}
& & K(m,\tilde{m},n,\tilde{n})\in\{m_1,\tilde{m}_1,n_1,\tilde{n}_1,m_2,\tilde{m}_2,
n_2,\tilde{n}_2,k_{\alpha\alpha},
k_{\alpha\beta}, \nonumber \\
& & \qquad \qquad \qquad \qquad \qquad \qquad \qquad 
k_{\beta\alpha},k_{\beta\beta},\tilde{k}_{\alpha\alpha},
\tilde{k}_{\alpha\beta},\tilde{k}_{\beta\alpha},\tilde{k}_{\beta\beta}
 = 0,1,2,\ldots \}
\label{eq:K(mn)}
\end{eqnarray}
with the restrictions:
\begin{eqnarray}
m_1+m_2&=&m \nonumber \\
n_1+n_2&=&n \nonumber \\
\tilde{m}_1+\tilde{m}_2&=&\tilde{m} \nonumber \\
\tilde{n}_1+\tilde{n}_2&=&\tilde{n}
\end{eqnarray}
and
\begin{eqnarray}\label{eq:CK}
C(K)&=&
\frac{i^{k_{\alpha\alpha}+k_{\beta\alpha}+\tilde{k}_{\alpha\alpha}
+\tilde{k}_{\alpha\beta}-
(k_{\alpha\beta}+k_{\beta\beta}+
\tilde{k}_{\beta\alpha}+\tilde{k}_{\beta\beta})}m!\tilde{m}!n!\tilde{n}!}
{m_1!m_2!n_1!n_2!
\tilde{m}_1!\tilde{m}_2!\tilde{n}_1!\tilde{n}_2!
k_{\alpha\alpha}!k_{\alpha \beta}!k_{\beta\alpha}!k_{\beta\beta}!
\tilde{k}_{\alpha\alpha}!\tilde{k}_{\alpha \beta}!
\tilde{k}_{\beta\alpha}!\tilde{k}_{\beta\beta}!}.
\end{eqnarray}
Of special interest is the exponential
\begin{eqnarray}
\kappa(y,z)& = &\exp(iy^{\alpha}z_{\alpha})=1+
i\eps^{\alpha\beta}y_{\beta}z_{\alpha}+\frac{i^2}{2!}
\eps^{\alpha(2)\beta(2)}y_{\beta(2)}z_{\alpha(2)}+ \ldots \\
\kappa^{\alpha(m),\beta(n)}& = &(-i)^n n!\eps^{\alpha(n)\beta(n)}\delta_{m,n}.
\end{eqnarray}
It has the property that
\begin{equation}\label{eq:kappa_prop}
\kappa^{\alpha(m),\beta(m)}\eps_{\alpha(k_1)\gamma(k_1)}
\eps_{\beta(k_2)\gamma(k_2)}=
0 \qquad \mbox{for} \qquad k_1\;\;\mbox{or}\;\; k_2 \geq m\;\;\mbox{and}\;\;k_1,k_2> 0
\end{equation}
which greatly simplifies the calculations below.

%------------------------------------------------------------

\section{Computation of $J_{\mu\nu}$}
\label{sect:result_apdx}

In this section we evaluate all the star products in the quantity
$J_{\mu\nu}$ given in (\ref{eq:Jnew}) using
eqs. (\ref{eq:gen_contr})--(\ref{eq:CK}). Moreover functions of the
oscillators are expanded using the convention in (\ref{expansion}),
except $e$ whose expansion contains an extra factor of $i/2$, as
defined in (\ref{eq:usefull2}). Upon using (\ref{eq:gen_contr}) one
obtains several different contributions corresponding to different
values of the index $K(m,\tilde m,n,\tilde n)$ defined in
(\ref{eq:K(mn)}). Below we shall give these quantities separately in
equations labelled by A{\footnotesize\#}, B{\footnotesize\#},
C{\footnotesize\#} and D{\footnotesize\#}, where A, B, C are the
contributions from the three different terms in (\ref{eq:curlL}),
respectively, and D those from the first term in (\ref{eq:curlJ}):
\begin{eqnarray}
\mathcal{J}_{\mu\nu}^{\alpha(2)}&=&\Big\{-\mbox{A1}+2\times\big(\mbox{B2}
                   +\mbox{B3}+\mbox{B4}+\mbox{B8}\big)\nonumber\\
& & \quad - \ 2\times\big(\mbox{C6}+\mbox{C7}+\mbox{C13}+\mbox{C14}
                   +\mbox{C15}+\mbox{C16}+\mbox{C17}\big)+\mbox{D1}\Big\}
                   \label{eq:Jaa} \\
\mathcal{J}_{\mu\nu}^{\alpha(1)\dot\alpha(1)}&=&
       \Big\{-\left(\mbox{A2}+\mbox{A3}\right)+2\times\big(\mbox{B5}
                    +\mbox{B6}+\mbox{B7}\big)\nonumber\\
& &\qquad - \ 2\times\big(\mbox{C1}+\mbox{C2}+\mbox{C8}+\mbox{C18}+\mbox{C19}
                  +\mbox{C20}+\mbox{C21}+\mbox{C22}\big) \nonumber \\
& &\qquad              + \ \mbox{D3}+\mbox{D4}\Big\} 
             \label{eq:Jaad} \\
\mathcal{J}_{\mu\nu}^{\dot\alpha(2)}&=&\Big\{-\mbox{A4}+2\times \mbox{B1}
- 2\times\big( \mbox{C3}+\mbox{C4}+\mbox{C5}+\mbox{C9} \nonumber\\
& &  \quad               +\ \mbox{C10}+\mbox{C11}+ \mbox{C12}\big)
                     +\mbox{D2}\Big\}
\label{eq:Jadad}
\end{eqnarray}
and 
\begin{eqnarray}
J_{\mu\nu}^{\alpha(2)} &= & \mathcal{J}_{\mu\nu}^{\alpha(2)} - 
\Big(\mathcal{J}_{\mu\nu}^{\dot\alpha(2)}\Big)^\dagger, \\
J_{\mu\nu}^{\alpha\dot\beta}&=&\mathcal{J}_{\mu\nu}^{\alpha\dot\beta}- 
\Big(\mathcal{J}_{\mu\nu}^{\beta\dot\alpha}\Big)^\dagger.
\end{eqnarray}
As we shall see, due to (\ref{eq:kappa_prop}) most of the
contributions obey
$k_{\alpha\alpha}=k_{\alpha\beta}=k_{\beta\alpha}=0$. In these cases
we indicate by subscripts the non-trivial values of $m_1,\tilde
m_1,n_1,\tilde n_1$. In the remaining cases, the subscripts indicate
non-trivial values of
$k_{\alpha\alpha},k_{\alpha\beta},k_{\beta\alpha}$.  Furthermore, in
listing the contributions below we split each one of them into a sum
of sub-contributions A{\footnotesize\#}.{\footnotesize\#} etc., where
the second entry labels distinct spinor index structures.

\subsection{Computation of $\mathcal{L}_{\mu\nu}$}

The quantity $\mathcal{L}_{\mu\nu}$, which is given in
(\ref{eq:curlL}), consists of the three structures given in
(\ref{eq:L1L1}), (\ref{eq:eL2}) and (\ref{eq:eL1L1}), which are
evaluated below.

\subsubsection{Evaluation of $[\wh{\mathcal{L}}^{(1)}(e_\mu),\wh{\mathcal{L}}^{(1)}(e_\nu)]$}

We have that 
\begin{equation}
\wh{\mathcal{L}}^{(1)}(e)=-\frac{b_1}{2}z^\alpha e_{\alpha\dot{\alpha}}
\bar{\p}_{(y)}^{\dot{\alpha}}\int_0^1\!\!\!\int_0^1 dt'tdt\Phi(-tt'z,\bar{y})
\kappa(tt'z,y).
\end{equation}
With our conventions it follows that
\begin{eqnarray}
&&
\big(\wh{\mathcal{L}}^{(1)}(e_{\mu})\big)^{\alpha(m)\dot{\alpha}(\tilde{m}),\beta(n)}=
-\frac{b_1}{2}
\frac{(-1)^{\tilde{m}+1}}{(n+1)n}\frac{n!}{m!(\tilde{m}+1)!}
\nonumber \\ &&
\qquad \qquad \qquad 
\times e^{\;\;\beta}_{\mu,\;\dot{\alpha}}
\Phi^{\beta(\tilde{m}+1)
\dot{\alpha}(\tilde{m}+1)}
\kappa^{\alpha(m),\beta(n-\tilde{m}-2)},\;\;n\geq 1.
\end{eqnarray}
Using the results of the previous section we find
\begin{eqnarray}
&&[\wh{\mathcal{L}}^{(1)}(e_\mu),\wh{\mathcal{L}}^{(1)}(e_\nu)]
^{\alpha(m)\dot{\alpha}(\tilde{m}),\beta(n)}
=b_1^2\sum_{k_{\alpha\alpha}+k_{\beta\beta}+\tilde{k}_{\alpha\alpha}=odd}
C(K(m,\tilde{m},n,0) \nonumber \\&&\times
\frac{(-1)^{m+k_{\alpha\beta}+k_{\beta\alpha}}}
{2(n_1+k_{\beta\alpha}+k_{\beta\beta}+1)(n_2+k_{\alpha\beta}+k_{\beta\beta}+1)
(n_1+k_{\beta\alpha}+k_{\beta\beta})(n_2+k_{\alpha\beta}+k_{\beta\beta})}
 \nonumber \\ &&\times
\frac{(n_1+k_{\beta\alpha}+k_{\beta\beta})!
(n_2+k_{\alpha\beta}+k_{\beta\beta})!}{(m_1+k_{\alpha\alpha}+k_{\alpha\beta})!
(m_2+k_{\alpha\alpha}+k_{\beta\alpha})!
(\tilde{m}_1+\tilde{k}_{\alpha\alpha}+1)!
(\tilde{m}_2+\tilde{k}_{\alpha\alpha}+1)!}
 \nonumber \\&&\times
e^{\;\;\beta_1}_{\mu,\;\dot{\alpha}_1}\Phi^{\beta_1(\tilde{m}_1+
\tilde{k}_{\alpha\alpha}+1)\dot{\alpha}_1(\tilde{m}_1+
\tilde{k}_{\alpha\alpha}+1)}
\kappa^{\alpha_1(m_1+k_{\alpha\alpha}+k_{\alpha\beta}),
\beta_1(n_1+k_{\beta\alpha}+k_{\beta\beta}-
\tilde{m}_1-\tilde{k}_{\alpha\alpha}-2)} \nonumber \\&&\times
e^{\;\;\beta_2}_{\nu,\;\dot{\alpha}_2}\Phi^{\beta_2(\tilde{m}_2+
\tilde{k}_{\alpha\alpha}+1)\dot{\alpha}_2(\tilde{m}_2+
\tilde{k}_{\alpha\alpha}+1)}
\kappa^{\alpha_2(m_2+k_{\alpha\alpha}+k_{\alpha\beta}),
\beta_2(n_2+k_{\beta\alpha}+k_{\beta\beta}-
\tilde{m}_2-\tilde{k}_{\alpha\alpha}-2)}
 \nonumber \\&&\times
\delta^{\alpha(m)}_{\alpha_1(m_1)\alpha_2(m_2)}
\delta^{\dot{\alpha}(\tilde{m})}_{\dot{\alpha}_1(\tilde{m}_1)
\dot{\alpha}_2(\tilde{m}_2)}\delta^{\beta(n)}_{\beta_1(n_1)\beta_2(n_2)}
 \nonumber \\&&\times\eps_{\alpha_1(k_{\alpha\alpha})\alpha_2(k_{\alpha\alpha})}
\eps_{\alpha_1(k_{\alpha\beta})\beta_2(k_{\alpha\beta})}
\eps_{\beta_1(k_{\beta\alpha})\alpha_2(k_{\beta\alpha})}
\eps_{\beta_1(k_{\beta\beta})\beta_2(k_{\beta\beta})}
\eps_{\dot{\alpha}_1(\tilde{k}_{\alpha\alpha})
\dot{\alpha}_2(\tilde{k}_{\alpha\alpha})}
\end{eqnarray}
A1.1-2:
\begin{eqnarray*}
&&[\wh{\mathcal{L}}^{(1)}(e_\mu),\wh{\mathcal{L}}^{(1)}(e_\nu)]
^{\dot{\alpha}(2)}
_{\tilde{m}_1=1}=-ib_1^2\sum_{k=0}^{\infty}
\frac{k}{(k+3)^2(k+2)^2(k+1)(k!)^2}\\&\times&\left\{
(\bar{e}_{\mu}e_{\nu})_{\dot{\gamma}_2(2)}
\Phi^{\dot{\alpha}(1)\dot{\gamma}_2}
_{\gamma_1(k+1)\dot{\gamma}_1(k-1)}
\Phi^{\dot{\alpha}(1)\dot{\gamma}_2
\gamma_1(k+1)\dot{\gamma}_1(k-1)}\right.\\
&+&\left.(k+1)e_{\mu,\gamma_2\dot{\gamma}_2}
e_{\nu,\gamma_3\dot{\gamma}_3}
\Phi^{\dot{\alpha}(1)\dot{\gamma}_2\gamma_3}
_{\gamma_1(k)\dot{\gamma}_1(k-1)}
\Phi^{\dot{\alpha}(1)\dot{\gamma}_3\gamma_2
\gamma_1(k)\dot{\gamma}_1(k-1)}\right\}
\end{eqnarray*} 
A2.1-3:
\begin{eqnarray*}
&&[\wh{\mathcal{L}}^{(1)}(e_\mu),\wh{\mathcal{L}}^{(1)}(e_\nu)]^{\alpha(1)\dot{\alpha}(1)}
_{m_1=1,\tilde{m}_1=0}
=-\frac{b_1^2}{2}\sum_{k=0}^{\infty}
\frac{k}{(k+3)^2(k+2)^2(k!)^2}
\\&\times&\left\{(\bar{e}_{\mu}e_{\nu})_{\dot{\gamma}_2(2)}
\Phi^{\dot{\gamma}_2}
_{\gamma_1(k)\dot{\gamma}_1(k-1)}
\Phi^{\alpha(1)\dot{\alpha}(1)\dot{\gamma}_2
\gamma_1(k)\dot{\gamma}_1(k-1)}\right.
\\&&-e_{\mu,\gamma_2\dot{\gamma}_2}e_{\nu,\dot{\gamma}_1}^{\alpha(1)}
\Phi^{\dot{\gamma}_2}
_{\gamma_1(k)\dot{\gamma}_1(k-1)}
\Phi^{\dot{\alpha}(1)\gamma_2
\gamma_1(k)\dot{\gamma}_1(k)}
\\&&+\left.
k e_{\mu,\gamma_3\dot{\gamma}_2}e_{\nu,\gamma_2\dot{\gamma}_1}
\Phi^{\dot{\gamma}_2\gamma_2}_{\gamma_1(k-1)\dot{\gamma}_1(k-1)}
\Phi^{\alpha(1)\dot{\alpha}(1)\gamma_3
\gamma_1(k-1)\dot{\gamma}_1(k)}\right\}
\end{eqnarray*}
A3.1-3:
\begin{eqnarray*}
&&[\wh{\mathcal{L}}^{(1)}(e_\mu),\wh{\mathcal{L}}^{(1)}(e_\nu)]^{\alpha(1)\dot{\alpha}(1)}
_{m_1=0,\tilde{m}_1=1}
=-[\wh{\mathcal{L}}^{(1)}(e_\nu),\wh{\mathcal{L}}^{(1)}(e_\mu)]^{\alpha(1)\dot{\alpha}(1)}
_{m_1=1,\tilde{m}_1=0}
\end{eqnarray*}
A4.1-4:
\begin{eqnarray*}
&&[\wh{\mathcal{L}}^{(1)}(e_\mu),\wh{\mathcal{L}}^{(1)}(e_\nu)]^{\alpha(2)}_{m_1=1}
=ib_1^2\sum_{k=0}^{\infty}
\frac{1}{(k+4)^2(k+3)^2(k+1)(k!)^2}\\&\times&
\left\{e_{\mu\dot{\gamma}_2}^{\alpha(1)}
e_{\nu\dot{\gamma}_1}^{\alpha(1)}
\Phi^{\dot{\gamma}_2}_{\gamma_1(k+1)\dot{\gamma}_1(k)}
\Phi^{\gamma_1(k+1)\dot{\gamma}_1(k+1)}\right.  
\\&&-\frac{2(k+1)}{k+2}
e_{[\mu,\dot{\gamma}_2}^{\alpha(1)}e_{\nu],\gamma_2\dot{\gamma}_1}
\Phi^{\dot{\gamma}_2\gamma_2}_{\gamma_1(k)\dot{\gamma}_1(k)}
\Phi^{\alpha(1)\gamma_1(k)\dot{\gamma}_1(k+1)}\\&&-
\frac{(k+1)k}{(k+2)}e_{\mu,\gamma_2\dot{\gamma}_2}
e_{\nu\gamma_3\dot{\gamma}_3}
\Phi^{\alpha(1)\dot{\gamma}_2\gamma_3}
_{\gamma_1(k-1)\dot{\gamma}_1(k)}
\Phi^{\alpha(1)\gamma_2\dot{\gamma}_3\gamma_1(k-1)
\dot{\gamma}_1(k)}
\\&&-\left.
(k+1)(\bar{e}_{\mu}e_{\nu})_{\dot{\gamma}_2(2)}
\Phi^{\alpha(1)\dot{\gamma}_2}_{\gamma_1(k)\dot{\gamma}_1(k)}
\Phi^{\alpha(1)\dot{\gamma}_2\gamma_1(k)\dot{\gamma}_1(k)}
\right\}
\end{eqnarray*}
In summary we find
\begin{eqnarray}
[\wh{\mathcal{L}}^{(1)}(e_\mu),\wh{\mathcal{L}}^{(1)}(e_\nu)]^{\dot{\alpha}(2)}
&=& \mbox{A1}\nonumber\\{}
[\wh{\mathcal{L}}^{(1)}(e_\mu),\wh{\mathcal{L}}^{(1)}(e_\nu)]^{\alpha(1)\dot{\alpha}(1)}
&=& \mbox{A2}+\mbox{A3} \nonumber\\{}
[\wh{\mathcal{L}}^{(1)}(e_\mu),\wh{\mathcal{L}}^{(1)}(e_\nu)]^{\alpha(2)}
&=& \mbox{A4}.
\end{eqnarray}

%==========================================================================

\subsubsection{Evaluation of $[e_\mu,\wh{\mathcal{L}}^{(2)}(e_\nu)]$}

From the definition of $e$ and $\wh{\mathcal{L}}^{(2)}$ one finds
\begin{eqnarray}
[e_\mu,\wh{\mathcal{L}}^{(2)}(e_\nu)]=-\frac{1}{2}e_\mu^{\beta\dot{\beta}}
e_\nu^{\alpha\dot{\alpha}}\left[y_\beta \bar{y}_{\dot{\beta}},
\int_0^1\frac{dt}{t}\bar{\p}^{(y)}_{\dot{\alpha}}A_\alpha^{(2)}(z\to tz)\right]
\label{eq:eL2def}
\end{eqnarray}
where
\begin{eqnarray}
A_\alpha^{(2)}=\frac{z_\alpha}{2}\eps^{\delta\gamma}
\int_0^1 tdt \left[A^{(1)}_\gamma,A^{(1)}_\delta\right](z\to tz)
-z_\alpha\int^1_0 tdt\frac{ib_1}{2}\hat{\Phi}^{(2)}\star\kappa(z\to tz).
\label{eq:A2def}
\end{eqnarray}
Substituting (\ref{eq:A2def}) into (\ref{eq:eL2def}) yields 
\begin{eqnarray}
&&[e_\mu,\wh{\mathcal{L}}^{(2)}(e_\nu)]^{\alpha(m)\dot{\alpha}(\tilde{m})}=
\frac{1}{2}
\eps^{\delta\gamma}(\bar{e}_{[\mu} e_{\nu]})_{\dot{\alpha}(2)}
\left[A^{(1)}_\gamma,A^{(1)}_\delta\right]^{\alpha(m)\dot{\alpha}(\tilde{m}+2)}
 \nonumber \\&& \qquad \qquad \qquad 
+ \frac{ib_1(-1)^{m+1}(m-1)!}{4}\eps^{\delta\beta(1)}
(\bar{e}_{[\mu} e_{\nu]})_{\dot{\alpha}(2)} \nonumber \\
&& \qquad \qquad \qquad 
\times \left(
\Phi\star \left\{A^{(1)}_\delta|_{\tilde y\to -\tilde y}\right\}
- A^{(1)}_\delta\star\Phi\right)
^{\dot\alpha(\tilde m +2),\beta(m-1)}\delta^{\alpha(m)}_{\beta(m)}.
\label{eq:eL2vidare}
\end{eqnarray}
Using
\begin{equation}
A_\alpha^{(1)}=-\frac{ib_1}{2}z_\alpha\int_0^1 t dt 
\Phi(-tz,\bar{y})\kappa(tz,y)
\end{equation}
and
\begin{eqnarray}
&& A_{\gamma}^{(1)\alpha(m)\dot{\alpha}(\tilde{m}),\beta(n)}=-\frac{ib_1}{2}
\frac{(-1)^{n-m-1}}{n+1}\frac{n!}{m!\tilde{m}!} \nonumber \\
&& \qquad \qquad \qquad \qquad 
\times \Phi^{\alpha'(\tilde{m})\dot{\alpha}(\tilde{m})}
\kappa^{\alpha(m),\beta'(n-\tilde{m}-1)}
\delta_{\gamma\alpha'(n-m-1)\beta'(m)}^{\beta(n)}
\end{eqnarray}
we determine the first term
\begin{eqnarray}\label{eq:AA_com}
&& \!\!\!\!\!\!\!\!\!\!\!\!\!\!\! [A_\gamma^{(1)},A_\delta^{(1)}]^{\alpha(m)\dot{\alpha}(\tilde{m}),\beta(n)}
=
\frac{-b_1^2(-1)^{n-m}}{2} \nonumber \\&\times&\sum_{k_{\alpha\alpha}+k_{\beta\beta}+
\tilde{k}_{\alpha\alpha}=odd}
\frac{C(K(m,\tilde{m},n,0))}{(n_1+k_{\beta\alpha}+k_{\beta\beta}+1)!
(n_2+k_{\alpha\beta}+k_{\beta\beta}+1)!}  \nonumber \\&\times&
\frac{(n_1+k_{\beta\alpha}+k_{\beta\beta})!
(n_2+k_{\alpha\beta}+k_{\beta\beta})!}
{(m_1+k_{\alpha\alpha}+k_{\alpha\beta})!
(m_2+k_{\alpha\alpha}+k_{\beta\alpha})!(\tilde{m}_1+\tilde{k}_{\alpha\alpha})!
(\tilde{m}_2+\tilde{k}_{\alpha\alpha})!} \nonumber \\&\times&
\Phi^{\alpha_1'(\tilde{m_1}+\tilde{k}_{\alpha\alpha})
\dot{\alpha}_1(\tilde{m_1})+\tilde{k}_{\alpha\alpha}}
\kappa^{\alpha_1(m_1+k_{\alpha\alpha}+k_{\alpha\beta}),
\beta_1'(n_1+k_{\beta\alpha}+k_{\beta\beta}-\tilde{m}_1-
\tilde{k}_{\alpha\alpha}-1)} \nonumber \\&\times&
\delta_{\gamma\alpha_1'(n_1+k_{\beta\alpha}+k_{\beta\beta}-
(m_1+k_{\alpha\alpha}+k_{\alpha\beta})-1)
\beta_1'(m_1+k_{\alpha\alpha}+k_{\alpha\beta})}
^{\beta_1(n_1+k_{\beta\alpha}+k_{\beta\beta})} \nonumber \\&\times&
\Phi^{\alpha_2'(\tilde{m_2}+\tilde{k}_{\alpha\alpha})
\dot{\alpha}_2(\tilde{m_2})+\tilde{k}_{\alpha\alpha}}
\kappa^{\alpha_2(m_2+k_{\alpha\alpha}+k_{\alpha\beta}),
\beta_2'(n_2+k_{\beta\alpha}+k_{\beta\beta}-\tilde{m}_2-
\tilde{k}_{\alpha\alpha}-1)} \nonumber \\&\times&
\delta_{\delta\alpha_2'(n_2+k_{\beta\alpha}+k_{\beta\beta}-
(m_2+k_{\alpha\alpha}+k_{\alpha\beta})-1)
\beta_2'(m_2+k_{\alpha\alpha}+k_{\alpha\beta})}
^{\beta_2(n_2+k_{\beta\alpha}+k_{\beta\beta})}
 \nonumber \\&\times&\delta^{\alpha(m)}_{\alpha_1(m_1)\alpha_2(m_2)}
\delta^{\dot{\alpha}(\tilde{m})}_{\dot{\alpha}_1(\tilde{m}_1)
\dot{\alpha}_2(\tilde{m}_2)}\delta^{\beta(n)}_{\beta_1(n_1)\beta_2(n_2)}
 \nonumber \\&\times&\eps_{\alpha_1(k_{\alpha\alpha})\alpha_2(k_{\alpha\alpha})}
\eps_{\alpha_1(k_{\alpha\beta})\beta_2(k_{\alpha\beta})}
\eps_{\beta_1(k_{\beta\alpha})\alpha_2(k_{\beta\alpha})}
\eps_{\beta_1(k_{\beta\beta})\beta_2(k_{\beta\beta})}
\eps_{\dot{\alpha}_1(\tilde{k}_{\alpha\alpha})
\dot{\alpha}_2(\tilde{k}_{\alpha\alpha})}.
\end{eqnarray}
In the end, the results are
\newline
B1.1-2:
\begin{eqnarray*}
&&\frac{1}{2}
\eps^{\delta\gamma}(\bar{e}_{[\mu} e_{\nu]})_{\dot{\alpha}(2)}
[A_\gamma^{(1)},A_\delta^{(1)}]^{\alpha(0)\dot{\alpha}(4)}
_{\tilde{m}_1=2}=
\frac12 ib_1^2\sum_{k=0}^{\infty}
\frac{1}{(k+4)(k+2)(k+1)(k!)^2}\\
&\times&\left(2(\bar{e}_{[\mu} e_{\nu]})_{\dot{\gamma}_2(2)}
\Phi^{\dot{\alpha}(1)\dot\gamma_2}_{\gamma_1(k+1)\dot{\gamma}_1(k-1)}
\Phi^{\dot{\alpha}(1)\dot\gamma_2\gamma_1(k+1)\dot{\gamma}_1(k-1)}\right.\\
&+&\left.(\bar{e}_{[\mu} e_{\nu]})_{\dot{\gamma}_2(2)}
\Phi^{\dot\gamma_2(2)}_{\gamma_1(k+1)\dot{\gamma}_1(k-1)}
\Phi^{\dot{\alpha}(2)\gamma_1(k+1)\dot{\gamma}_1(k-1)}\right)
\end{eqnarray*}
B2:
\begin{eqnarray*}
&&\frac{1}{2}
\eps^{\delta\gamma}(\bar{e}_{[\mu} e_{\nu]})_{\dot{\alpha}_2(2)}
[A_\gamma^{(1)},A_\delta^{(1)}]^{\alpha(2)\dot{\alpha}(2)}
_{m_1=\tilde{m}_1=1}=
ib_1^2\sum_{k=0}^{\infty}
\frac{1}{(k+4)(k!)^2}\\
&\times&(\bar{e}_{[\mu} e_{\nu]})_{\dot{\gamma}_2(2)}
\Phi^{\alpha(1)\dot{\gamma}_2(1)}_{\gamma_1(k)\dot{\gamma}_1(k)}
\Phi^{\alpha(1)\dot{\gamma}_2(1)\gamma_1(k)\dot{\gamma}_1(k)}
\end{eqnarray*}
B3:
\begin{eqnarray*}
&&\frac{1}{2}
\eps^{\delta\gamma}(\bar{e}_{[\mu} e_{\nu]})_{\dot{\alpha}(2)}
[A_\gamma^{(1)},A_\delta^{(1)}]^{\alpha(2)\dot{\alpha}(2)}
_{m_1=2,\tilde{m}_1=0}=
-\frac{ib_1^2}{2}\sum_{k=0}^{\infty}
\frac{1}{(k+4)(k!)^2}\\
&\times&(\bar{e}_{[\mu} e_{\nu]})_{\dot{\gamma}_2(2)}\Phi_{\gamma_1(k)\dot{\gamma}_1(k)}
\Phi^{\alpha(2)\dot{\gamma_2(2)}\gamma_1(k)\dot{\gamma}_1(k)}
\end{eqnarray*}
B4:
\begin{eqnarray*}
&&\frac{1}{2}
\eps^{\delta\gamma}(\bar{e}_{[\mu} e_{\nu]})_{\dot{\alpha}(2)}
[A_\gamma^{(1)},A_\delta^{(1)}]^{\alpha(2)\dot{\alpha}(2)}
_{m_1=0,\tilde{m}_1=2}=
-\frac{1}{2}
\eps^{\delta\gamma}(\bar{e}_{[\mu} e_{\nu]})_{\dot{\alpha}(2)}
\eps^{\delta\gamma}[A_\delta^{(1)},A_\gamma^{(1)}]^{\alpha(2)\dot{\alpha}(2)}
_{m_1=2,\tilde{m}_1=0}
\end{eqnarray*}
B5.1-2:
\begin{eqnarray*}
&&\frac{1}{2}
\eps^{\delta\gamma}(\bar{e}_{[\mu} e_{\nu]})_{\dot{\alpha}(2)}
[A_\gamma^{(1)},A_\delta^{(1)}]^{\alpha(1)\dot{\alpha}(3)}
_{m_1=1}=
\frac{b_1^2}{4}
\sum_{k=0}^{\infty} 
\frac{k}{(k+3)(k!)^2}\\&\times&\left(2(\bar{e}_{[\mu} e_{\nu]})_{\dot{\gamma}_2(2)}e^{\kappa\dot{\delta}}
\Phi^{\dot{\gamma}_2(1)}_{\gamma_1(k)\dot{\gamma}_1(k-1)}
\Phi^{\alpha(1)\dot{\alpha}(1)\dot\gamma_2(1)\gamma_1(k)\dot{\gamma}_1(k-1)}\right.\\
&+&\left.(\bar{e}_{[\mu} e_{\nu]})_{\dot{\gamma}_2(2)}e^{\kappa\dot{\delta}}
\Phi^{\dot{\alpha}(1)}_{\gamma_1(k)\dot{\gamma}_1(k-1)}
\Phi^{\alpha(1)\dot\gamma_2(2)\gamma_1(k)\dot{\gamma}_1(k-1)}\right)
\end{eqnarray*}
B6.1-2:
\begin{eqnarray*}
&&\frac{1}{2}
\eps^{\delta\gamma}(\bar{e}_{[\mu} e_{\nu]})_{\dot{\alpha}(2)}
[A_\gamma^{(1)},A_\delta^{(1)}]^{\alpha(1)\dot{\alpha}(3)}
_{m_1=0}=-\frac{1}{2}
\eps^{\delta\gamma}(\bar{e}_{[\mu} e_{\nu]})_{\dot{\alpha}(2)}
[A_\delta^{(1)},A_\gamma^{(1)}]^{\alpha(1)\dot{\alpha}(3)}
_{m_1=1}.
\end{eqnarray*}
Now to the second term. Define the projections
\begin{eqnarray}
P_\pm\equiv\frac{1\pm\bar\pi}{2},
\end{eqnarray}
then it follows
\begin{eqnarray}
\left(
\Phi\star \left\{A^{(1)}_\delta|_{\tilde y\to -\tilde y}\right\}
- A^{(1)}_\delta\star\Phi\right)=
[\Phi,P_+ A^{(1)}_{\delta}]-\{\Phi,P_- A^{(1)}_{\delta}\}.
\end{eqnarray}
We also have
\begin{eqnarray}
&&[\Phi,P_+ A^{(1)}_{\delta}]^{\dot\alpha(\tilde
m),\beta(n)}= 2 \!\!\!\!\!\!\sum_{k_{\alpha\alpha}+\tilde k_{\alpha\alpha}=odd}
C(K(0,\tilde m,n,0))\nonumber\\
&&\times\Phi^{\alpha_1(m_1+k_{\alpha\alpha}+k_{\alpha\beta})
\dot\alpha_1(m_1+\tilde k_{\alpha\alpha})}
\left(P_+A_\delta^{(1)}\right)^{\alpha_2(m_2+k_{\alpha\alpha})
\dot\alpha_2(\tilde m_2+\tilde k_{\alpha\alpha}),\beta_2(n+k_{\alpha\beta})}
\nonumber\\
&&\times
\delta^{\dot{\alpha}(\tilde{m})}_{\dot{\alpha}_1(\tilde{m}_1)
\dot{\alpha}_2(\tilde{m}_2)}
\delta^{\beta(n)}_{\beta_1(n_1)\beta_2(n_2)}
\eps_{\alpha_1(k_{\alpha\alpha})\alpha_2(k_{\alpha\alpha})}
\eps_{\alpha_1(k_{\alpha\beta})\beta_2(k_{\alpha\beta})}
\eps_{\dot{\alpha}_1(\tilde{k}_{\alpha\alpha})
\dot{\alpha}_2(\tilde{k}_{\alpha\alpha})}
\end{eqnarray}
\begin{eqnarray}
&&\{\Phi,P_- A^{(1)}_{\delta}\}^{\dot\alpha(\tilde
m),\beta(n)}= 2 \!\!\!\!\!\!\sum_{k_{\alpha\alpha}+\tilde k_{\alpha\alpha}=even}
C(K(0,\tilde m,n,0))\nonumber\\
&&\times\Phi^{\alpha_1(m_1+k_{\alpha\alpha}+k_{\alpha\beta})
\dot\alpha_1(m_1+\tilde k_{\alpha\alpha})}
\left(P_-A_\delta^{(1)}\right)^{\alpha_2(m_2+k_{\alpha\alpha})
\dot\alpha_2(\tilde m_2+\tilde k_{\alpha\alpha}),\beta_2(n+k_{\alpha\beta})}
\nonumber\\
&&\times
\delta^{\dot{\alpha}(\tilde{m})}_{\dot{\alpha}_1(\tilde{m}_1)
\dot{\alpha}_2(\tilde{m}_2)}
\delta^{\beta(n)}_{\beta_1(n_1)\beta_2(n_2)}
\eps_{\alpha_1(k_{\alpha\alpha})\alpha_2(k_{\alpha\alpha})}
\eps_{\alpha_1(k_{\alpha\beta})\beta_2(k_{\alpha\beta})}
\eps_{\dot{\alpha}_1(\tilde{k}_{\alpha\alpha})
\dot{\alpha}_2(\tilde{k}_{\alpha\alpha})}
\end{eqnarray}
where
\begin{equation}
\left(P_+A_\delta^{(1)}\right)^{\alpha(m)
\dot\alpha(\tilde m),\beta(n)}=\left\{\begin{array}{ccc}A_\delta^{(1),\alpha(m)
\dot\alpha(\tilde m),\beta(n)}\;\;&\mbox{for}&\;\;\tilde m=\mbox{even}\\
0\;\;&\mbox{for}&\;\;\tilde m=\mbox{odd}\end{array}\right.
\end{equation}
\begin{equation}
\left(P_-A_\delta^{(1)}\right)^{\alpha(m)
\dot\alpha(\tilde m),\beta(n)}=\left\{\begin{array}{ccc}A_\delta^{(1),\alpha(m)
\dot\alpha(\tilde m),\beta(n)}\;\;&\mbox{for}&\;\;\tilde m=\mbox{odd}\\
0\;\;&\mbox{for}&\;\;\tilde m=\mbox{even}\end{array}\right.
\end{equation}
Altogether this amounts to
\newline
B7.1-2:
\begin{eqnarray*}
&&\frac{ib_1}{4}(\bar{e}_{[\mu} e_{\nu]})_{\dot{\alpha}_2(2)}\eps^{\delta\beta(1)}
\left(\Phi\star \left\{A^{(1)}_\delta|_{\bar y\to -\bar y}\right\}
- A^{(1)}_\delta\star\Phi\right)^{\dot\alpha(3)}\delta^{\alpha(1)}_{\beta(1)}\\
&=&\frac14 b_1^2
\sum_{k=0}^{\infty}\frac{k}{(k!)^2}\left(2\eps^{\delta\alpha(1)}(\bar{e}_{[\mu} e_{\nu]})_{\dot{\gamma}_2(2)}
\Phi^{\dot\alpha(1)\dot\gamma_2(1)}_{\delta\gamma_1(k)\dot\gamma_1(k-1)}
\Phi^{\dot\gamma_2(1)\gamma_1(k)\dot\gamma_1(k-1)}\right.\\&+&\left.
\eps^{\delta\alpha(1)}(\bar{e}_{[\mu} e_{\nu]})_{\dot{\gamma}_2(2)}
\Phi^{\dot\gamma_2(2)}_{\delta\gamma_1(k)\dot\gamma_1(k-1)}
\Phi^{\dot\alpha(1)\gamma_1(k)\dot\gamma_1(k-1)}\right)
\end{eqnarray*}
B8.1-2:
\begin{eqnarray*}
&&-\frac{ib_1}{4}(\bar{e}_{[\mu} e_{\nu]})_{\dot{\alpha}_2(2)}\eps^{\delta\beta(1)}
\left(\Phi\star \left\{A^{(1)}_\delta|_{\bar y\to -\bar y}\right\}
- A^{(1)}_\delta\star\Phi\right)^{\dot\alpha(2),\beta(1)}\delta^{\alpha(2)}_{\beta(2)}\\&=&\frac{i}{2}b_1^2
\sum_{k=0}^{\infty}\frac{1}{(k+3)(k!)^2}(\bar{e}_{[\mu} e_{\nu]})_{\dot{\gamma}_2(2)}\\&\times&
\left\{
\eps^{\delta\alpha(1)}\Phi^{\dot\gamma_2(1)}_{\delta\gamma_1(k)\dot\gamma_1(k)}
\Phi^{\alpha(1)\dot\gamma_2(1)\gamma_1(k)\dot\gamma_1(k)}+
\eps^{\delta\alpha(1)}\Phi^{\alpha(1)\dot\gamma_2(2)}_{\delta\gamma_1(k)\dot\gamma_1(k)}
\Phi^{\gamma_1(k)\dot\gamma_1(k)}\right\}.
\end{eqnarray*}
In summary we find
\begin{eqnarray}
[e_\mu,\wh{\mathcal{L}}^{(2)}(e_\nu)]^{\alpha(2)}&=&\mbox{B2}+\mbox{B3}+\mbox{B4}+\mbox{B8}\nonumber\\{}
[e_\mu,\wh{\mathcal{L}}^{(2)}(e_\nu)]^{\alpha(1)\dot{\alpha}(1)}
&=&\mbox{B5}+\mbox{B6}+\mbox{B7}\nonumber\\{}
[e_\mu,\wh{\mathcal{L}}^{(2)}(e_\nu)]^{\dot{\alpha}(2)}&=&\mbox{B1}
\end{eqnarray}

%-------------------------------------------------------------

\subsubsection{Evaluation of $[e_\mu,\wh{\mathcal{L}}^{(1)} \circ
\wh{\mathcal{L}}^{(1)}(e_\nu)]$}

From the definition of $\wh{\mathcal{L}}^{(1)}$ one obtains
\begin{eqnarray}
&& \wh{\mathcal{L}}^{(1)} \circ \wh{\mathcal{L}}^{(1)}(e)
=-i\eps^{\delta\gamma}\int_0^1\frac{dt}{t}
\Big(\left\{A^{(1)}_\gamma ,
\p_\delta^{(z)}\wh{\mathcal{L}}^{(1)}(e)\right\} \nonumber \\
&& \qquad \qquad \qquad \qquad \qquad \qquad 
-\left[A^{(1)}_\gamma ,
\p_\delta^{(y)}\wh{\mathcal{L}}^{(1)}(e)\right]\Big)(z\to tz). 
\end{eqnarray}
The commutator in (\ref{eq:eL1L1}) is
\begin{eqnarray}
[e_\mu,\wh{\mathcal{L}}^{(1)} \circ \wh{\mathcal{L}}^{(1)}(e_\nu)]^{\alpha(m)\dot{\alpha}(\tilde{m})}&=&
-ie_{\mu,\beta\dot{\alpha}}\left(\wh{\mathcal{L}}^{(1)} \circ 
\wh{\mathcal{L}}^{(1)}(e_\nu)\right)^{\alpha(m)\dot{\alpha}(\tilde{m}+1)\beta}
\end{eqnarray}
with the components
\begin{eqnarray}
&& \left(\wh{\mathcal{L}}^{(1)} \circ \wh{\mathcal{L}}^{(1)}(e_\nu)\right)^
{\alpha(m)\dot{\alpha}(\tilde{m}+1),\beta}=-i\eps^{\delta\gamma}
\Big(\left\{A^{(1)}_\gamma ,
\p_\delta^{(z)}\wh{\mathcal{L}}^{(1)}(e)\right\} \nonumber \\
&& \qquad \qquad \qquad \qquad \qquad \qquad 
-\left[A^{(1)}_\gamma ,
\p_\delta^{(y)}\wh{\mathcal{L}}^{(1)}(e)\right]\Big)^{\alpha(m)\dot{\alpha}(\tilde{m}+1),\beta}
\end{eqnarray}
where
\begin{eqnarray}
&& \left(\p_{\delta}^{(z)}\wh{\mathcal{L}}^{(1)}(e)\right)
^{\alpha(m)\dot{\alpha}(\tilde{m}),\beta(n)}
=-\frac{i}{n+1}e^{\kappa\dot{\delta}}
\left(\bar{\p}^{(y)}_{\dot{\delta}}\p_\delta^{(z)}A^{(1)}_\kappa\right)
^{\alpha(m)\dot{\alpha}(\tilde{m}),\beta(n)} \nonumber \\
&& = -\frac{i}{n+1}e^{\kappa\dot{\delta}}\eps_{\dot{\delta}\dot{\alpha}}
\eps_{\delta\beta}A_\kappa^{(1)\alpha(m)\dot{\alpha}(\tilde{m}+1),\beta(n+1)} 
 =-\frac{b_1}{2}
\frac{(-1)^{n-m}}{n+2}\frac{n!}{m!(\tilde{m}+1)!}
e^{\kappa\dot{\delta}}\eps_{\delta\beta} \nonumber \\
&& 
\qquad \qquad \qquad \qquad \qquad 
\times \Phi^{\alpha'(\tilde{m}+1)\dot{\alpha}(\tilde{m})}_{\dot{\delta}}
\kappa^{\alpha(m),\beta'(n-\tilde{m}-1)} 
\delta_{\kappa\alpha'(n-m)\beta'(m)}^{\beta(n+1)}. 
\end{eqnarray}
and
\begin{eqnarray}
&& \left(\p_{\delta}^{(y)}\wh{\mathcal{L}}^{(1)}(e)\right)
^{\alpha(m)\dot{\alpha}(\tilde{m}),\beta(n)}
=-\frac{i}{n}e^{\kappa\dot{\delta}}
\left(\bar{\p}^{(y)}_{\dot{\delta}}\p_\delta^{(y)}A^{(1)}_\kappa\right)
^{\alpha(m)\dot{\alpha}(\tilde{m}),\beta(n)} \nonumber \\
&& = -\frac{i}{n}e^{\kappa\dot{\delta}}\eps_{\dot{\delta}\dot{\alpha}}
\eps_{\delta\alpha}A_\kappa^{(1)\alpha(m+1)\dot{\alpha}(\tilde{m}+1),\beta(n)}
= -\frac{b_1}{2}
\frac{(-1)^{n-m}}{n+1}\frac{(n-1)!}{(m+1)!(\tilde{m}+1)!}
e^{\kappa\dot{\delta}}\eps_{\delta\alpha}
 \nonumber \\
&& 
\qquad \qquad \qquad \qquad \qquad 
\times \Phi^{\alpha'(\tilde{m}+1)\dot{\alpha}(\tilde{m})}_{\dot{\delta}}
\kappa^{\alpha(m+1),\beta'(n-\tilde{m}-2)}
\delta_{\kappa\alpha'(n-m-2)\beta'(m+1)}^{\beta(n)}.
\end{eqnarray}
The general expressions needed then follows as 
\begin{eqnarray}
&&\!\!\!\!\!\!\!\!\!\!\!\!\!\left\{A^{(1)}_\gamma ,
\p_\delta^{(z)}\wh{\mathcal{L}}^{(1)}(e)\right\}^{\alpha(m)
\dot{\alpha}(\tilde{m}),\beta(n)}=
\frac{ib_1^2(-1)^{n-m-1}}{2} \nonumber \\&\times&
\sum_{k_{\alpha\alpha}+k_{\beta\beta}+
\tilde{k}_{\alpha\alpha}=even} 
\frac{C(K(m,\tilde{m},n,0))(n_1+k_{\beta\alpha}+k_{\beta\beta})!
(n_2+k_{\alpha\beta}+k_{\beta\beta})!}
{(n_1+k_{\beta\alpha}+k_{\beta\beta}+1)
(n_2+k_{\alpha\beta}+k_{\beta\beta}+2)}
 \nonumber \\&\times&
\frac{1}{(m_1+k_{\alpha\alpha}+k_{\alpha\beta})!
(m_2+k_{\alpha\alpha}+k_{\beta\alpha})!
(\tilde{m}_1+\tilde{k}_{\alpha\alpha})!
(\tilde{m}_2+\tilde{k}_{\alpha\alpha}+1)!} \nonumber \\
&\times&e^{\kappa\dot{\delta}}
\Phi^{\alpha_1'(\tilde{m_1}+\tilde{k}_{\alpha\alpha})
\dot{\alpha}_1(\tilde{m}_1+\tilde{k}_{\alpha\alpha})}
\kappa^{\alpha_1(m_1+k_{\alpha\alpha}+k_{\alpha\beta}),
\beta_1'(n_1+k_{\beta\alpha}+k_{\beta\beta}
-\tilde{m}_1-\tilde{k}_{\alpha\alpha}-1)} \nonumber \\&\times&
\delta_{\gamma\alpha_1'(n_1+k_{\beta\alpha}+k_{\beta\beta}
-m_1-k_{\alpha\alpha}-k_{\alpha\beta}-1)
\beta_1'(m_1+k_{\alpha\alpha}+k_{\alpha\beta})}^{\beta_1(n_1+k_{\beta\alpha}+
k_{\beta\beta})}
 \nonumber \\&\times&\eps_{\delta\beta_2}
\Phi^{\alpha_2'(\tilde{m}_2+\tilde{k}_{\alpha\alpha}+1)
\dot{\alpha}_2(\tilde{m}_2+\tilde{k}_{\alpha\alpha})}_{\dot{\delta}}
\kappa^{\alpha_2(m_2+k_{\alpha\alpha}+k_{\beta\alpha}),
\beta_2'(n_2+k_{\alpha\beta}+k_{\beta\beta}-\tilde{m}_2-
\tilde{k}_{\alpha\alpha}-1)} \nonumber \\&\times&
\delta_{\kappa\alpha_2'(n_2+k_{\alpha\beta}+k_{\beta\beta}-
m_2-k_{\alpha\alpha}-k_{\beta\alpha})
\beta_2'(m_2+k_{\alpha\alpha}+k_{\beta\alpha})}
^{\beta_2(n_2+k_{\alpha\beta}+k_{\beta\beta}+1)}
 \nonumber \\&\times&\delta^{\alpha(m)}_{\alpha_1(m_1)\alpha_2(m_2)}
\delta^{\dot{\alpha}(\tilde{m})}_{\dot{\alpha}_1(\tilde{m}_1)
\dot{\alpha}_2(\tilde{m}_2)}\delta^{\beta(n)}_{\beta_1(n_1)\beta_2(n_2)} \nonumber \\
&\times&\eps_{\alpha_1(k_{\alpha\alpha})\alpha_2(k_{\alpha\alpha})}
\eps_{\alpha_1(k_{\alpha\beta})\beta_2(k_{\alpha\beta})}
\eps_{\beta_1(k_{\beta\alpha})\alpha_2(k_{\beta\alpha})}
\eps_{\beta_1(k_{\beta\beta})\beta_2(k_{\beta\beta})}
\eps_{\dot{\alpha}_1(\tilde{k}_{\alpha\alpha})
\dot{\alpha}_2(\tilde{k}_{\alpha\alpha})}
\end{eqnarray}
and
\begin{eqnarray}
&&\!\!\!\!\!\!\!\!\!\!\!\!\!\!\left[A^{(1)}_\gamma ,
\p_\delta^{(y)}\wh{\mathcal{L}}^{(1)}(e)\right]^{\alpha(m)
\dot{\alpha}(\tilde{m}),\beta(n)}=\frac{ib_1^2(-1)^{n-m-1}}{2}
\sum_{k_{\alpha\alpha}+k_{\beta\beta}+\tilde{k}_{\alpha\alpha}=odd} \nonumber \\&\times& 
\frac{C(K(m,\tilde{m},n,0))(n_1+k_{\beta\alpha}+k_{\beta\beta})!
(n_2+k_{\alpha\beta}+k_{\beta\beta}-1)!}{(n_1+k_{\beta\alpha}+k_{\beta\beta}+1)
(n_2+k_{\alpha\beta}+k_{\beta\beta}+1)}
 \nonumber \\&\times&\frac{1}{(m_1+k_{\alpha\alpha}+k_{\alpha\beta})!
(\tilde{m}_1+\tilde{k}_{\alpha\alpha})!
(\tilde{m}_2+\tilde{k}_{\alpha\alpha}+1)!
(m_2+k_{\alpha\alpha}+k_{\beta\alpha}+1)!}
 \nonumber \\&\times&e^{\kappa\dot{\delta}}
\Phi^{\alpha_1'(\tilde{m}_1+\tilde{k}_{\alpha\alpha})
\dot{\alpha}_1(\tilde{m}_1+\tilde{k}_{\alpha\alpha})}
\kappa^{\alpha_1(m_1+k_{\alpha\alpha}+k_{\alpha\beta}),
\beta_1'(n_1+k_{\beta\alpha}+k_{\beta\beta}-
\tilde{m}_1-\tilde{k}_{\alpha\alpha}-1)}
 \nonumber \\&\times&
\delta_{\gamma\alpha_1'(n_1+k_{\beta\alpha}+k_{\beta\beta}
-m_1-k_{\alpha\alpha}-k_{\alpha\beta}-1)\beta_1'
(m_1+k_{\alpha\alpha}+k_{\alpha\beta})}^
{\beta_1(n_1+k_{\beta\alpha}+k_{\beta\beta})}
 \nonumber \\&\times&
\eps_{\delta\alpha_2}
\Phi^{\alpha_2'(\tilde{m}_2+\tilde{k}_{\alpha\alpha}+1)
\dot{\alpha}(\tilde{m}_2+\tilde{k}_{\alpha\alpha})}_{\dot{\delta}}
\kappa^{\alpha_2(m_2+k_{\alpha\alpha}+k_{\beta\alpha}+1),
\beta_2'(n_2+k_{\alpha\beta}+k_{\beta\beta}-
\tilde{m}_2-\tilde{k}_{\alpha\alpha}-2)}
 \nonumber \\&\times&
\delta_{\kappa\alpha_2'(n_2+k_{\alpha\beta}+k_{\beta\beta}-
m_2-k_{\alpha\alpha}-k_{\beta\alpha}-2)
\beta_2'(m_2+k_{\alpha\alpha}+k_{\beta\alpha}+1)}
^{\beta_2(n_2+k_{\alpha\beta}+k_{\beta\beta})}
 \nonumber \\&\times&\delta^{\alpha(m)}_{\alpha_1(m_1)\alpha_2(m_2)}
\delta^{\dot{\alpha}(\tilde{m})}_{\dot{\alpha}_1(\tilde{m}_1)
\dot{\alpha}_2(\tilde{m}_2)}\delta^{\beta(n)}_{\beta_1(n_1)\beta_2(n_2)} \nonumber \\
&\times&\eps_{\alpha_1(k_{\alpha\alpha})\alpha_2(k_{\alpha\alpha})}
\eps_{\alpha_1(k_{\alpha\beta})\beta_2(k_{\alpha\beta})}
\eps_{\beta_1(k_{\beta\alpha})\alpha_2(k_{\beta\alpha})}
\eps_{\beta_1(k_{\beta\beta})\beta_2(k_{\beta\beta})}
\eps_{\dot{\alpha}_1(\tilde{k}_{\alpha\alpha})
\dot{\alpha}_2(\tilde{k}_{\alpha\alpha})}.
\end{eqnarray}
Projecting onto the particular components we are interested in,
starting with terms not satisfying
$k_{\alpha\alpha}=k_{\alpha\beta}=k_{\beta\alpha}=0$, one finds
eight anti-commutators and two commutators:
\newline\newline
C1.1-4:
\begin{eqnarray*}
&&-e_{\mu,\beta\dot\alpha}\eps^{\delta\gamma}\left\{A^{(1)}_\gamma ,
\p_\delta^{(z)}\wh{\mathcal{L}}^{(1)}(e_\nu)\right\}^{\alpha(1)
\dot{\alpha}(2),\beta(1)}_{k_{\beta\alpha=1}}
=\frac{b_1^2}{4}\sum_{k=0}^{\infty}
\frac{(k+2)k^2(k-1)}
{(k+4)(k+3)^2(k!)^2}\\&\times&
e_{\mu,\gamma_2\dot\gamma_2}e_{\nu}^{\kappa\dot{\delta}}
\Phi_{\kappa\gamma_1(k-1)\dot{\gamma}_1(k-1)}
^{\alpha(1)\dot{\alpha}(1)\dot\gamma_2}
\Phi_{\dot{\delta}}
^{\gamma_2\gamma_1(k-1)\dot{\gamma}_1(k-1)}
+\frac{5b_1^2}{4}\sum_{k=0}^{\infty}
\frac{(k+2)k}{(k+4)(k+3)^2(k!)^2}\\&\times&\left\{
e_{\mu,\gamma_2\dot\gamma_2}e_{\nu}^{\kappa\dot{\delta}}
\Phi_{\kappa\gamma_1(k-1)\dot{\gamma}_1(k-1)}
^{\alpha(1)\dot{\alpha}(1)\dot\gamma_2}
\Phi_{\dot{\delta}}
^{\gamma_2\gamma_1(k-1)\dot{\gamma}_1(k-1)}
+e_{\mu,\gamma_2\dot\gamma_2}e_{\nu}^{\kappa\dot{\delta}}\eps^{\alpha(1)\gamma_2}
\Phi^{\dot{\alpha}(1)\dot\gamma_2}_{\kappa\gamma_1(k)\dot{\gamma}_1(k-1)}
\Phi_{\dot{\delta}}
^{\gamma_1(k)\dot{\gamma}_1(k-1)}\right.\\&+&\left.
e_{\mu,\gamma_2\dot\gamma_2}e_{\nu}^{\gamma_2\dot{\delta}}
\Phi_{\gamma_1(k)\dot{\gamma}_1(k-1)}
^{\alpha(1)\dot{\alpha}(1)\dot\gamma_2}
\Phi_{\dot{\delta}}
^{\gamma_1(k)\dot{\gamma}_1(k-1)}
\right\}
\end{eqnarray*}
C2.1-3:
\begin{eqnarray*}
&&-e_{\mu,\beta\dot\alpha}\eps^{\delta\gamma}\left\{A^{(1)}_\gamma ,
\p_\delta^{(z)}\wh{\mathcal{L}}^{(1)}(e_\nu)\right\}^{\alpha(1)
\dot{\alpha}(2),\beta(1)}_{k_{\beta\alpha}=2}=\\&&
\frac{b_1^2}{3}\sum_{k=0}^{\infty} 
\frac{k}{(k+4)^2(k!)^2}
\left\{-k e_{\mu,\gamma_2\dot\gamma_2}e_{\nu}^{\kappa\dot{\delta}}
\Phi_{\kappa\gamma_1(k-1)\dot{\gamma}_1(k-1)}
^{\gamma_2\dot{\alpha}(1)\dot\gamma_2}
\Phi_{\dot{\delta}}
^{\alpha(1)\gamma_1(k-1)\dot{\gamma}_1(k-1)}
\right.\\&+&\left.e_{\mu,\gamma_2\dot\gamma_2}
e_{\nu}^{\kappa\dot{\delta}}\eps^{\alpha(1)\gamma_2}
\Phi^{\dot{\alpha}(1)\dot\gamma_2}_{\kappa\gamma_1(k)\dot{\gamma}_1(k-1)}
\Phi_{\dot{\delta}}
^{\gamma_1(k)\dot{\gamma}_1(k-1)}-
e_{\mu,\gamma_2\dot\gamma_2}e_{\nu}^{\alpha(1)\dot{\delta}}
\Phi_{\gamma_1(k)\dot{\gamma}_1(k-1)}
^{\gamma_2\dot{\alpha}(1)\dot\gamma_2}
\Phi_{\dot{\delta}}
^{\gamma_1(k)\dot{\gamma}_1(k-1)}\right\}
\end{eqnarray*}
C3.1-4:
\begin{eqnarray*}
&&-e_{\mu,\beta\dot\alpha}\eps^{\delta\gamma}\left\{A^{(1)}_\gamma ,
\p_\delta^{(z)}\wh{\mathcal{L}}^{(1)}(e_\nu)\right\}^{
\dot{\alpha}(3),\beta(1)}_{k_{\alpha\alpha}=1}=
-\frac{ib_1^2}{2}\sum_{k=0}^{\infty} \frac{k}{(k+4)(k+1)(k!)^2}
\\&\times&\left\{2e_{\mu,\gamma_2\dot\gamma_2}
e_{\nu}^{\gamma_2\dot{\delta}}
\Phi_{\gamma_1(k+1)\dot{\gamma}_1(k-1)}
^{\dot{\alpha}(1)\dot\gamma_2}
\Phi_{\dot{\delta}}
^{\dot{\alpha}(1)\gamma_1(k+1)\dot{\gamma}_1(k-1)}\right. \\ &+& 2
(k+1)e_{\mu,\gamma_2\dot\gamma_2} e_{\nu}^{\kappa\dot{\delta}}
\Phi_{\kappa\gamma_1(k)\dot{\gamma}_1(k-1)}
^{\dot{\alpha}(1)\dot\gamma_2}
\Phi_{\dot{\delta}}
^{\dot{\alpha}(1)\gamma_2\gamma_1(k)\dot{\gamma}_1(k-1)}\\&+&\left.
e_{\mu,\gamma_2\dot\gamma_2}e_{\nu}^{\gamma_2\dot{\delta}}
\Phi_{\gamma_1(k+1)\dot{\gamma}_1(k-1)}
^{\dot{\alpha}(2)}
\Phi_{\dot{\delta}}
^{\dot\gamma_2\gamma_1(k+1)\dot{\gamma}_1(k-1)}+
(k+1) e_{\mu,\gamma_2\dot\gamma_2}e_{\nu}^{\kappa\dot{\delta}}
\Phi_{\kappa\gamma_1(k)\dot{\gamma}_1(k-1)}
^{\dot{\alpha}(2)}
\Phi_{\dot{\delta}}
^{\gamma_2\dot\gamma_2\gamma_1(k)\dot{\gamma}_1(k-1)}\right\}
\end{eqnarray*}
C4.1-4:
\begin{eqnarray*}
&&-e_{\mu,\beta\dot\alpha}\eps^{\delta\gamma}\left\{A^{(1)}_\gamma ,
\p_\delta^{(z)}\wh{\mathcal{L}}^{(1)}(e_\nu)\right\}^{
\dot{\alpha}(3),\beta(1)}_{k_{\alpha\beta}=k_{\beta\alpha}=1}=
2ib_1^2\sum_{k=0}^{\infty} 
\frac{k}
{(k+1)(k+4)^2(k!)^2}
\\&\times&\left\{2e_{\mu,\gamma_2\dot\gamma_2}
e_{\nu}^{\gamma_2\dot{\delta}}
\Phi_{\gamma_1(k+1)\dot{\gamma}_1(k-1)}
^{\dot{\alpha}(1)\dot\gamma_2}
\Phi_{\dot{\delta}}
^{\dot{\alpha}(1)\gamma_1(k+1)\dot{\gamma}_1(k-1)} \right.\\&+&
2(k+1)e_{\mu,\gamma_2\dot\gamma_2}e_{\nu}^{\kappa\dot{\delta}}
\Phi_{\kappa\gamma_1(k)\dot{\gamma}_1(k-1)}
^{\dot{\alpha}(1)\dot\gamma_2}
\Phi_{\dot{\delta}}
^{\dot{\alpha}(1)\gamma_2\gamma_1(k)\dot{\gamma}_1(k-1)}\\&+&
\left.e_{\mu,\gamma_2\dot\gamma_2}
e_{\nu}^{\gamma_2\dot{\delta}}
\Phi_{\gamma_1(k+1)\dot{\gamma}_1(k-1)}
^{\dot{\alpha}(2)}
\Phi_{\dot{\delta}}
^{\dot\gamma_2\gamma_1(k+1)\dot{\gamma}_1(k-1)}+
(k+1)e_{\mu,\gamma_2\dot\gamma_2}e_{\nu}^{\kappa\dot{\delta}}
\Phi_{\kappa\gamma_1(k)\dot{\gamma}_1(k-1)}
^{\dot{\alpha}(2)}
\Phi_{\dot{\delta}}
^{\dot\gamma_2\gamma_2\gamma_1(k)\dot{\gamma}_1(k-1)}
\right\}
\end{eqnarray*}
C5:
\begin{eqnarray*}
&&-e_{\mu,\beta\dot\alpha}\eps^{\delta\gamma}\left\{A^{(1)}_\gamma ,
\p_\delta^{(z)}\wh{\mathcal{L}}^{(1)}(e_\nu)\right\}^{
\dot{\alpha}(3),\beta(1)}_{k_{\beta\alpha}=2}=
\frac{ib_1^2}{4}\sum_{k=0}^{\infty}
\frac{k(k+5)}{(k+4)^2(k+3)(k!)^2} \\&\times&
e_{\mu,\gamma_2\dot\gamma_2}e_{\nu}^{\kappa\dot{\delta}}
\Phi_{\kappa\gamma_1(k)\dot{\gamma}_1(k-1)}
^{\gamma_2\dot{\alpha}(2)\dot\gamma_2}
\Phi_{\dot{\delta}}
^{\gamma_1(k)\dot{\gamma}_1(k-1)}
\end{eqnarray*}
C6.1-5:
\begin{eqnarray*}
&&-e_{\mu,\beta\dot\alpha}\eps^{\delta\gamma}\left\{A^{(1)}_\gamma ,
\p_\delta^{(z)}\wh{\mathcal{L}}^{(1)}(e_\nu)\right\}^{\alpha(2)
\dot{\alpha}(1),\beta(1)}_{k_{\beta\alpha=1}}=
\frac{-ib_1^2}{2}\sum_{k=0}^{\infty}
\frac{(2k+7)}{(k+4)^2(k+5)(k+1)(k!)^2}\\&\times&
\left\{e_{\mu,\gamma_2\dot\gamma_2}e_{\nu}^{\alpha(1)\dot{\delta}}\eps^{\alpha(1)\gamma_2}
\Phi_{\gamma_1(k+1)\dot{\gamma}_1(k)}
^{\dot\gamma_2}
\Phi_{\dot{\delta}}
^{\gamma_1(k+1)\dot{\gamma}_1(k)} \right.\\&+& 
(k+1)e_{\mu,\gamma_2\dot\gamma_2} e_{\nu}^{\kappa\dot{\delta}} \eps^{\alpha(1)\gamma_2}
\Phi_{\kappa\gamma_1(k)\dot{\gamma}_1(k)}
^{\dot\gamma_2}
\Phi_{\dot{\delta}}
^{\alpha(1)\gamma_1(k)\dot{\gamma}_1(k)}\\&+&
e_{\mu,\gamma_2\dot\gamma_2}e_{\nu}^{\alpha(1)\dot{\delta}}
\Phi_{\gamma_1(k)\dot{\gamma}_1(k)}
^{\alpha(1)\dot\gamma_2}
\Phi_{\dot{\delta}}
^{\gamma_2\gamma_1(k)\dot{\gamma}_1(k)}+
e_{\mu,\gamma_2\dot\gamma_2}e_{\nu}^{\gamma_2\dot{\delta}}
\Phi_{\gamma_1(k)\dot{\gamma}_1(k)}
^{\alpha(1)\dot\gamma_2}
\Phi_{\dot{\delta}}
^{\alpha(1)\gamma_1(k)\dot{\gamma}_1(k)}\\&+&\left.
 ke_{\mu,\gamma_2\dot\gamma_2}e_{\nu}^{\kappa\dot{\delta}}
\Phi_{\kappa\gamma_1(k-1)\dot{\gamma}_1(k)}
^{\alpha(1)\dot\gamma_2}
\Phi_{\dot{\delta}}
^{\alpha(1)\gamma_2\gamma_1(k-1)\dot{\gamma}_1(k)}\right\}
\end{eqnarray*}
C7.1-7:
\begin{eqnarray*}
&&-e_{\mu,\beta\dot\alpha}\eps^{\delta\gamma}\left\{A^{(1)}_\gamma ,
\p_\delta^{(z)}\wh{\mathcal{L}}^{(1)}(e_\nu)\right\}^{\alpha(2)
\dot{\alpha}(1),\beta(1)}_{k_{\beta\alpha}=2}=
\frac{ib_1^2}{4}\sum_{k=0}^{\infty}
\frac{1}{(k+5)^2(k+4)(k+2)(k+1)(k!)^2}
\\&\times&
\left\{2(k+1)\left(e_{\mu,\gamma_2\dot\gamma_2}
e_{\nu}^{\alpha(1)\dot{\delta}}
\Phi_{\gamma_1(k)\dot{\gamma}_1(k)}
^{\alpha(1)\dot\gamma_2}
\Phi_{\dot{\delta}}
^{\gamma_2\gamma_1(k)\dot{\gamma}_1(k)}
+e_{\mu,\gamma_2\dot\gamma_2}e_{\nu}^{\gamma_2\dot{\delta}}
\Phi_{\gamma_1(k)\dot{\gamma}_1(k)}
^{\alpha(1)\dot\gamma_2}
\Phi_{\dot{\delta}}
^{\alpha(1)\gamma_1(k)\dot{\gamma}_1(k)}\right)\right.\\
&+&2(k+1)ke_{\mu,\gamma_2\dot\gamma_2}
e_{\nu}^{\kappa\dot{\delta}}
\Phi_{\kappa\gamma_1(k-1)\dot{\gamma}_1(k)}
^{\alpha(1)\dot\gamma_2}
\Phi_{\dot{\delta}}
^{\alpha(1)\gamma_2\gamma_1(k-1)\dot{\gamma}_1(k)}
\\&+&(k+5)\left(e_{\mu,\gamma_2\dot\gamma_2}
e_{\nu}^{\alpha(1)\dot{\delta}}\eps^{\alpha(1)\gamma_2}
\Phi_{\gamma_1(k+1)\dot{\gamma}_1(k)}
^{\dot\gamma_2}
\Phi_{\dot{\delta}}
^{\gamma_1(k+1)\dot{\gamma}_1(k)}\right.\\
&+&\left.\left. (k+1)e_{\mu,\gamma_2\dot\gamma_2} e_{\nu}^{\kappa\dot{\delta}}\eps^{\alpha(1)\gamma_2}
\Phi_{\kappa\gamma_1(k)\dot{\gamma}_1(k)}
^{\dot\gamma_2}
\Phi_{\dot{\delta}}
^{\alpha(1)\gamma_1(k)\dot{\gamma}_1(k)}\right)\right\}
\\&+&\frac{ib_1^2}{4}\sum_{k=0}^{\infty}
\frac{(k+3)}{(k+5)^2(k+4)(k+2)(k!)^2}
\left\{e_{\mu,\gamma_2\dot\gamma_2}
e_{\nu}^{\alpha(1)\dot{\delta}}
\Phi_{\gamma_1(k)\dot{\gamma}_1(k)}
^{\gamma_2\dot\gamma_2}
\Phi_{\dot{\delta}}
^{\alpha(1)\gamma_1(k)\dot{\gamma}_1(k)}\right.\\&+&
\left.  \frac{k(k+1)}{2}e_{\mu,\gamma_2\dot\gamma_2}e_{\nu}^{\kappa\dot{\delta}}
\Phi_{\kappa\gamma_1(k-1)\dot{\gamma}_1(k)}
^{\gamma_2\dot\gamma_2}
\Phi_{\dot{\delta}}
^{\alpha(2)\gamma_1(k-1)\dot{\gamma}_1(k)}\right\}
\end{eqnarray*}
C8.1-3:
\begin{eqnarray*}
&&e_{\mu,\beta\dot\alpha}\eps^{\delta\gamma}\left[A^{(1)}_\gamma ,
\p_\delta^{(y)}\wh{\mathcal{L}}^{(1)}(e_\nu)\right]_{k_{\alpha\beta}=1}^{\alpha(1)
\dot{\alpha}(2),\beta(1)}=\\&&
\frac{b_1^2}{4}
\sum_{k=0}^{\infty}\frac{k}{(k+4)(k+3)(k!)^2}
\left\{e_{\mu,\gamma_2\dot\gamma_2}e_{\nu}^{\kappa\dot{\delta}}\eps^{\alpha(1)\gamma_2}
\Phi_{\kappa\gamma_1(k)\dot{\gamma}_1(k-1)}
^{\dot{\alpha}(1)\dot\gamma_2}
\Phi_{\dot{\delta}}
^{\gamma_1(k)\dot{\gamma}_1(k-1)}\right.\\&+&\left.
e_{\mu,\gamma_2\dot\gamma_2}e_{\nu}^{\gamma_2\dot{\delta}}
\Phi_{\gamma_1(k)\dot{\gamma}_1(k-1)}
^{\alpha(1)\dot{\alpha}(1)\dot\gamma_2}
\Phi_{\dot{\delta}}
^{\gamma_1(k)\dot{\gamma}_1(k-1)}+
ke_{\mu,\gamma_2\dot\gamma_2} e_{\nu}^{\kappa\dot{\delta}}
\Phi_{\kappa\gamma_1(k-1)\dot{\gamma}_1(k-1)}
^{\alpha(1)\dot{\alpha}(1)\dot\gamma_2}
\Phi_{\dot{\delta}}
^{\gamma_2\gamma_1(k-1)\dot{\gamma}_1(k-1)}\right\}
\end{eqnarray*}
C9.1-4:
\begin{eqnarray*}
&&e_{\mu,\beta\dot\alpha}\eps^{\delta\gamma}\left[A^{(1)}_\gamma ,
\p_\delta^{(y)}\wh{\mathcal{L}}^{(1)}(e_\nu)\right]_{k_{\alpha\beta}=1}^{
\dot{\alpha}(3),\beta(1)}
=\\&&
-\frac{ib_1^2}{2}\sum_{k=0}^{\infty} 
\frac{(2k+5)k}{(k+4)^2(k+3)(k+1)(k!)^2}
\left\{2e_{\mu,\gamma_2\dot\gamma_2}e_{\nu}^{\gamma_2\dot{\delta}}
\Phi_{\gamma_1(k+1)\dot{\gamma}_1(k-1)}
^{\dot{\alpha}(1)\dot\gamma_2}
\Phi_{\dot{\delta}}
^{\dot{\alpha}(1)\gamma_1(k+1)\dot{\gamma}_1(k-1)}\right.
\\&+&
2(k+1)e_{\mu,\gamma_2\dot\gamma_2}e_{\nu}^{\kappa\dot{\delta}}
\Phi_{\kappa\gamma_1(k)\dot{\gamma}_1(k-1)}
^{\dot{\alpha}(1)\gamma_2}
\Phi_{\dot{\delta}}
^{\gamma_2\dot{\alpha}(1)\gamma_1(k)\dot{\gamma}_1(k-1)}
+e_{\mu,\gamma_2\dot\gamma_2}e_{\nu}^{\gamma_2\dot{\delta}}
\Phi_{\gamma_1(k+1)\dot{\gamma}_1(k-1)}
^{\dot{\alpha}(2)}
\Phi_{\dot{\delta}}
^{\dot\gamma_2\gamma_1(k+1)\dot{\gamma}_1(k-1)}\\&+&\left.
e_{\mu,\gamma_2\dot\gamma_2}e_{\nu}^{\kappa\dot{\delta}}
\Phi_{\kappa\gamma_1(k)\dot{\gamma}_1(k-1)}
^{\dot{\alpha}(2)}
\Phi_{\dot{\delta}}
^{\gamma_2\dot\gamma_2\gamma_1(k)\dot{\gamma}_1(k-1)}
\right\}
\end{eqnarray*}
\newline
The remaining terms, obeying $k_{\alpha\alpha}=k_{\alpha\beta}=k_{\beta\alpha}=0$,
are (projecting first onto $yy$, then $\bar y\bar y$ and $y\bar y$):
\newline
\newline
C10.1-4:
\begin{eqnarray*}
&&-e_{\mu,\beta\dot\alpha}\eps^{\delta\gamma}\left\{A^{(1)}_\gamma ,
\p_\delta^{(z)}\wh{\mathcal{L}}^{(1)}(e_\nu)\right\}^{
\dot{\alpha}(3),\beta(1)}_{n_1=1}=ib_1^2\sum_{k=0}^{\infty} 
\frac{k}{(k+3)^2(k+2)(k+1)(k!)^2}\\
&\times&\left(2e_{\mu,\gamma_2\dot\gamma_2}e_{\nu}^{\gamma_2\dot{\delta}}\Phi^{\dot{\alpha}(1)\dot\gamma_2}
_{\gamma_1(k+1)\dot{\gamma}_1(k-1)}
\Phi^{\dot{\alpha}(1)\gamma_1(k+1)\dot{\gamma}_1(k-1)}
_{\dot{\delta}}\right.\\&+&
2(k+1)e_{\mu,\gamma_2\dot\gamma_2} e_{\nu}^{\kappa\dot{\delta}}\Phi^{\dot{\alpha}(1)\dot\gamma_2}
_{\kappa\gamma_1(k)\dot{\gamma}_1(k-1)}
\Phi^{\dot{\alpha}(1)\gamma_2\gamma_1(k)\dot{\gamma}_1(k-1)}
_{\dot{\delta}}\\&+&\left.
e_{\mu,\gamma_2\dot\gamma_2}e_{\nu}^{\gamma_2\dot{\delta}}\Phi^{\dot{\alpha}(2)}
_{\gamma_1(k+1)\dot{\gamma}_1(k-1)}
\Phi^{\dot\gamma_2\gamma_1(k+1)\dot{\gamma}_1(k-1)}
_{\dot{\delta}}+
(k+1)e_{\mu,\gamma_2\dot\gamma_2} e_{\nu}^{\kappa\dot{\delta}}\Phi^{\dot{\alpha}(2)}
_{\kappa\gamma_1(k)\dot{\gamma}_1(k-1)}
\Phi^{\dot\gamma_2\gamma_2\gamma_1(k)\dot{\gamma}_1(k-1)}
_{\dot{\delta}}
\right)
\end{eqnarray*}
C11:
\begin{eqnarray*}
&&e_{\mu,\beta\dot\alpha}\eps^{\delta\gamma}\left[A^{(1)}_\gamma ,
\p_\delta^{(y)}\wh{\mathcal{L}}^{(1)}(e_\nu)\right]
^{\dot{\alpha}(3),\beta(1)}_{n_1=1}=
ib_1^2\sum_{k=0}^{\infty}
\frac{k(2k+5)}{(k+4)(k+3)(k+2)(k!)^2}
\\&\times&e_{\mu,\gamma_2\dot\gamma_2}e_{\nu}^{\kappa\dot{\delta}}
\Phi^{\dot{\alpha}(2)\dot\gamma_2\gamma_2}
_{\kappa\gamma_1(k)\dot{\gamma}_1(k-1)}
\Phi^{\gamma_1(k)\dot{\gamma}_1(k-1)}_{\dot{\delta}}
\end{eqnarray*}
C12.1-4:
\begin{eqnarray*}
&&e_{\mu,\beta\dot\alpha}\eps^{\delta\gamma}\left[A^{(1)}_\gamma ,
\p_\delta^{(y)}\wh{\mathcal{L}}^{(1)}(e_\nu)\right]
^{\dot{\alpha}(3),\beta(1)}_{n_1=0}=
ib_1^2\sum_{k=0}^{\infty}
\frac{k(2k+5)}{(k+1)(k+2)(k+3)(k!)^2}
\\&\times&
\left(2e_{\mu,\gamma_2\dot\gamma_2}e_{\nu}^{\gamma_2\dot{\delta}}\Phi^{\dot{\alpha}(1)\dot\gamma_2}
_{\gamma_1(k+1)\dot{\gamma}_1(k-1)}
\Phi^{\dot{\alpha}(1)\gamma_1(k+1)\dot{\gamma}_1(k-1)}
_{\dot{\delta}}\right.\\&+&
2(k+1)e_{\mu,\gamma_2\dot\gamma_2}e_\nu^{\kappa\dot{\delta}}\Phi^{\dot{\alpha}(1)\dot\gamma_2}_
{\kappa\gamma_1(k)\dot{\gamma}_1(k-1)}
\Phi^{\dot{\alpha}(1)\gamma_2\gamma_1(k)\dot{\gamma}_1(k-1)}_
{\dot{\delta}}\\&+&\left.
e_{\mu,\gamma_2\dot\gamma_2}e_{\nu}^{\gamma_2\dot{\delta}}\Phi^{\dot{\alpha}(2)}
_{\gamma_1(k+1)\dot{\gamma}_1(k-1)}
\Phi^{\dot\gamma_2\gamma_1(k+1)\dot{\gamma}_1(k-1)}
_{\dot{\delta}}
+(k+1)e_{\mu,\gamma_2\dot\gamma_2}e_\nu^{\kappa\dot{\delta}}\Phi^{\dot{\alpha}(2)}_
{\kappa\gamma_1(k)\dot{\gamma}_1(k-1)}
\Phi^{\dot\gamma_2\gamma_2\gamma_1(k)\dot{\gamma}_1(k-1)}_{\dot{\delta}}\right)
\end{eqnarray*}
C13.1-5:
\begin{eqnarray*}
&&-e_{\mu,\beta\dot\alpha}\eps^{\delta\gamma}\left\{A^{(1)}_\gamma ,
\p_\delta^{(z)}\wh{\mathcal{L}}^{(1)}(e_\nu)\right\}^{\alpha(2)
\dot{\alpha}(1),\beta(1)}_{n_1=\tilde{m}_1=1}=
ib_1^2\sum_{k=0}^{\infty}
\frac{1}{(k+4)^2(k+3)(k+1)(k!)^2}\\
&\times&\left\{
(k+1)e_{\mu,\gamma_2\dot\gamma_2} e_{\nu}^{\gamma_2\dot{\delta}}\Phi^{\alpha(1)\dot\gamma_2}
_{\gamma_1(k)\dot{\gamma}_1(k)}
\Phi^{\alpha(1)\gamma_1(k)\dot{\gamma}_1(k)}_{\dot{\delta}}
+e_{\mu,\gamma_2\dot\gamma_2}e_{\nu}^{\alpha(1)\dot{\delta}}\eps^{\alpha(1)\gamma_2}
\Phi^{\dot\gamma_2}
_{\gamma_1(k+1)\dot{\gamma}_1(k)}
\Phi^{\gamma_1(k+1)\dot{\gamma}_1(k)}_{\dot{\delta}}
\right.\\&+&(k+1)e_{\mu,\gamma_2\dot\gamma_2} e_{\nu}^{\kappa\dot{\delta}}
\eps^{\alpha(1)\gamma_2}\Phi^{\dot\gamma_2}
_{\kappa\gamma_1(k)\dot{\gamma}_1(k)}
\Phi^{\alpha(1)\gamma_1(k)\dot{\gamma}_1(k)}_{\dot{\delta}}
+(k+1)e_{\mu,\gamma_2\dot\gamma_2} e_{\nu}^{\alpha(1)\dot{\delta}}\Phi^{\alpha(1)\dot\gamma_2}
_{\gamma_1(k)\dot{\gamma}_1(k)}
\Phi^{\gamma_2\gamma_1(k)\dot{\gamma}_1(k)}_{\dot{\delta}}
\\&+&\left.(k+1)ke_{\mu,\gamma_2\dot\gamma_2} e_{\nu}^{\kappa\dot{\delta}}
\Phi^{\alpha(1)\dot\gamma_2}
_{\kappa\gamma_1(k-1)\dot{\gamma}_1(k)}
\Phi^{\alpha(1)\gamma_2\gamma_1(k-1)\dot{\gamma}_1(k)}
_{\dot{\delta}}\right\}
\end{eqnarray*}
C14.1-3:
\begin{eqnarray*}
&&-e_{\mu,\beta\dot\alpha}\eps^{\delta\gamma}\left\{A^{(1)}_\gamma ,
\p_\delta^{(z)}\wh{\mathcal{L}}^{(1)}(e_\nu)\right\}^{\alpha(2)
\dot{\alpha}(1),\beta(1)}_{n_1=1,\tilde{m}_1=0}=
-i\frac{b_1^2}{4}\sum_{k=0}^{\infty}
\frac{1}{(k+4)(k+3)^2(k!)^2}
\\&\times&\left\{e_{\mu,\gamma_2\dot\gamma_2}e_{\nu}^{\gamma_2\dot{\delta}}
\Phi_{\gamma_1(k)\dot{\gamma}_1(k)}
\Phi^{\alpha(2)\dot\gamma_2\gamma_1(k)\dot{\gamma}_1(k)}_
{\dot{\delta}}\right.
\\&+& ke_{\mu,\gamma_2\dot\gamma_2} e_{\nu}^{\kappa\dot{\delta}}
\Phi_{\kappa\gamma_1(k-1)\dot{\gamma}_1(k)}
\Phi^{\alpha(2)\dot\gamma_2\gamma_2\gamma_1(k-1)
\dot{\gamma}_1(k)}_{\dot{\delta}}
\\&+&\left.2e_{\mu,\gamma_2\dot\gamma_2}e_{\nu}^{\kappa\dot{\delta}}
\Phi_{\gamma_1(k)\dot{\gamma}_1(k)}
\Phi^{\alpha(1)\dot\gamma_2\gamma_2\gamma_1(k)
\dot{\gamma}_1(k)}_{\dot{\delta}}\right\}
\end{eqnarray*}
C15.1-4:
\begin{eqnarray*}
&&e_{\mu,\beta\dot\alpha}\eps^{\delta\gamma}\left[A^{(1)}_\gamma ,
\p_\delta^{(y)}\wh{\mathcal{L}}^{(1)}(e_\nu)\right]^{\alpha(2)
\dot{\alpha}(1),\beta(1)}_{n_1=\tilde{m}_1=1}=
-\frac{ib_1^2}{2}\sum_{k=0}^{\infty}
\frac{1}{(k+3)(k+4)(k+5)(k+1)(k!)^2}
\\&\times& \left(-(k+5)e_{\mu,\gamma_2\dot\gamma_2}e_{\nu}^{\alpha(1)\dot{\delta}}
\eps^{\alpha(1)\gamma_2}\Phi^{\dot\gamma_2}_{\gamma_1(k+1)
\dot{\gamma}_1(k)}\Phi^{\gamma_1(k+1)
\dot{\gamma}_1(k)}_{\dot{\delta}}\right.
\\&+&(k+1)^2e_{\mu,\gamma_2\dot\gamma_2}e_{\nu}^{\alpha(1)\dot{\delta}}
\Phi^{\dot\gamma_2\gamma_2}_{\gamma_1(k)
\dot{\gamma}_1(k)}\Phi^{\alpha(1)\gamma_1(k)
\dot{\gamma}_1(k)}_{\dot{\delta}}
\\&-&(k+1)(k+5)e_{\mu,\gamma_2\dot\gamma_2}e_{\nu}^{\kappa\dot{\delta}}\eps^{\alpha(1)\gamma_2}
\Phi^{\dot\gamma_2}_{\kappa\gamma_1(k)
\dot{\gamma}_1(k)}\Phi^{\alpha(1)\gamma_1(k)
\dot{\gamma}_1(k)}_{\dot{\delta}}\\&+&
\left.3(k+1)ke_{\mu,\gamma_2\dot\gamma_2}e_{\nu}^{\kappa\dot{\delta}}
\Phi^{\dot\gamma_2\gamma_2}_{\kappa\gamma_1(k-1)
\dot{\gamma}_1(k)}\Phi^{\alpha(2)\gamma_1(k-1)
\dot{\gamma}_1(k)}_{\dot{\delta}}\right)
\end{eqnarray*}
C16.1-5:
\begin{eqnarray*}
&&e_{\mu,\beta\dot\alpha}\eps^{\delta\gamma}\left[A^{(1)}_\gamma ,
\p_\delta^{(y)}\wh{\mathcal{L}}^{(1)}(e_\nu)\right]^{\alpha(2)
\dot{\alpha}(1),\beta(1)}_{n_1=0,\tilde{m}_1=1}=
ib_1^2\sum_{k=0}^{\infty}
\frac{(k+3)}{(k+4)^2(k+5)(k+1)(k!)^2}
\\&\times&
\left(2e_{\mu,\gamma_2\dot\gamma_2}e_{\nu}^{\alpha(1)\dot{\delta}}\eps^{\alpha(1)\gamma_2}
\Phi^{\dot\gamma_2}_{\gamma_1(k+1)
\dot{\gamma}_1(k)}\Phi^{\gamma_1(k+1)
\dot{\gamma}_1(k)}_{\dot{\delta}}\right.
\\&+&2(k+1)e_{\mu,\gamma_2\dot\gamma_2} e_{\nu}^{\kappa\dot{\delta}}\eps^{\alpha(1)\gamma_2}
\Phi^{\dot\gamma_2}_{\kappa\gamma_1(k)
\dot{\gamma}_1(k)}\Phi^{\alpha(1)\gamma_1(k)
\dot{\gamma}_1(k)}_{\dot{\delta}}
\\&+&(k+1)e_{\mu,\gamma_2\dot\gamma_2} e_{\nu}^{\alpha\dot{\delta}}
\Phi^{\alpha(1)\gamma_2}_{\gamma_1(k)
\dot{\gamma}_1(k)}\Phi^{\gamma_2\gamma_1(k)
\dot{\gamma}_1(k)}_{\dot{\delta}}
\\&+&(k+1)e_{\mu,\gamma_2\dot\gamma_2} e_{\nu}^{\gamma_2\dot{\delta}}
\Phi^{\alpha(1)\dot\gamma_2}_{\gamma_1(k)
\dot{\gamma}_1(k)}\Phi^{\alpha(1)\gamma_1(k)
\dot{\gamma}_1(k)}_{\dot{\delta}}
\\&+&\left.(k+1)ke_{\mu,\gamma_2\dot\gamma_2} e_{\nu}^{\kappa\dot{\delta}}
\Phi^{\alpha(1)\dot\gamma_2}_{\kappa\gamma_1(k-1)
\dot{\gamma}_1(k)}\Phi^{\alpha(1)\gamma_2\gamma_1(k-1)
\dot{\gamma}_1(k)}_{\dot{\delta}}\right)
\end{eqnarray*}
C17.1-3:
\begin{eqnarray*}
&&e_{\mu,\beta\dot\alpha}\eps^{\delta\gamma}\left[A^{(1)}_\gamma ,
\p_\delta^{(y)}\wh{\mathcal{L}}^{(1)}(e_\nu)\right]^{\alpha(2)
\dot{\alpha}(1),\beta(1)}_{n_1=\tilde{m}_1=0}=
-\frac{ib_1^2}{2}
\sum_{k=0}^{\infty}
\frac{1}{(k+4)^2(k+5)(k+3)(k!)^2}\\
&\times&
\left((2k+5)e_{\mu,\gamma_2\dot\gamma_2}e_{\nu}^{\alpha(1)\dot{\delta}}
\Phi_{\gamma_1(k)\dot{\gamma}_1(k)}
\Phi^{\alpha(1)\gamma_2\dot\gamma_2
\gamma_1(k)\dot{\gamma}_1(k)}
_{\dot{\delta}}\right.
\\&+&\frac{2k+5}{2}e_{\mu,\gamma_2\dot\gamma_2}e_{\nu}^{\gamma_2\dot{\delta}}
\Phi_{\gamma_1(k)\dot{\gamma}_1(k)}
\Phi^{\alpha(2)\dot\gamma_2
\gamma_1(k)\dot{\gamma}_1(k)}
_{\dot{\delta}}
\\&+&\left.(k+2)ke_{\mu,\gamma_2\dot\gamma_2}e_{\nu}^{\kappa\dot{\delta}}
\Phi_{\kappa\gamma_1(k-1)
\dot{\gamma}_1(k)}\Phi^{\alpha(2)\gamma_2
\dot\gamma_2\gamma_1(k-1)
\dot{\gamma}_1(k)}_{\dot{\delta}}\right)
\end{eqnarray*}
C18.1-6:
\begin{eqnarray*}
&&-e_{\mu,\beta\dot\alpha}\eps^{\delta\gamma}\left\{A^{(1)}_\gamma ,
\p_\delta^{(z)}\wh{\mathcal{L}}^{(1)}(e_\nu)\right\}^{\alpha(1)
\dot{\alpha}(2),\beta(1)}_{n_1=m_1=1}=
\frac12b_1^2\sum_{k=0}^{\infty}
\frac{k}{(k+3)^2(k+2)(k!)^2}
\\&\times&\left(e_{\mu,\gamma_2\dot\gamma_2}e_{\nu}^{\gamma_2\dot{\delta}}
\Phi^{\dot\gamma_2}_{\gamma_1(k)\dot{\gamma_1}(k-1)}
\Phi^{\alpha(1)\dot{\alpha}(1)\gamma_1(k)\dot{\gamma_1}(k-1)}
_{\dot{\delta}}
+e_{\mu,\gamma_2\dot\gamma_2}e_{\nu}^{\alpha(1)\dot{\delta}}
\Phi^{\dot\gamma_2}_{\gamma_1(k)\dot{\gamma_1}(k-1)}
\Phi^{\gamma_2\dot{\alpha}(1)\gamma_1(k)\dot{\gamma_1}(k-1)}
_{\dot{\delta}}\right.
\\&+&\left.ke_{\mu,\gamma_2\dot\gamma_2} e_{\nu}^{\kappa\dot{\delta}}
\Phi^{\dot\gamma_2}_{\kappa\gamma_1(k-1)\dot{\gamma_1}(k-1)}
\Phi^{\alpha(1)\gamma_2\dot{\alpha}(1)\gamma_1(k-1)
\dot{\gamma_1}(k-1)}_{\dot{\delta}}+e_{\mu,\gamma_2\dot\gamma_2}e_{\nu}^{\gamma_2\dot{\delta}}
\Phi^{\dot{\alpha}(1)}_{\gamma_1(k)\dot{\gamma_1}(k-1)}
\Phi^{\alpha(1)\dot\gamma_2\gamma_1(k)\dot{\gamma_1}(k-1)}
_{\dot{\delta}}\right.\\&+&\left.
e_{\mu,\gamma_2\dot\gamma_2}e_{\nu}^{\alpha(1)\dot{\delta}}
\Phi^{\dot{\alpha}(1)}_{\gamma_1(k)\dot{\gamma_1}(k-1)}
\Phi^{\gamma_2\dot\gamma_2\gamma_1(k)\dot{\gamma_1}(k-1)}
_{\dot{\delta}}+ke_{\mu,\gamma_2\dot\gamma_2} e_{\nu}^{\kappa\dot{\delta}}
\Phi^{\dot{\alpha}(1)}_{\kappa\gamma_1(k-1)\dot{\gamma_1}(k-1)}
\Phi^{\alpha(1)\gamma_2\dot\gamma_2\gamma_1(k-1)
\dot{\gamma_1}(k-1)}_{\dot{\delta}}\right)
\end{eqnarray*}
C19.1-3:
\begin{eqnarray*}
&&-e_{\mu,\beta\dot\alpha}\eps^{\delta\gamma}\left\{A^{(1)}_\gamma ,
\p_\delta^{(z)}\wh{\mathcal{L}}^{(1)}(e_\nu)\right\}^{\alpha(1)
\dot{\alpha}(2),\beta(1)}_{n_1=1,m_1=0}=
-\frac{b_1^2}{2}\sum_{k=0}^{\infty}
\frac{k}{(k+3)^2(k+2)(k!)^2}
\\&\times&\left(e_{\mu,\gamma_2\dot\gamma_2}e_\nu^{\kappa\dot{\delta}}\eps^{\alpha(1)\gamma_2}
\Phi^{\dot{\alpha}(1)\dot\gamma_2}
_{\kappa\gamma_1(k)\dot{\gamma_1}(k-1)}
\Phi^{\gamma_1(k)\dot{\gamma_1}(k-1)}
_{\dot{\delta}}+e_{\mu,\gamma_2\dot\gamma_2}e_\nu^{\gamma_2\dot{\delta}}
\Phi^{\alpha(1)\dot{\alpha}(1)\dot\gamma_2}
_{\gamma_1(k)\dot{\gamma_1}(k-1)}
\Phi^{\gamma_1(k)\dot{\gamma_1}(k-1)}
_{\dot{\delta}}\right.
\\&+&\left.ke_{\mu,\gamma_2\dot\gamma_2} e_\nu^{\kappa\dot{\delta}}\Phi^{\alpha(1)\dot{\alpha}(1)\dot\gamma_2}
_{\kappa\gamma_1(k-1)\dot{\gamma_1}(k-1)}
\Phi^{\gamma_2\gamma_1(k-1)\dot{\gamma_1}(k-1)}
_{\dot{\delta}}\right)
\end{eqnarray*}
C20.1-3:
\begin{eqnarray*}
&&e_{\mu,\beta\dot\alpha}\eps^{\delta\gamma}\left[A^{(1)}_\gamma ,
\p_\delta^{(y)}\wh{\mathcal{L}}^{(1)}(e_\nu)\right]^{\alpha(1)
\dot{\alpha}(2),\beta(1)}_{n_1=m_1=1}=
-\frac{b_1^2}{2}\sum_{k=0}^{\infty}
\frac{k}{(k+3)(k+2)(k!)^2}
\\&\times&\left(e_{\mu,\gamma_2\dot\gamma_2}e_\nu^{\kappa\dot{\delta}}\eps^{\alpha(1)\gamma_2}
\Phi^{\dot{\alpha}(1)\dot\gamma_2}
_{\kappa\gamma_1(k)\dot{\gamma_1}(k-1)}
\Phi^{\gamma_1(k)\dot{\gamma_1}(k-1)}
_{\dot{\delta}}\right.
\\&-&e_{\mu,\gamma_2\dot\gamma_2}e_\nu^{\alpha(1)\dot{\delta}}
\Phi^{\gamma_2\dot{\alpha}(1)\dot\gamma_2}
_{\gamma_1(k)\dot{\gamma_1}(k-1)}
\Phi^{\gamma_1(k)\dot{\gamma_1}(k-1)}
_{\dot{\delta}}
\\&-&\left.k e_{\mu,\gamma_2\dot\gamma_2}e_\nu^{\kappa\dot{\delta}}\Phi^{\gamma_2\dot{\alpha}(1)\dot\gamma_2}
_{\kappa\gamma_1(k-1)\dot{\gamma_1}(k-1)}
\Phi^{\alpha(1)\gamma_1(k-1)\dot{\gamma_1}(k-1)}
_{\dot{\delta}}\right)
\end{eqnarray*}
C21.1-6:
\begin{eqnarray*}
&&e_{\mu,\beta\dot\alpha}\eps^{\delta\gamma}\left[A^{(1)}_\gamma ,
\p_\delta^{(y)}\wh{\mathcal{L}}^{(1)}(e_\nu)\right]^{\alpha(1)
\dot{\alpha}(2),\beta(1)}_{n_1=0,m_1=1}=
\frac12 b_1^2\sum_{k=0}^{\infty}
\frac{k}{(k+3)^2(k!)^2}\\&\times&
\left(e_{\mu,\gamma_2\dot\gamma_2}e_\nu^{\gamma_2\dot{\delta}}
\Phi^{\dot\gamma_2}_{\gamma_1(k)\dot{\gamma_1}(k-1)}
\Phi^{\alpha(1)\dot{\alpha}(1)\gamma_1(k)\dot{\gamma_1}(k-1)}
_{\dot{\delta}}\right.
\\&+&e_{\mu,\gamma_2\dot\gamma_2}e_\nu^{\alpha(1)\dot{\delta}}
\Phi^{\dot\gamma_2}_{\gamma_1(k)\dot{\gamma_1}(k-1)}
\Phi^{\gamma_2\dot{\alpha}(1)\gamma_1(k)\dot{\gamma_1}(k-1)}
_{\dot{\delta}}
\\&+&\left.ke_{\mu,\gamma_2\dot\gamma_2} e_\nu^{\kappa\dot{\delta}}
\Phi^{\dot\gamma_2}_{\kappa\gamma_1(k-1)\dot{\gamma_1}(k-1)}
\Phi^{\alpha(1)\gamma_2\dot{\alpha}(1)\gamma_1(k-1)
\dot{\gamma_1}(k-1)}_{\dot{\delta}}\right.\\&+&\left.
e_{\mu,\gamma_2\dot\gamma_2}e_\nu^{\gamma_2\dot{\delta}}
\Phi^{\dot{\alpha}(1)}_{\gamma_1(k)\dot{\gamma_1}(k-1)}
\Phi^{\alpha(1)\dot\gamma_2\gamma_1(k)\dot{\gamma_1}(k-1)}
_{\dot{\delta}}\right.
\\&+&e_{\mu,\gamma_2\dot\gamma_2}e_\nu^{\alpha(1)\dot{\delta}}
\Phi^{\dot{\alpha}(1)}_{\gamma_1(k)\dot{\gamma_1}(k-1)}
\Phi^{\gamma_2\dot\gamma_2\gamma_1(k)\dot{\gamma_1}(k-1)}
_{\dot{\delta}}
\\&+&\left.ke_{\mu,\gamma_2\dot\gamma_2} e_\nu^{\kappa\dot{\delta}}
\Phi^{\dot{\alpha}(1)}_{\kappa\gamma_1(k-1)\dot{\gamma_1}(k-1)}
\Phi^{\alpha(1)\gamma_2\dot\gamma_2\gamma_1(k-1)
\dot{\gamma_1}(k-1)}_{\dot{\delta}}
\right)
\end{eqnarray*}
C22.1-3:
\begin{eqnarray*}
&&e_{\mu,\beta\dot\alpha}\eps^{\delta\gamma}\left[A^{(1)}_\gamma ,
\p_\delta^{(y)}\wh{\mathcal{L}}^{(1)}(e_\nu)\right]^{\alpha(1)
\dot{\alpha}(2),\beta(1)}_{n_1=m_1=0}=
-\frac{b_1^2}{4}\sum_{k=0}^{\infty}\frac{k}{(k+3)^2(k!)^2}
\\&\times&\left(
e_{\mu,\gamma_2\dot\gamma_2}e_\nu^{\gamma_2\dot{\delta}}
\Phi^{\alpha(1)\dot{\alpha}(1)\dot\gamma_2}_{\gamma_1(k)\dot{\gamma_1}(k-1)}
\Phi^{\gamma_1(k)\dot{\gamma_1}(k-1)}
_{\dot{\delta}}\right.
\\&+&e_{\mu,\gamma_2\dot\gamma_2}e_\nu^{\kappa\dot{\delta}}\eps^{\alpha(1)\gamma_2}
\Phi^{\dot{\alpha}(1)\dot\gamma_2}_{\kappa\gamma_1(k)\dot{\gamma_1}(k-1)}
\Phi^{\gamma_1(k)\dot{\gamma_1}(k-1)}
_{\dot{\delta}}
\\&+&\left.k e_{\mu,\gamma_2\dot\gamma_2}e_\nu^{\kappa\dot{\delta}}
\Phi^{\alpha(1)\dot{\alpha}(1)\dot\gamma_2}_{\kappa\gamma_1(k-1)
\dot{\gamma_1}(k-1)}
\Phi^{\gamma_2\gamma_1(k-1)\dot{\gamma_1}(k-1)}
_{\dot{\delta}}\right)
\end{eqnarray*}
In summary we have
\begin{eqnarray}
[e_\mu,\wh{\mathcal{L}}^{(1)} \circ \wh{\mathcal{L}}^{(1)}(e_\nu)]^{\alpha(2)}
&=&\mbox{C6}+\mbox{C7}+\mbox{C13}+\mbox{C14}+\mbox{C15}+\mbox{C16}+\mbox{C17}\nonumber\\{}
[e_\mu,\wh{\mathcal{L}}^{(1)} \circ \wh{\mathcal{L}}^{(1)}(e_\nu)]^{\alpha(1)\dot{\alpha}(1)}
&=&\mbox{C1}+\mbox{C2}+\mbox{C8}+\mbox{C18}+\mbox{C19}+\mbox{C20}+\mbox{C21}+\mbox{C22}\nonumber\\{}
[e_\mu,\wh{\mathcal{L}}^{(1)} \circ \wh{\mathcal{L}}^{(1)}(e_\nu)]^{\dot{\alpha}(2)}
&=&\mbox{C3}+\mbox{C4}+\mbox{C5}+\mbox{C9}+\mbox{C10}+\mbox{C11}+\mbox{C12}
\end{eqnarray}

%=============================================================

\subsection{Evaluation of $A_\gamma^{(1)}\star A_\delta^{(1)}$}

Apart from $\mathcal{L}_{\mu\nu}$, $\mathcal{J}_{\mu\nu}$ also contain
a term involving $A_{(\gamma}^{(1)}\star A_{\delta)}^{(1)}$ (see
eq. (\ref{eq:curlJ})).  It is convenient rewriting this as an
anti-commutator, $\{A_\gamma^{(1)},A_\delta^{(1)}\}$. The
anti-commutator is evaluated using the prescription for the commutator
(\ref{eq:AA_com}), but replacing odd with even in the sum. The
resulting structures that we are interested in read:
\newline
\newline
D1.1-3:
\begin{eqnarray*}
&&i(e_{[\mu}\bar e_{\nu]})^{\gamma\delta}
\{A_\gamma^{(1)},A_\delta^{(1)}\}^{\alpha(2)\dot{\alpha}(0)}
_{m_1=1}=ib_1^2\sum_{k=0}^{\infty}\frac{1}{(k+4)^2(k+1)^2(k!)^2}
\\&\times&\left\{2(k+1)(e_{[\mu}\bar e_{\nu]})^{\gamma\delta}\delta^{\alpha(1)}_{(\gamma}
\Phi_{\delta)\gamma_1(k)\dot{\gamma}_1(k+1)}
\Phi^{\alpha(1)\gamma_1(k)\dot{\gamma}_1(k+1)}\right.
\\&-&(k-1)k(e_{[\mu}\bar e_{\nu]})^{\gamma\delta}
\Phi^{\alpha(1)}_{\gamma\gamma_1(k-1)\dot{\gamma}_1(k+1)}
\Phi^{\alpha(1)\gamma_1(k-1)\dot{\gamma}_1(k)}_{\delta}
\\&+&\left.(e_{[\mu}\bar e_{\nu]})^{\gamma\delta} \delta^{\alpha(2)}_{\gamma\delta}
\Phi_{\gamma_1(k+1)\dot{\gamma}_1(k+1)}
\Phi^{\gamma_1(k+1)\dot{\gamma}_1(k+1)}\right\}
\end{eqnarray*}
D2:
\begin{eqnarray*}
&&i(e_{[\mu}\bar e_{\nu]})^{\gamma\delta}\{A_\gamma^{(1)},A_\delta^{(1)}\}^{\alpha(0)\dot{\alpha}(2)}
_{\tilde{m}_1=1}=
-ib_1^2\sum_{k=0}^{\infty}\frac{1}{(k+3)^2(k!)^2}
 (e_{[\mu}\bar e_{\nu]})^{\gamma\delta}\Phi^{\dot{\alpha}(1)}_{\gamma\gamma_1(k)\dot{\gamma}_1(k)}
\Phi^{\dot{\alpha}(1)\gamma_1(k)\dot{\gamma}_1(k)}
_{\delta}
\end{eqnarray*}
D3.1-2:
\begin{eqnarray*}
&&i(e_{[\mu}\bar e_{\nu]})^{\gamma\delta}\{A_\gamma^{(1)},A_\delta^{(1)}\}^{\alpha(1)\dot{\alpha}(1)}
_{m_1=1,\tilde{m}_1=0}=
-\frac{b_1^2}2\sum_{k=0}^{\infty}
\frac{1}{(k+3)^2(k!)^2}
\\&\times&\left\{(e_{[\mu}\bar e_{\nu]})^{\gamma\delta}\delta^{\alpha(1)}_{\delta}
\Phi_{\gamma_1(k)\dot{\gamma}_1(k)}
\Phi_{\gamma}^{\dot{\alpha}(1)\gamma_1(k)\dot{\gamma}_1(k)}
\right.\\&+& k\left.(e_{[\mu}\bar e_{\nu]})^{\gamma\delta}
\Phi_{\delta\gamma_1(k-1)\dot{\gamma}_1(k)}
\Phi_{\gamma}
^{\dot{\alpha}(1)\alpha(1)\gamma_1(k-1)\dot{\gamma}_1(k)}
\right\}
\end{eqnarray*}
D4.1-2:
\begin{eqnarray*}
&&i(e_{[\mu}\bar e_{\nu]})^{\gamma\delta}\{A_\gamma^{(1)},A_\delta^{(1)}\}^{\alpha(1)\dot{\alpha}(1)}
_{m_1=0,\tilde{m}_1=1}=
i(e_{[\mu}\bar e_{\nu]})^{\gamma\delta}\{A_\delta^{(1)},A_\gamma^{(1)}\}^{\alpha(1)\dot{\alpha}(1)}
_{m_1=1,\tilde{m}_1=0}
\end{eqnarray*}

\begin{table}[t]
\begin{tabular}{@{\qquad\qquad}c@{ }c@{ }c}
\begin{tabular}{|l|l|}
\hline
\multicolumn{2}{|c|}{\it Table of structures}\\
\hline
A1.1&a1\\
A1.2&a2\\
A2.1&c1\\
A2.2&$-$c2\\
A2.3&c3\\
A3.1&c1\\
A3.2&$-$c2\\
A3.3&c3\\
A4.1&a4\\
A4.2&a7\\
A4.3&$-$a2\\
A4.4&a6\\
B1.1&a1\\
B1.2&a3\\
B2&a6\\
B3&b2\\
B4&b2\\
B5.1&c1\\
B5.2&c4\\
B6.1&c1\\
B6.2&c4\\
B7.1&c1\\
B7.2&c4\\
B8.1&a6\\
B8.2&b2\\
C1.1-2&c3\\
C1.3&$-$c2\\
C1.4&$-$c1\\
C2.1&c5\\
C2.2&$-$c2\\
C2.3&$-$c6\\
C3.1&a1\\
C3.2&a2\\
C3.3&a3\\
C3.4&$-$a5\\
C4.1&a1\\
C4.2&a2\\
\hline
\end{tabular}
&
\begin{tabular}{|l|l|}
\hline
C4.3&a3 \\
C4.4&$-$a5\\
C5&b1\\
C6.1&$-$a4\\
C6.2-3&a7\\
C6.4&$-$a6\\
C6.5&$-$a2\\
C7.1&a7\\
C7.2&$-$a6\\
C7.3&$-$a2\\
C7.4&$-$a4\\
C7.5&a7\\
C7.6&$-$a8\\
C7.7&a5\\
C8.1&$-$c2\\
C8.2&$-$c1\\
C8.3&c3\\
C9.1&a1\\
C9.2&a2\\
C9.3&a3\\
C9.4&$-$a5\\
C10.1&a1\\
C10.2&a2\\
C10.3&a3\\
C10.4&$-$a5\\
C11&b1\\
C12.1&a1\\
C12.2&a2\\
C12.3&a3\\
C12.4&$-$a5\\
C13.1&$-$a6\\
C13.2&$-$a4\\
C13.3-4&a7\\
C13.5&$-$a2\\
C14.1&$-$b2\\
C14.2&$-$b1\\
C14.3&$-$b3\\
C15.1&$-$a4\\
\hline
\end{tabular}
&
\begin{tabular}{|l|l|}
\hline
C15.2&$-$a8\\
C15.3&a7\\
C15.4&a5\\
C16.1&$-$a4\\
C16.2-3&a7\\
C16.4&$-$a6\\
C16.5&$-$a2\\
C17.1&$-$b3\\
C17.2&$-$b2\\
C17.3&$-$b1\\
C18.1&c1\\
C18.2&c2\\
C18.3&$-$c3\\
C18.4&c4\\
C18.5&c7\\
C18.6&c8\\
C19.1&$-$c2\\
C19.2&$-$c1\\
C19.3&c3\\
C20.1&$-$c2\\
C20.2&$-$c6\\
C20.3&c5\\
C21.1&c1\\
C21.2&c2\\
C21.3&$-$c3\\
C21.4&c4\\
C21.5&c7\\
C21.6&c8\\
C22.1&$-$c1\\
C22.2&$-$c2\\
C22.3&c3\\
D2&d1\\
D3.1&e1\\
D3.2&e2\\
D4.1&e1\\
D4.2&e2\\
&\\
&\\
\hline
\end{tabular}
\end{tabular}
\caption{{\small Here we tabulate the transformation of terms of
appendix C into structures of appendix D.  The signs indicate that the
two index patterns differ by a sign (due to flipping of spinor
indices). The D1.1-3 structures only contribute to the antisymmetric
part of the traced eq. (\ref{eq:R=J1}). See the discussion following
(\ref{eq:Ricci_contr}). }}
\label{table:1}
\end{table}

%============================================================

\section{Conversion into derivatives of the scalar}
\label{sect:derivs_apdx}

The different sub-contributions A{\footnotesize\#}.{\footnotesize\#}
idem B, C, D, listed in the previous section contain a number of
different contraction patterns of the spinor indices. Below we label
these patterns by a{\footnotesize\#}, b{\footnotesize\#},
c{\footnotesize\#}, d1 or e{\footnotesize\#}, and rewrite them as
bilinears in derivatives of the scalar, using
(\ref{eq:deriv_scalar}). Here a, b arise in the contributions from
$L_{\mu\nu}$ to the Ricci tensor, c arises in the contributions from
$L_{\mu\nu}$ to the contorsion, and d and e arise in the contributions
from $\widehat A^{(1)}_{\alpha} \star \widehat A^{(1)}_{\beta}$ to the
Ricci tensor and the contorsion, respectively. The result, which will
be finally assembled in the next section, is that the contorsion
tensor is made up by two distinct bilinear structures and the Ricci
tensor by three distinct bilinear structures, as shown in
(\ref{eq:smakprov}).

%------------------------------------------------------------

\subsection{Contributions from $L_{\mu\nu}$ to the Ricci tensor}

The Ricci tensor $\Ric_{\rho\mu}$ receives contribution from
$L_{\mu\nu}$ of the form $(\sigma_\rho{}^\nu)_{\dot\alpha\dot\beta}
L_{\mu\nu}^{\dot\alpha\dot\beta}$. Terms with an equal number of
derivatives hitting the two scalars arise from the following basic
structures:
\begin{eqnarray}
& \mbox{a1.} & (\sigma_\rho{}^\nu)^{\dot\alpha\dot\beta}
(\sigma_\mu)^{\beta\dot\gamma(1)}
(\sigma_\nu)_\beta{}^{\dot\gamma'}
\Phi_{\gamma(k+1) \dot\gamma(k) \dot\alpha}
\Phi^{\gamma(k+1) \dot\gamma(k-1)}{}_{\dot\gamma' \dot\beta} \ = 
\nonumber \\
& & \qquad \qquad \qquad \qquad \qquad \qquad \qquad
= \ -2 \cdot 2^k \, g_{\rho\mu}
\nabla_{\mu\{k+1\}} \phi \ 
\nabla^{\mu\{k+1\}} \phi \label{eq:a1}
\\
& \mbox{a2.} & (\sigma_\rho{}^\nu)^{\dot\alpha\dot\beta}
(\sigma_\mu)^{\gamma' \dot\gamma(1)}
(\sigma_\nu)^{\gamma(1) \dot\gamma'}
\Phi_{\gamma(k+1)\dot\gamma(k) \dot\alpha}
\Phi^{\gamma(k)}{}_{\gamma'}{}^{\dot\gamma(k-1)}
{}_{\dot\gamma'\dot\beta} \ =
\nonumber \\
& & \qquad \qquad \qquad \qquad \qquad \qquad \qquad
= \ - 2 \cdot 2^k \, g_{\rho\mu} \nabla_{\mu\{k+1\}} \phi \ 
\nabla^{\mu\{k+1\}} \phi \nonumber \\
& & \qquad \qquad \qquad \qquad \qquad \qquad \qquad
 \qquad + 4 \cdot 2^k \, \nabla_{\mu\{k\} \rho} \phi \ 
\nabla^{\mu\{k\}}{}_\mu \phi 
\\
& \mbox{a3.} & (\sigma_\rho{}^\nu)^{\dot\alpha\dot\beta}
(\sigma_\mu)^{\beta\dot\gamma(1)}
(\sigma_\nu)_\beta{}^{\dot\gamma(1)}
\Phi_{\gamma(k+1) \dot\gamma(k+1)}
\Phi^{\gamma(k+1) \dot\gamma(k-1)}{}_{\dot\alpha \dot\beta} =
\nonumber \\
& & \qquad \qquad \qquad \qquad \qquad \qquad \qquad
= \ 4 \cdot 2^k \, g_{\rho\mu}
\nabla_{\mu\{k+1\}} \phi \ 
\nabla^{\mu\{k+1\}} \phi
\\
& \mbox{a4.} & (\sigma_\rho{}^\nu)^{\dot\alpha\dot\beta}
(\sigma_\mu)^{\gamma(1)}{}_{\dot\alpha}
(\sigma_\nu)^{\gamma'}{}_{\dot\beta}
\Phi_{\gamma(k+1) \dot\gamma(k+1)}
\Phi^{\gamma(k)}{}_{\gamma'}{}^{\dot\gamma(k+1)} = \nonumber \\
& & \qquad \qquad \qquad \qquad \qquad \qquad \qquad
= \ -6 \cdot 2^k \, g_{\rho\mu}
\nabla_{\mu\{k+1\}} \phi \ 
\nabla^{\mu\{k+1\}} \phi 
\\
& \mbox{a5.} & (\sigma_\rho{}^\nu)^{\dot\alpha\dot\beta}
(\sigma_\mu)^{\gamma(1) \dot\gamma(1)}
(\sigma_\nu)^{\gamma' \dot\gamma(1)}
\Phi_{\gamma(k+1) \dot\gamma(k+1)}
\Phi^{\gamma(k)}{}_{\gamma'}{}^{\dot\gamma(k-1)}{}_{\dot\alpha 
\dot\beta} = \nonumber \\
& & \qquad \qquad \qquad \qquad \qquad \qquad \qquad
= \ -2 \cdot 2^k \, g_{\rho\mu}
\nabla_{\mu\{k+1\}} \phi \ 
\nabla^{\mu\{k+1\}} \phi 
\\
& \mbox{a6.} & (\sigma_\rho{}^\nu)^{\dot\alpha\dot\beta}
(\sigma_\mu)^{\gamma(1) \dot\delta}
(\sigma_\nu)^{\gamma'}{}_{\dot\delta}
\Phi_{\gamma(k+1) \dot\gamma(k) \dot\alpha}
\Phi^{\gamma(k)}{}_{\gamma'}{}^{\dot\gamma(k)}{}_{\dot\beta} = 
\nonumber \\
& & \qquad \qquad \qquad \qquad \qquad \qquad \qquad
= \ 2 \cdot 2^k \, g_{\rho\mu} \nabla_{\mu\{k+1\}} \phi \ 
\nabla^{\mu\{k+1\}} \phi \nonumber \\
& & \qquad \qquad \qquad \qquad \qquad \qquad \qquad
 \qquad -8 \cdot 2^k \, \nabla_{\mu\{k\} \rho} \phi \ 
\nabla^{\mu\{k\}}{}_\mu \phi 
\\
& \mbox{a7.} & (\sigma_\rho{}^\nu)^{\dot\alpha\dot\beta}
(\sigma_\mu)^{\gamma' \dot\gamma(1)}
(\sigma_\nu)^{\gamma(1)}{}_{\dot\alpha}
\Phi_{\gamma(k+1) \dot\gamma(k+1)}
\Phi^{\gamma(k)}{}_{\gamma'}{}^{\dot\gamma(k)}{}_{\dot\beta} = 
\nonumber \\
& & \qquad \qquad \qquad \qquad \qquad \qquad \qquad
= \ 6 \cdot 2^k \, g_{\rho\mu} \nabla_{\mu\{k+1\}} \phi \ 
\nabla^{\mu\{k+1\}} \phi \nonumber \\
& & \qquad \qquad \qquad \qquad \qquad \qquad \qquad
 \qquad - 4 \cdot 2^k \, \nabla_{\mu\{k\} \rho} \phi \ 
\nabla^{\mu\{k\}}{}_\mu \phi 
\\
& \mbox{a8.} & (\sigma_\rho{}^\nu)^{\dot\alpha\dot\beta}
(\sigma_\mu)^{\gamma(1)}{}_{\dot\alpha}
(\sigma_\nu)^{\gamma' \dot\gamma'}
\Phi_{\gamma(k+1) \dot\gamma(k)\dot\beta}
\Phi^{\gamma(k)}{}_{\gamma'}{}^{\dot\gamma(k)}{}_{\dot\gamma'} = 
\nonumber \\
& & \qquad \qquad \qquad \qquad \qquad \qquad \qquad
= \ 2 \cdot 2^k \, g_{\rho\mu} \nabla_{\mu\{k+1\}} \phi \ 
\nabla^{\mu\{k+1\}} \phi \nonumber \\
& & \qquad \qquad \qquad \qquad \qquad \qquad \qquad
 \qquad + 4 \cdot 2^k \, \nabla_{\mu\{k\} \rho} \phi \ 
\nabla^{\mu\{k\}}{}_\mu \phi,
\end{eqnarray}
where we use the notation defined in (\ref{eq:traceless}).  Terms in
which the number of derivatives hitting the two scalars differs by two
arise from the following structures:
\begin{eqnarray}
& \mbox{b1.} & (\sigma_\rho{}^\nu)^{\dot\alpha\dot\beta}
(\sigma_\mu)^{\gamma(1) \dot\gamma(1)}
(\sigma_\nu)^{\gamma(1) \dot\gamma'}
\Phi_{\gamma(k+2) \dot\gamma(k) \dot\alpha \dot\beta}
\Phi^{\gamma(k) \dot\gamma(k-1)}{}_{\dot\gamma'} = \nonumber \\
& & \qquad \qquad \qquad \qquad \qquad \qquad \qquad \qquad
= \ 4 \cdot 2^k \, \nabla_{\mu\{k\}\rho\mu} \, \phi \ 
\nabla^{\mu\{k\}} \phi 
\\
& \mbox{b2.} & (\sigma_\rho{}^\nu)^{\dot\alpha\dot\beta}
(\sigma_\mu)^{\gamma(1) \dot\delta}
(\sigma_\nu)^{\gamma(1)}{}_{\dot\delta}
\Phi_{\gamma(k+2) \dot\gamma(k) \dot\alpha \dot\beta}
\Phi^{\gamma(k) \dot\gamma(k)} = \nonumber \\
& & \qquad \qquad \qquad \qquad \qquad \qquad \qquad \qquad
= \ -8 \cdot 2^k \, \nabla_{\mu\{k\}\rho\mu} \, \phi \ 
\nabla^{\mu\{k\}} \phi 
\\
& \mbox{b3.} & (\sigma_\rho{}^\nu)^{\dot\alpha\dot\beta}
(\sigma_\mu)^{\gamma(1)}{}_{\dot\alpha}
(\sigma_\nu)^{\gamma(1) \dot\gamma(1)}
\Phi_{\gamma(k+2) \dot\gamma(k+1) \dot\beta}
\Phi^{\gamma(k) \dot\gamma(k)}
 = \nonumber \\
& & \qquad \qquad \qquad \qquad \qquad \qquad \qquad \qquad
= \ 4 \cdot 2^k \, \nabla_{\mu\{k\}\rho\mu} \, \phi \ 
\nabla^{\mu\{k\}} \phi 
\end{eqnarray}

There are no more structures appearing in this sector.

%------------------------------------------------

\subsection{Contributions from $L_{\mu\nu}$ to the contorsion}
\label{sec:d2}

The contorsion tensor receives contributions from $L_{\mu\nu}$ of the
form $(\sigma^\rho)_{\alpha\dot\beta}
L_{\mu\nu}^{\alpha\dot\beta}$. In all the contributions the number of
derivatives hitting the two scalars differs by one, and they arise
from the following structures%
\footnote{Here we have omitted terms containing 
$\epsilon^{\mu\nu\rho\tau}$ since they are traced away in the
Einstein equation}:
\begin{eqnarray}
& \mbox{c1.} & (\sigma^\rho)^{\alpha\dot\beta}
(\sigma_\mu)^{\delta \dot\gamma(1)}
(\sigma_\nu)_\delta{}^{\dot\gamma'}
\Phi_{\gamma(k) \dot\gamma(k)}
\Phi^{\gamma(k)}{}_\alpha{}^{\dot\gamma(k-1)}{}_{\dot\gamma'\dot\beta} 
= \nonumber \\
& & \qquad \qquad \qquad \qquad \qquad \qquad \qquad
= \ 4 i \cdot 2^k \, \nabla_{\mu\{k-1\}[\mu|} \, \phi \ 
\nabla^{\mu\{k-1\}\rho}{}_{|\nu]} \, \phi 
\\
& \mbox{c2.} & (\sigma^\rho)^{\alpha\dot\beta}
(\sigma_\mu)^{\gamma' \dot\gamma(1)}
(\sigma_\nu)_\alpha{}^{\dot\gamma'}
\Phi_{\gamma(k) \dot\gamma(k)}
\Phi^{\gamma(k)}{}_{\gamma'}{}^{\dot\gamma(k-1)}{}_{\dot\gamma'
\dot\beta} = \nonumber \\
& & \qquad \qquad \qquad \qquad \qquad \qquad \qquad
= \ 2 i \cdot 2^k \, \nabla_{\mu\{k-1\}[\mu|} \, \phi \ 
\nabla^{\mu\{k-1\}\rho}{}_{|\nu]} \, \phi \nonumber \\
& & \qquad \qquad \qquad \qquad \qquad \qquad \qquad \qquad
  - 2 i \cdot 2^k g^\rho{}_{[\mu|} \, 
\nabla_{\mu\{k\}}  \, \phi \  \nabla^{\mu\{k\}}{}_{|\nu]} \, \phi
\\
& \mbox{c3.} & (\sigma^\rho)^{\alpha\dot\beta}
(\sigma_\mu)^{\gamma' \dot\gamma(1)}
(\sigma_\nu)^{\gamma(1) \dot\gamma'}
\Phi_{\gamma(k) \dot\gamma(k)}
\Phi^{\gamma(k-1)}{}_{\gamma'\alpha}{}^{\dot\gamma(k-1)}{}_{\dot\gamma'
\dot\beta} = \nonumber \\
& & \qquad \qquad \qquad \qquad \qquad \qquad \qquad
= \ 0 
\\
& \mbox{c4.} & (\sigma^\rho)^{\alpha\dot\beta}
(\sigma_\mu)^{\delta \dot\gamma(1)'}
(\sigma_\nu)_\delta{}^{\dot\gamma(1)'}
\Phi_{\gamma(k) \dot\gamma(k-1)\dot\beta}
\Phi^{\gamma(k)}{}_{\alpha}{}^{\dot\gamma(k-1)}{}_{
\dot\gamma(2)'} = \nonumber \\
& & \qquad \qquad \qquad \qquad \qquad \qquad \qquad
= \ 4 i \cdot 2^k \, \nabla_{\mu\{k-1\}[\mu|} \, \phi \ 
\nabla^{\mu\{k-1\}\rho}{}_{|\nu]} \, \phi \nonumber \\
& & \qquad \qquad \qquad \qquad \qquad \qquad \qquad \qquad
  - 4 i \cdot 2^k g^\rho{}_{[\mu|} \, 
\nabla_{\mu\{k\}}  \, \phi \  \nabla^{\mu\{k\}}{}_{|\nu]} \, \phi
\\
& \mbox{c5.} & (\sigma^\rho)^{\alpha\dot\beta}
(\sigma_\mu)^{\gamma(1)' \dot\gamma'}
(\sigma_\nu)^{\gamma(1)' \dot\gamma(1)}
\Phi_{\gamma(k-1) \alpha \dot\gamma(k)}
\Phi^{\gamma(k-1)}{}_{\gamma(2)'}
      {}^{\dot\gamma(k-1)}{}_{\dot\gamma'\dot\beta} = \nonumber \\
& & \qquad \qquad \qquad \qquad \qquad \qquad \qquad
= \ 2 i \cdot 2^k \, \nabla_{\mu\{k-1\}[\mu|} \, \phi \ 
\nabla^{\mu\{k-1\}\rho}{}_{|\nu]} \, \phi \nonumber \\
& & \qquad \qquad \qquad \qquad \qquad \qquad \qquad \qquad
  - 2 i \cdot 2^k g^\rho{}_{[\mu|} \, 
\nabla_{\mu\{k\}}  \, \phi \  \nabla^{\mu\{k\}}{}_{|\nu]} \, \phi
\\
& \mbox{c6.} & (\sigma^\rho)^{\alpha\dot\beta}
(\sigma_\mu)^{\gamma' \dot\gamma'}
(\sigma_\nu)_\alpha{}^{\dot\gamma(1)}
\Phi_{\gamma(k) \dot\gamma(k)}
\Phi^{\gamma(k)}{}_{\gamma'}{}^{\dot\gamma(k-1)}{}_{\dot\gamma'
\dot\beta} = \nonumber \\
& & \qquad \qquad \qquad \qquad \qquad \qquad \qquad
= \ 2 i \cdot 2^k \, \nabla_{\mu\{k-1\}[\mu|} \, \phi \ 
\nabla^{\mu\{k-1\}\rho}{}_{|\nu]} \, \phi \nonumber \\
& & \qquad \qquad \qquad \qquad \qquad \qquad \qquad \qquad
  + 2 i \cdot 2^k g^\rho{}_{[\mu|} \, 
\nabla_{\mu\{k\}}  \, \phi \  \nabla^{\mu\{k\}}{}_{|\nu]} \, \phi
\\
& \mbox{c7.} & (\sigma^\rho)^{\alpha\dot\beta}
(\sigma_\mu)^{\gamma' \dot\gamma(1)'}
(\sigma_\nu)_\alpha{}^{\dot\gamma(1)'}
\Phi_{\gamma(k) \dot\gamma(k-1)\dot\beta}
\Phi^{\gamma_{k}}{}_{\gamma'}{}^{\dot\gamma(k-1)}{}_{\dot\gamma(2)'}
= \nonumber \\
& & \qquad \qquad \qquad \qquad \qquad \qquad \qquad
= \ 2 i \cdot 2^k \, \nabla_{\mu\{k-1\}[\mu|} \, \phi \ 
\nabla^{\mu\{k-1\}\rho}{}_{|\nu]} \, \phi \nonumber \\
& & \qquad \qquad \qquad \qquad \qquad \qquad \qquad \qquad
  - 2 i \cdot 2^k g^\rho{}_{[\mu|} \, 
\nabla_{\mu\{k\}}  \, \phi \  \nabla^{\mu\{k\}}{}_{|\nu]} \, \phi
\\
& \mbox{c8.} & (\sigma^\rho)^{\alpha\dot\beta}
(\sigma_\mu)^{\gamma' \dot\gamma(1)'}
(\sigma_\nu)^{\gamma(1) \dot\gamma(1)'}
\Phi_{\gamma(k) \dot\gamma(k-1)\dot\beta}
\Phi^{\gamma(k-1)}{}_{\gamma' \alpha}
      {}^{\dot\gamma(k-1)}{}_{\dot\gamma(2)'} = \nonumber \\
& & \qquad \qquad \qquad \qquad \qquad \qquad \qquad
= \ 2 i \cdot 2^k \, \nabla_{\mu\{k-1\}[\mu|} \, \phi \ 
\nabla^{\mu\{k-1\}\rho}{}_{|\nu]} \, \phi \nonumber \\
& & \qquad \qquad \qquad \qquad \qquad \qquad \qquad \qquad
  - 2 i \cdot 2^k g^\rho{}_{[\mu|} \, 
\nabla_{\mu\{k\}}  \, \phi \  \nabla^{\mu\{k\}}{}_{|\nu]} \, \phi
\end{eqnarray}

%-----------------------------------------------------------

\subsection{Contributions from $\widehat A^{(1)}_{\alpha} \star
\widehat A^{(1)}_{\beta}$ to Ricci tensor and contorsion}\label{sec:d3}

The contributions from $(\widehat A^{(1)}_{\alpha} \star \widehat
A^{(1)}_{\beta})^{\dot\alpha\dot\beta}$ to the Ricci tensor contain
the following structures:
\begin{eqnarray}
& \mbox{d1.} & 
(\sigma_\mu)^{(\alpha| \dot\alpha}
(\sigma_\nu)^{|\beta) \dot\beta}
\Phi_{\gamma(k) \alpha \dot\gamma(k) \dot\alpha}
\Phi^{\gamma(k)}{}_\beta
      {}^{\dot\gamma(k)}{}_{\dot\beta} = \nonumber \\
& & \qquad \qquad \qquad \qquad \qquad \qquad \qquad
= \ -4 \cdot 2^k \, \nabla_{\mu\{k\}\mu} \, \phi \ 
\nabla^{\mu\{k\}}{}_\nu \, \phi \nonumber \\
& & \qquad \qquad \qquad \qquad \qquad \qquad \qquad \qquad
   + 2^k g_{\mu\nu} \, 
\nabla_{\mu\{k+1\}}  \, \phi \  \nabla^{\mu\{k+1\}} \, \phi
\end{eqnarray}
The contributions from $(\widehat A^{(1)}_{\alpha} \star \widehat
A^{(1)}_{\beta})^{\dot\alpha\dot\beta}$ to the contorsion contain
the following structures:
\begin{eqnarray}
& \mbox{e1.} & (\sigma^\rho)_{\alpha\dot\beta}
(\sigma_{\mu\nu})^{\delta \gamma}
\epsilon_\gamma{}^\alpha
\Phi_{\gamma(k) \dot\gamma(k)}
\Phi^{\gamma(k)}{}_\delta{}^{\dot\gamma(k)\dot\beta} 
= \nonumber \\
& & \qquad \qquad \qquad \qquad \qquad \qquad \qquad
= \ 4 i \cdot 2^k g^\rho{}_{[\mu|} \, 
\nabla_{\mu\{k\}}  \, \phi \  \nabla^{\mu\{k\}}{}_{|\nu]} \,
\phi \\
& \mbox{e2.} & (\sigma^\rho)_{\alpha\dot\beta}
(\sigma_{\mu\nu})^{\delta \gamma}
\Phi_{\gamma(k-1)\gamma \dot\gamma(k)}
\Phi^{\gamma(k-1)}{}_\delta{}^{\alpha\dot\gamma(k)\dot\beta} 
= \nonumber \\
& & \qquad \qquad \qquad \qquad \qquad \qquad \qquad
= \ 4 i \cdot 2^k \, \nabla_{\mu\{k-1\}[\mu|} \, \phi \ 
\nabla^{\mu\{k-1\}\rho}{}_{|\nu]} \, \phi \label{eq:e2}
\end{eqnarray}

%------------------------------------------------------------

\subsection{Derivative of $\kappa$}

In order to complete the computation of the Ricci tensor we need to
differentiate the contorsion tensor. In doing so we make use of
\begin{equation}
\nabla \Phi_{\gamma_1 \ldots \gamma_k \dot\gamma_1 \ldots
\dot\gamma_k} = ie^{\alpha\dot\alpha} \Phi_{\gamma_1 \ldots \gamma_k \alpha
\dot\gamma_1 \ldots \dot\gamma_k \dot\alpha} - i k^2 e_{(\gamma_k|
(\dot\gamma_k| } \Phi_{|\gamma_1 \ldots \gamma_{k-1}) |\dot\gamma_1 \ldots
\dot\gamma_{k-1})}.
\end{equation}
which follows from (\ref{eq:nablamuphi}).
Using (\ref{eq:deriv_scalar}) we conclude that
\begin{eqnarray}
\nabla_\nu \nabla_{\mu\{k\}} \, \phi & = & 
\nabla_{\nu \mu\{k\}} \, \phi - 
\frac{k^2}2 \, g_{\nu\{\mu_k} \nabla_{\mu\{k-1\}\}} \, \phi
\nonumber \\
& = & \nabla_{\nu \mu\{k\}} \, \phi - \frac{k^2}2 \, 
g_{\nu(\mu_k} \nabla_{\mu\{k-1\})} \, \phi + \nonumber \\
& & \qquad \qquad + \frac{k(k-1)}{4} \, 
g_{(\mu_k \mu_{k-1}} \nabla_{\mu\{k-2\})\nu}\, \phi.
\end{eqnarray}
From this we can see that the content of
$\nabla_{[\rho}\kappa_{\mu]\nu}{}^\rho$ is of the same form as the
content of
$i(\sigma_\mu{}^\rho)_{\alpha\beta}J_{\nu\rho}{}^{\alpha\beta}+\mbox{h.c}$.
Indeed, since the structures c{\footnotesize\#} and e{\footnotesize\#}
goes into $\kappa_{\mu\nu}{}^{\rho}$, we find that
$\nabla_{[\rho}\kappa_{\mu]\nu}{}^{\rho}$ contains
\begin{eqnarray}
& & \nabla_{[\rho|}\Big( 
\nabla_{\mu\{k-1\}\nu}  \, \phi \  
\nabla^{\mu\{k\}}{}_{|\mu]}{}^\rho \, \phi
- \nabla_{\mu\{k-1\}}{}^\rho  \, \phi \  
\nabla^{\mu\{k\}}{}_{|\mu]\nu} \, \phi
+ \{\mbox{perm. acc. to (\ref{eq:kappa_def})}\} \Big) \ = \nonumber \\ 
& & \qquad \qquad 
= \ 2 \nabla_{[\rho|} \Big( 
\nabla_{\mu\{k-1\}\nu}{}  \, \phi \  
\nabla^{\mu\{k\}}{}_{|\mu]}{}^\rho \, \phi
- 
\nabla_{\mu\{k-1\}}{}^\rho  \, \phi \  
\nabla^{\mu\{k\}}{}_{|\mu]\nu} \, \phi
\Big) \ = \nonumber \\ 
& & \qquad \qquad 
= \ 2 \nabla_{\mu\{k\}\mu}  \, \phi \  
\nabla^{\mu\{k\}}{}_\nu \, \phi \ - \ 
k(k+2) \nabla_{\mu\{k-1\}\mu}  \, \phi \  
\nabla^{\mu\{k-1\}}{}_\nu \, \phi \ +
\nonumber \\ 
& & \qquad \qquad \qquad 
- \ k \nabla_{\mu\{k-1\}}  \, \phi \  
\nabla^{\mu\{k-1\}}{}_{\mu\nu} \, \phi
\ - \ \frac{k+2}2
g_{\mu\nu} \, \nabla_{\mu\{k\}}  \, \phi \  
\nabla^{\mu\{k\}} \, \phi
\label{eq:kappa_part1}
\end{eqnarray}
and
\begin{eqnarray}
& & \nabla_{[\rho|}\Big(g_{|\mu]\nu} \,
\nabla_{\mu\{k\}}  \, \phi \  \nabla^{\mu\{k\}\rho} \, \phi  -
g_{|\mu]}{}^\rho \,
\nabla_{\mu\{k\}}  \, \phi \  \nabla^{\mu\{k\}}{}_\nu \, \phi
+ \{\mbox{perm. acc. to (\ref{eq:kappa_def})}\} \Big) \ = \nonumber \\ 
& & \qquad \qquad 
= \ 2 \nabla_{[\rho|} \Big( 
g_{|\mu]\nu}
\,\nabla_{\mu\{k\}}  \, \phi \  \nabla^{\mu\{k\}\rho} \, \phi
- g_{|\mu]}{}^\rho
\,\nabla_{\mu\{k\}}  \, \phi \  \nabla^{\mu\{k\}}{}_\nu \, \phi
\Big) \ = \nonumber \\ 
& & \qquad \qquad 
= \ 2\nabla_{\mu} \big( \nabla_{\mu\{k\}}  \, \phi \ 
\nabla^{\mu\{k\}}{}_\nu \, \phi \big) 
\ + \ g_{\mu\nu} \, \nabla_{\rho} \big(\nabla_{\mu\{k\}}  \, \phi \  
\nabla^{\mu\{k\}\rho} \, \phi \big) 
 \ = \nonumber \\ 
& & \qquad \qquad 
= \ 2 \nabla_{\mu\{k\}\mu}  \, \phi \  
\nabla^{\mu\{k\}}{}_\nu \, \phi
   \ + \ 2 \nabla_{\mu\{k\}}  \, \phi \  
\nabla^{\mu\{k\}}{}_{\mu\nu} \, \phi \ +
\nonumber \\ 
& & \qquad \qquad \qquad 
+ \ g_{\mu\nu} \, \nabla_{\mu\{k+1\}}  \, \phi \ 
\nabla^{\mu\{k+1\}} \, \phi
\ + \nonumber \\ 
& & \qquad \qquad \qquad 
- \ \frac12 (k^2+2k+2) 
g_{\mu\nu} \, \nabla_{\mu\{k\}}  \, \phi \  
\nabla^{\mu\{k\}} \, \phi \ + \nonumber \\ 
& & \qquad \qquad \qquad 
- \ k^2 \nabla_{\mu\{k-1\}\mu}  \, \phi \  
\nabla^{\mu\{k-1\}}{}_\nu \, \phi
\ - \ k^2 \nabla_{\mu\{k-1\}}  \, \phi \  
\nabla^{\mu\{k-1\}}{}_{\mu\nu} \, \phi.
\label{eq:kappa_part2}
\end{eqnarray}
Note that terms from $\nabla_{[\rho}\kappa_{\mu]\nu}{}^{\rho}$
contributes on two levels in the sum, since the number of derivatives
differ among the terms in (\ref{eq:kappa_part1}) and
(\ref{eq:kappa_part2}).

%====================================================================

\section{Computation of the Ricci tensor}

In this section we use the results of section \ref{sect:result_apdx}
and \ref{sect:derivs_apdx} to evaluate the expression for
$\Ric_{\mu\nu}+3g_{\mu\nu}$ given in (\ref{eq:Ricci_contr}). We divide
the computation into the evaluation of the contorsion term, the
$\widehat A^{(1)}_{\alpha} \star \widehat A^{(1)}_{\beta}$ term, and
the $L_{\mu\nu}$ term.

%---------------------------------------------------------------

\subsection{The contorsion term}

\label{sec:E1}

From (\ref{eq:kappa_def}) it follows that the contorsion term is given by
\begin{equation}
2\nabla_{[\rho} \kappa_{\mu]\nu}{}^{\rho} = i\nabla_{[\rho} \Big(
(\sigma_{\mu]})^{\alpha\dot\beta} J_\nu{}^\rho{}_{\alpha\dot\beta} + \{\mbox{perm. acc. to (\ref{eq:kappa_def})}\} \Big)
\end{equation}
The contributions to $J_\nu{}^\rho{}_{\alpha\dot\beta}$ are summarised
in (\ref{eq:Jaad}). To perform the contraction by
$(\sigma_{\mu})^{\alpha\dot\beta}$ we use (\ref{eq:usefull}) and
expand the result in terms of the c{\footnotesize\#} and
e{\footnotesize\#} structures defined in Section \ref{sec:d2} and
Section \ref{sec:d3}. We write this as
\begin{eqnarray}
(\sigma_{\mu})^{\alpha\dot\beta}\,
[\mbox{A2.1}]_\nu{}^\rho{}_{\alpha\dot\beta} & = & 
\frac14 \sum_{k=0}^\infty \frac{-b_1^2 k\,
[\mbox{c1}]_{\mu\nu}{}^\rho(k)}{2(k+3)^2(k+2)^2(k!)^2}
  \nonumber \\
& \equiv & b_1^2 \sum_{k=0}^\infty \frac{f_{\mathrm{A2.1}}(k)}{(k!)^2}
 \,
[\mbox{c1}]_{\mu\nu}{}^\rho(k)  \label{eq:A21c1}
\\
(\sigma_{\mu})^{\alpha\dot\beta}\,
[\mbox{A2.2}]_\nu{}^\rho{}_{\alpha\dot\beta} & = & 
-\frac14 \sum_{k=0}^\infty \frac{b_1^2 k \,
[\mbox{c2}]_{\mu\nu}{}^\rho(k)}{2(k+3)^2(k+2)^2(k!)^2}
  \nonumber \\
& \equiv & b_1^2 \sum_{k=0}^\infty \frac{f_{\mathrm{A2.2}}(k)}{(k!)^2}  \,
[\mbox{c2}]_{\mu\nu}{}^\rho(k) 
\end{eqnarray}
and so on schematically as
\begin{equation}
(\sigma_{\mu})^{\alpha\dot\beta}\,
[\mbox{A{\footnotesize\#}.{\footnotesize\#}}]_\nu{}^\rho{}_{\alpha\dot\beta}  =
b_1^2 \sum_{k=0}^\infty \frac{f_{\mathrm{A\#.\#}}(k)}{(k!)^2}  \,
[\mbox{c{\footnotesize\#}$'$ or e{\footnotesize\#}$'$}]_{\mu\nu}{}^\rho(k) 
\label{eq:Ace}
\end{equation}
where $f_{\mathrm{A\#.\#}}(k)$ is the coefficient of $e_\mu
e_\nu\Phi\Phi$ in A{\footnotesize\#}.{\footnotesize\#}, which is a
rational function of $k$ which is easy to read off. An analogous
notation is used for contributions of type B, C and D. All the
required matches between the index structures can be found in Table
\ref{table:1}. We can now write
\begin{eqnarray}
(\sigma_{\mu})^{\alpha\dot\beta} J_\nu{}^\rho{}_{\alpha\dot\beta} 
 & = &
b_1^2 \sum_{k=0}^\infty \frac{t_{\mathrm{c}i}(k) 
\, [\mbox{c$i$}]_{\mu\nu}{}^\rho(k) + t_{\mathrm{e}j}(k) 
\, [\mbox{e$j$}]_{\mu\nu}{}^\rho(k)}{(k!)^2} \ - \ \mbox{h.c.}
\nonumber
\\
& = & 2 \mbox{Re}\{b_1^2\} \sum_{k=0}^\infty \frac{t_{\mathrm{c}i}(k) 
\,[\mbox{c$i$}]_{\mu\nu}{}^\rho(k) + t_{\mathrm{e}j}(k) 
\,[\mbox{e$j$}]_{\mu\nu}{}^\rho(k)}{(k!)^2}
\end{eqnarray}
where summation over $i=1,\ldots,8$ and $j=1,2$ is assumed and
\begin{eqnarray}
t_{\mathrm{c1}}(k) & = & - \Big(f_{\mathrm{A2.1}}(k) 
+ f_{\mathrm{A3.1}}(k) - 2f_{\mathrm{B5.1}}(k) - 2f_{\mathrm{B6.1}}(k) 
   \nonumber \\
& & \quad - \ 2f_{\mathrm{B7.1}}(k) + 2f_{\mathrm{C1.4}}(k) 
+ 2f_{\mathrm{C8.2}}(k) + 2f_{\mathrm{C18.1}}(k) 
   \nonumber \\
& & \quad + \ 2f_{\mathrm{C19.2}}(k) + 2f_{\mathrm{C21.1}}(k) 
+ 2f_{\mathrm{C22.1}}(k) \Big), 
\\
t_{\mathrm{c2}}(k) & = & - \Big( f_{\mathrm{A2.2}}(k)
+ f_{\mathrm{A3.2}}(k) + 2f_{\mathrm{C1.3}}(k) + 2f_{\mathrm{C2.2}}(k) 
\nonumber \\
& & \quad + \ 2f_{\mathrm{C8.1}}(k) + 2f_{\mathrm{C18.2}}(k) 
+ 2f_{\mathrm{C19.1}}(k) + 2f_{\mathrm{C20.1}}(k) \nonumber \\
& & \quad + \ 2f_{\mathrm{C21.2}}(k) + 2f_{\mathrm{C22.2}}(k) \Big),
\\
t_{\mathrm{c3}}(k) & = & - \Big( f_{\mathrm{A2.3}}(k)
+ f_{\mathrm{A3.3}}(k) + 2f_{\mathrm{C1.1}}(k) + 2f_{\mathrm{C1.2}}(k) 
\nonumber \\
& & \quad + \ 2f_{\mathrm{C8.3}}(k) + 2f_{\mathrm{C18.3}}(k) 
+ 2f_{\mathrm{C19.3}}(k) + 2f_{\mathrm{C21.3}}(k) \nonumber \\
& & \quad 
+ 2f_{\mathrm{C22.3}}(k) \Big),
\\
t_{\mathrm{c4}}(k) & = & - \Big( -2f_{\mathrm{B5.2}}(k)
- 2f_{\mathrm{B6.2}}(k) - 2f_{\mathrm{B7.2}}(k) + 2f_{\mathrm{C18.4}}(k) 
\nonumber \\
& & \quad + \ 2f_{\mathrm{C21.4}}(k) \Big),
\\
t_{\mathrm{c5}}(k) & = & - \Big( 2f_{\mathrm{C2.1}}(k)
+ 2f_{\mathrm{C20.3}}(k) \Big),
\\
t_{\mathrm{c6}}(k) & = & - \Big( 2f_{\mathrm{C2.3}}(k)
+ 2f_{\mathrm{C20.2}}(k) \Big),
\\
t_{\mathrm{c7}}(k) & = & - \Big( 2f_{\mathrm{C18.5}}(k)
+ 2f_{\mathrm{C21.5}}(k) \Big),
\\
t_{\mathrm{c8}}(k) & = & - \Big( 2f_{\mathrm{C18.6}}(k)
+ 2f_{\mathrm{C21.6}}(k) \Big),
\\
t_{\mathrm{e1}}(k) & = & - \Big( f_{\mathrm{D3.1}}(k)
+ f_{\mathrm{D4.1}}(k) \Big),
\\
t_{\mathrm{e2}}(k) & = & - \Big( f_{\mathrm{D3.2}}(k)
+ f_{\mathrm{D4.2}}(k) \Big).
\end{eqnarray}
Finally, using (\ref{eq:kappa_part1}) and (\ref{eq:kappa_part2}) we compute
\begin{eqnarray}
2\nabla_{[\rho} \kappa_{\mu]\nu}{}^{\rho} & = & - 2 \mbox{Re}\{b_1^2\}
\sum_{k=0}^\infty \frac{2^k}{(k!)^2} \bigg\{  4\Big( t_{\mathrm{c1}}(k) 
+ t_{\mathrm{c6}}(k) + t_{\mathrm{e1}}(k) 
+ t_{\mathrm{e2}}(k) \Big) \nonumber \\
& & \qquad \qquad \qquad \qquad \qquad \qquad \qquad \qquad \qquad \qquad 
\times \ \nabla_{\mu\{k\}\mu}  \, \phi \  
\nabla^{\mu\{k\}}{}_\nu \, \phi \nonumber \\
& &  - \ \Big( t_{\mathrm{c2}}(k) +
2t_{\mathrm{c4}}(k) + t_{\mathrm{c5}}(k) - t_{\mathrm{c6}}(k)
+ t_{\mathrm{c7}}(k) + t_{\mathrm{c8}}(k) 
- 2t_{\mathrm{e1}}(k) \Big) \nonumber \\
& & \qquad \qquad \qquad \qquad 
\times \ \Big( g_{\mu\nu} \, \nabla_{\mu\{k+1\}}  \, \phi \ 
\nabla^{\mu\{k+1\}} \, \phi  + 
2 \nabla_{\mu\{k\}}  \, \phi \  
\nabla^{\mu\{k\}}{}_{\mu\nu} \, \phi \Big)
\nonumber \\
& & - \ \Big( 2k(k+2) \big( t_{\mathrm{c1}}(k) 
+ t_{\mathrm{e2}}(k) \big)  + 2k(k+1)t_{\mathrm{c6}}(k) \nonumber \\
& & \qquad + \ 2k \big( t_{\mathrm{c2}}(k) 
+ 2t_{\mathrm{c4}}(k) + t_{\mathrm{c5}}(k) + t_{\mathrm{c7}}(k) +
t_{\mathrm{c8}}(k) + kt_{\mathrm{e1}}(k) \big) \Big) \nonumber \\
& & \qquad \qquad \qquad \qquad \qquad \qquad \qquad \qquad \qquad 
\times \ \nabla_{\mu\{k-1\}\mu}  \, \phi \  
\nabla^{\mu\{k-1\}}{}_\nu \, \phi 
\nonumber \\
& & \!\!\!\!\!\!\!\!\!\!\!\!\!\!\!\!\!\!\!\! 
- \ \Big( (k+2) \big( t_{\mathrm{c1}}(k) 
+ t_{\mathrm{e2}}(k) \big)  + \frac12 (k^2+3k+4)t_{\mathrm{c6}}(k) 
+ (k^2+2k+2)t_{\mathrm{e1}}(k)\nonumber \\
& & \qquad - \ \frac12 k(k+1) \big( t_{\mathrm{c2}}(k) 
+ 2t_{\mathrm{c4}}(k) + t_{\mathrm{c5}}(k) + t_{\mathrm{c7}}(k) +
t_{\mathrm{c8}}(k)\big)   
\Big) \nonumber \\
& & \qquad \qquad \qquad \qquad \qquad \qquad \qquad \qquad \qquad \quad 
\times \ g_{\mu\nu} \, \nabla_{\mu\{k\}}  \, \phi \  
\nabla^{\mu\{k\}} \, \phi \nonumber \\
& & - \ \Big( 2k \big( t_{\mathrm{c1}}(k) 
+ kt_{\mathrm{e1}}(k) + t_{\mathrm{e2}}(k) \big) 
+ k(k+1) t_{\mathrm{c6}}(k) \nonumber \\
& & \qquad - \ k(k-1) \big( t_{\mathrm{c2}}(k) 
+ 2t_{\mathrm{c4}}(k) + t_{\mathrm{c5}}(k) + t_{\mathrm{c7}}(k) +
t_{\mathrm{c8}}(k) \big) \Big) \nonumber \\
& & \qquad \qquad \qquad \qquad \qquad \qquad \qquad \qquad 
\times \ \nabla_{\mu\{k-1\}}  \, \phi \  
\nabla^{\mu\{k-1\}}{}_{\mu\nu} \, \phi \bigg\}.
\label{eq:term1_kappa}
\end{eqnarray}

%----------------------------------------------------------------

\subsection{The $\widehat 
A^{(1)}_{\alpha} \star \widehat A^{(1)}_{\beta}$ term}
\label{sec:E2}

In order to compute the $\widehat A^{(1)}_{\alpha} \star \widehat
A^{(1)}_{\beta}$ term in (\ref{eq:Ricci_contr}) we note that
\begin{eqnarray}
& & \!\!\!\!\!\!\!\!\!\!\!\!\!
 -\frac12 \Big\{ (\sigma_\mu)^{\gamma\dot\alpha}
(\sigma_\nu)^{\delta\dot\beta} \big(\widehat A^{(1)}_{\dot\alpha} 
\star \widehat A^{(1)}_{\dot\beta} \big)_{\gamma\delta}
\Big\}_{z=0}  \ + \ \mbox{h.c.}  \nonumber \\[5pt]
& = & 
 \frac{i}2 (\sigma_\mu{}^\rho)^{\gamma\delta} \ 2i \Big\{
e_{[\nu}{}^{\alpha\dot\alpha} e_{\rho]}{}^\beta{}_{\dot\alpha} 
\big( \widehat A^{(1)}_\alpha
\star \widehat A^{(1)}_\beta \big)_{\gamma\delta} 
+  e_{[\nu}{}^{\alpha\dot\alpha} e_{\rho]}{}_\alpha{}^{\dot\beta} 
\big( \widehat A^{(1)}_{\dot\alpha}
\star \widehat A^{(1)}_{\dot\beta} \big)_{\gamma\delta} \Big\}_{z=0}
\ + \ \mbox{h.c.}   \nonumber \\[5pt]
& = & - \frac{i}2 (\sigma_\mu{}^\rho)^{\gamma\delta}\,
\Big([\mbox{D2}]_{\nu\rho,
\dot\gamma\dot\delta}\Big)^\dagger
 + \mbox{h.c.} \ 
= \ \frac1{4} \sum_{k=0}^\infty \frac{b_1^2  \,
[\mbox{d1}]_{\mu\nu}(k)}{(k+3)^2(k!)^2} \ + \ \mbox{h.c.}
\nonumber \\[5pt]
& = &
 2 \mbox{Re}\{b_1^2\} \sum_{k=0}^\infty \frac{2^k
t_{\mathrm{d1}}(k)}{(k!)^2} \Big( -4 \, \nabla_{\mu\{k\}\mu}  \, \phi \  
\nabla^{\mu\{k\}}{}_\nu \, \phi \ + \ 
g_{\mu\nu} \, \nabla_{\mu\{k+1\}}  \, \phi \ 
\nabla^{\mu\{k+1\}} \, \phi \Big)
\label{eq:term2_A*A}
\end{eqnarray}
where $t_{\mathrm{d1}}(k) = (k+3)^{-2}/4$. (Here we do not bother to
introduce $f_{\mathrm{D2}}(k)$.)

%----------------------------------------------------------------

\subsection{The $L_{\mu\nu}$ term}
\label{sec:E3}

The contributions to $\mathcal{L}_{\nu\rho}{}^{\alpha\beta}$ and
$\mathcal{L}_{\nu\rho}{}^{\dot\alpha\dot\beta}$ are the A, B and C
terms given in (\ref{eq:Jaa}) and (\ref{eq:Jadad}). Contracting by
$(\sigma_{\mu}{}^{\rho})_{\alpha\beta}$ or
$(\sigma_{\mu}{}^{\rho})_{\dot\alpha\dot\beta}$ we get
\begin{eqnarray}
(\sigma_{\mu}{}^{\rho})_{\alpha\beta}\,
[\mbox{A1.1}]_{\nu\rho}{}^{\alpha\beta} & = & 
\frac14 \sum_{k=0}^\infty \frac{-ib_1^2 k \,
[\mbox{a1}]_{\mu\nu}(k)}{(k+3)^2(k+2)^2(k+1)(k!)^2}
  \nonumber \\
& \equiv & b_1^2 \sum_{k=0}^\infty \frac{f_{\mathrm{A1.1}}(k)}{(k!)^2} \,
[\mbox{a1}]_{\mu\nu}(k)  
\\
(\sigma_{\mu}{}^{\rho})_{\dot\alpha\dot\beta}\,
[\mbox{A4.1}]_{\nu\rho}{}^{\dot\alpha\dot\beta} & = & 
\frac14 \sum_{k=0}^\infty \frac{ib_1^2 k \,
[\mbox{a4}]_{\mu\nu}(k)}{(k+4)^2(k+3)^2(k+1)(k!)^2}
  \nonumber \\
& \equiv & b_1^2 \sum_{k=0}^\infty  \frac{f_{\mathrm{A4.1}}(k)}{(k!)^2}  \,
[\mbox{a4}]_{\mu\nu}(k)  
\\
(\sigma_{\mu}{}^{\rho})_{\alpha\beta}\,
[\mbox{C14.1}]_{\nu\rho}{}^{\alpha\beta} & = & 
-\frac14 \sum_{k=0}^\infty \frac{-ib_1^2  \,
[\mbox{b2}]_{\mu\nu}(k)}{4(k+4)(k+3)^2(k!)^2}
  \nonumber \\
& \equiv & b_1^2 \sum_{k=0}^\infty \frac{f_{\mathrm{C14.1}}(k)}{(k!)^2}  \,
[\mbox{b2}]_{\mu\nu}(k)  
\end{eqnarray}
and so on schematically as
\begin{equation}
(\sigma_{\mu}{}^{\rho})_{\alpha\beta}\,
[\mbox{A{\footnotesize\#}.{\footnotesize\#}}]_{\nu\rho}{}^{\alpha\beta} =
 b_1^2 \sum_{k=0}^\infty \frac{f_{\mathrm{A\#.\#}}(k)}{(k!)^2}  \,
[\mbox{a{\footnotesize\#$'$} or b{\footnotesize\#$'$}}]_{\mu\nu}(k) 
\end{equation}
using the same conventions as before (see the discussion following
(\ref{eq:Ace}) for details). Collecting the
a{\footnotesize\#} and b{\footnotesize\#} contributions we have
\begin{equation}
\frac{i}2 (\sigma_{\mu}{}^{\rho})_{\alpha\beta}
L_{\nu\rho}{}^{\alpha\beta} \ + \ \mathrm{h.c.} 
 =
2 \mbox{Re}\{b_1^2\} \sum_{k=0}^\infty \frac{t_{\mathrm{a}i}(k) 
\, [\mbox{a$i$}]_{\mu\nu}{}(k) + t_{\mathrm{b}j}(k) 
\, [\mbox{b$j$}]_{\mu\nu}{}(k)}{(k!)^2}
\label{eq:L=2Reb^2}
\end{equation}
where summation over $i=1,\ldots,8$ and $j=1,3$ is assumed and
\begin{eqnarray}
t_{\mathrm{a1}}(k) & = & - \frac{i}2 \Big(f_{\mathrm{A1.1}}(k) 
- 2f_{\mathrm{B1.1}}(k) + 2f_{\mathrm{C3.1}}(k) + 2f_{\mathrm{C4.1}}(k) 
   \nonumber \\
& & \quad + \ 2f_{\mathrm{C9.1}}(k) + 2f_{\mathrm{C10.1}}(k) 
+ 2f_{\mathrm{C12.1}}(k) \Big), 
\\
t_{\mathrm{a2}}(k) & = & -\frac{i}2 \Big( f_{\mathrm{A1.2}}(k)
+ f_{\mathrm{A4.3}}(k) + 2f_{\mathrm{C3.2}}(k) + 2f_{\mathrm{C4.2}}(k) 
\nonumber \\
& & \quad + \ 2f_{\mathrm{C6.5}}(k) + 2f_{\mathrm{C7.3}}(k) 
+ 2f_{\mathrm{C9.2}}(k) + 2f_{\mathrm{C10.2}}(k) \nonumber \\
& & \quad + \ 2f_{\mathrm{C12.2}}(k) + 2f_{\mathrm{C13.5}}(k) 
+ f_{\mathrm{C16.5}}(k) \Big),
\\
t_{\mathrm{a3}}(k) & = & -\frac{i}2 \Big( -2f_{\mathrm{B1.2}}(k)
+ 2f_{\mathrm{C3.3}}(k) + 2f_{\mathrm{C4.3}}(k) + 2f_{\mathrm{C9.3}}(k) 
\nonumber \\
& & \quad + \ 2f_{\mathrm{C10.3}}(k) + 2f_{\mathrm{C12.3}}(k) \Big),
\\
t_{\mathrm{a4}}(k) & = & -\frac{i}2 \Big( f_{\mathrm{A4.1}}(k)
+ 2f_{\mathrm{C6.1}}(k) + 2f_{\mathrm{C7.4}}(k) + 2f_{\mathrm{C13.2}}(k) 
\nonumber \\
& & \quad + \ 2f_{\mathrm{C15.1}}(k) +  2f_{\mathrm{C16.1}}(k) \Big),
\\
t_{\mathrm{a5}}(k) & = & -\frac{i}2 \Big( 2f_{\mathrm{C3.4}}(k)
+ 2f_{\mathrm{C4.4}}(k) + 2f_{\mathrm{C7.7}}(k) + 2f_{\mathrm{C9.4}}(k) 
\nonumber \\
& & \quad + 2\ f_{\mathrm{C10.4}}(k) +  2f_{\mathrm{C12.4}}(k) 
+ 2f_{\mathrm{C15.4}}(k) \Big),
\\
t_{\mathrm{a6}}(k) & = & -\frac{i}2 \Big( f_{\mathrm{A4.4}}(k)
- 2f_{\mathrm{B2}}(k) - 2f_{\mathrm{B8.1}}(k) + 2f_{\mathrm{C6.4}}(k) 
\nonumber \\
& & \quad + \ 2f_{\mathrm{C7.2}}(k) +  2f_{\mathrm{C13.1}}(k) 
+ 2f_{\mathrm{C16.4}}(k) \Big),
\\
t_{\mathrm{a7}}(k) & = & -\frac{i}2 \Big( f_{\mathrm{A4.2}}(k)
+ 2f_{\mathrm{C6.2}}(k) + 2f_{\mathrm{C6.3}}(k) + 2f_{\mathrm{C7.1}}(k) 
\nonumber \\
& & \quad + \ 2f_{\mathrm{C7.5}}(k) + 2f_{\mathrm{C13.3}}(k) 
+ 2f_{\mathrm{C13.4}}(k) + 2f_{\mathrm{C15.3}}(k) \nonumber \\
& & \quad + \ 2f_{\mathrm{C16.2}}(k) + 2f_{\mathrm{C16.3}}(k) \Big),
\\
t_{\mathrm{a8}}(k) & = & -\frac{i}2 \Big( 2f_{\mathrm{C7.6}}(k)
+ 2f_{\mathrm{C15.2}}(k) \Big),
\\
t_{\mathrm{b1}}(k) & = & -\frac{i}2 \Big( 2f_{\mathrm{C5}}(k)
+ 2f_{\mathrm{C11}}(k) + 2f_{\mathrm{C14.2}}(k) + 2f_{\mathrm{C17.3}}(k) \Big),
\\
t_{\mathrm{b2}}(k) & = & -\frac{i}2 \Big( -2f_{\mathrm{B3}}(k)
- 2f_{\mathrm{B4}}(k) - 2f_{\mathrm{B8.2}}(k) 
+ 2f_{\mathrm{C14.1}}(k) \nonumber \\
& & \quad + \ 2f_{\mathrm{C17.2}}(k) \Big),
\\
t_{\mathrm{b3}}(k) & = & -\frac{i}2 \Big(+ 2f_{\mathrm{C14.3}}(k)
+ 2f_{\mathrm{C17.1}}(k) \Big).
\end{eqnarray}
From (\ref{eq:L=2Reb^2}) it follows that
\begin{eqnarray}
& & \!\!\!\!\!\!\!\!\!\!\!\!\!\!\!\!\!\!\!\!\!\!\!
\frac{i}2 (\sigma_{\mu}{}^{\rho})_{\alpha\beta}
L_{\nu\rho}{}^{\alpha\beta} \ + \ \mathrm{h.c.}  \ = \nonumber \\
& = & 2 \mbox{Re}\{b_1^2\}
\sum_{k=0}^\infty \frac{2^k}{(k!)^2} \bigg\{ 4\,\Big( t_{\mathrm{a2}}(k) 
- 2t_{\mathrm{a6}}(k) - t_{\mathrm{a7}}(k) 
+ t_{\mathrm{a8}}(k) \Big) \nonumber \\
& & \qquad \qquad \qquad \qquad \qquad \qquad \quad \qquad \quad \quad \quad 
\times \ \nabla_{\mu\{k\}\mu}  \, \phi \  
\nabla^{\mu\{k\}}{}_\nu \, \phi \nonumber \\
& & \quad \quad \quad  + \ 2\Big( -t_{\mathrm{a1}}(k) -t_{\mathrm{a2}}(k) 
+ 2t_{\mathrm{a3}}(k) - 3t_{\mathrm{a4}}(k) \nonumber \\
& & \qquad \qquad \qquad \qquad \qquad \qquad 
- \ t_{\mathrm{a5}}(k) + t_{\mathrm{a6}}(k) 
+ 3t_{\mathrm{a7}} + t_{\mathrm{a8}}(k) \Big) \nonumber \\
& & \qquad \qquad \qquad \qquad \qquad \qquad \qquad \qquad \qquad 
\times \ g_{\mu\nu} \, \nabla_{\mu\{k+1\}}  \, \phi \ 
\nabla^{\mu\{k+1\}} \, \phi
\nonumber \\
& & \qquad \qquad \qquad + \ 4\Big( t_{\mathrm{b1}}(k) 
 - 2t_{\mathrm{b2}}(k) \big)  + t_{\mathrm{b3}}(k)  \Big) \nonumber \\
& & \qquad \qquad \qquad \qquad \quad \qquad \qquad \qquad 
\times \   + 
2 \nabla_{\mu\{k\}}  \, \phi \  
\nabla^{\mu\{k\}}{}_{\mu\nu} \, \phi \bigg\} .
\label{eq:term3_L}
\end{eqnarray}

%----------------------------------------------------------------

\subsection{The $\xi$, $\eta$ and $\zeta$ functions}

We are now ready to express $\xi(k)$, $\eta(k)$ and $\zeta(k)$,
defined in (\ref{eq:smakprov}), in terms of the $t$-functions. To this
end we begin by rewriting (\ref{eq:smakprov}) as
\begin{eqnarray}
\Ric_{\mu\nu} + 3g_{\mu\nu} & = & \mbox{Re}\{b_1^2\} \bigg[ 
\sum_{k=0}^\infty \frac{2^k}{(k!)^2}
\bigg( \Big(-\xi(k) -\frac12 \eta(k) \Big) 
\, g_{\mu\nu} \nabla_{\rho\{k+1\}}  \, \phi \  
\nabla^{\rho\{k+1\}} \, \phi + \nonumber \\
& & \qquad \qquad \qquad \qquad 
+ \ \eta(k) \nabla_{\rho\{k\}\mu}  \, \phi \  
\nabla^{\rho\{k\}}{}_\nu \, \phi  + \nonumber \\
& & \qquad \qquad \qquad \quad 
+ \ \zeta(k) \nabla_{\rho\{k\}\mu\nu}  \, \phi \  
\nabla^{\rho\{k\}} \, \phi \bigg)  \
- \ \frac12\theta \ g_{\mu\nu}\, \phi\,\phi \bigg]
\label{eq:Ricci_xietazeta},
\end{eqnarray}
where the right hand side can now be identified with the sum of
(\ref{eq:term1_kappa}), (\ref{eq:term2_A*A}) and
(\ref{eq:term3_L}). We find
\begin{eqnarray}
-\xi(k)-\frac12 \eta(k) & = & 
4\,\Big(-t_\mathrm{a1}(k) -t_\mathrm{a2}(k) 
+2t_\mathrm{a3}(k) -3t_\mathrm{a4}(k) -t_\mathrm{a5}(k) \nonumber \\
& & \qquad  +t_\mathrm{a6}(k) +3t_\mathrm{a7}(k)
+t_\mathrm{a8}(k) + \frac12 t_\mathrm{d1}(k) \Big) \nonumber \\
& & + \ 2\, \Big( t_\mathrm{c2}(k) + 
2t_\mathrm{c4}(k)  +
t_\mathrm{c5}(k) - t_\mathrm{c6}(k)  \nonumber \\
& & \qquad \qquad
+ t_\mathrm{c7}(k) +
t_\mathrm{c8}(k) -2t_\mathrm{e1}(k) \Big) \nonumber \\
& & + \ 4\,\frac{k+3}{(k+1)^2}\Big( t_\mathrm{c1}(k+1) 
 +t_\mathrm{e2}(k+1)  \Big) \nonumber \\
& & - \ 2 \,\frac{k+2}{k+1} \Big( 
t_\mathrm{c2}(k+1) + 2t_\mathrm{c4}(k+1) +
t_\mathrm{c5}(k+1) \nonumber \\
& & \qquad \qquad \qquad + t_\mathrm{c7}(k+1) +
t_\mathrm{c8}(k+1) \Big) \nonumber \\
& & + \ 2 \, \frac{k^2+5k+8}{(k+1)^2}  
t_\mathrm{c6}(k+1) \nonumber \\
& & + \ 4 \, \frac{k^2+4k+5}{(k+1)^2}  t_\mathrm{e1}(k+1),
\end{eqnarray}
\begin{eqnarray}
\eta(k) & = & 8\,\Big( t_\mathrm{a2}(k) - 2t_\mathrm{a6}(k)
- t_\mathrm{a7}(k) + t_\mathrm{a8}(k) 
- t_\mathrm{d1}(k) \Big) \nonumber \\
& & - \ 8 \,\Big( t_\mathrm{c1}(k) +t_\mathrm{c6}(k)
 +t_\mathrm{e1}(k) +t_\mathrm{e2}(k)\Big) \nonumber \\
& & + \ 8 \,\frac{k+3}{k+1} \Big( 
t_\mathrm{c1}(k+1) +t_\mathrm{e2}(k+1) \Big) \nonumber \\
& & + \ 8 \,\frac1{k+1} \Big( 
t_\mathrm{c2}(k+1) + 2t_\mathrm{c4}(k+1) +
t_\mathrm{c5}(k+1) \nonumber \\
& & \qquad \qquad \qquad + t_\mathrm{c7}(k+1) +
t_\mathrm{c8}(k+1) \Big)  \nonumber \\
& & + \ 8 \,\frac{k+2}{k+1} t_\mathrm{c6}(k+1)
\ + \ 8 \,t_\mathrm{e1}(k+1),
\end{eqnarray}
\begin{eqnarray}
\zeta(k) & = & 8\,\Big(t_\mathrm{b1}(k) - 2t_\mathrm{b2}(k) + 
t_\mathrm{b3}(k) \Big)  \nonumber \\
& & + \ 4 \,\Big( t_\mathrm{c2}(k) +2t_\mathrm{c4}(k) 
+ t_\mathrm{c5}(k) -  t_\mathrm{c6}(k) \nonumber \\
& & \qquad \qquad + t_\mathrm{c7}(k)
+ t_\mathrm{c8}(k) - 2t_\mathrm{e1}(k)  \Big)  \nonumber \\
& & + \ 8 \,\frac{1}{k+1}\Big( t_\mathrm{c1}(k+1) 
 +t_\mathrm{e2}(k+1)   \Big)  \nonumber \\
& & - \ 4 \,\frac{k}{k+1} \Big( 
t_\mathrm{c2}(k+1) + 2t_\mathrm{c4}(k+1) +
t_\mathrm{c5}(k+1) \nonumber \\
& & \qquad \qquad \qquad + t_\mathrm{c7}(k+1) +
t_\mathrm{c8}(k+1) \Big) \nonumber \\
& & + \ 4\, \frac{k+2}{k+1} t_\mathrm{c6}(k+1) \ + \ 
8 \, t_\mathrm{e1}(k+1).
\end{eqnarray}
and
\begin{equation}
-\frac12 \theta = 4 t_\mathrm{e1}(0)
\end{equation}
Upon substituting the explicit values for the $t$-functions given in
Sections \ref{sec:E1}, \ref{sec:E2} and \ref{sec:E3}, we arrive at
(\ref{eq:xi})--(\ref{eq:zeta}).

%==========================================================

\end{document}